\documentclass[prl,twocolumn,aps,superscriptaddress,10pt]{revtex4-1}
\pdfoutput=1
\usepackage[utf8]{inputenc}
\usepackage{graphicx}
\usepackage{color}
\usepackage{xcolor}
\usepackage{float}
\usepackage[bookmarks=false,linkcolor=blue,urlcolor=blue,colorlinks,citecolor=blue]{hyperref}
\usepackage{amsmath,mathtools}
\usepackage{dsfont}
\usepackage{amssymb}
\usepackage{soul}
\usepackage{silence}
\usepackage{multirow}
\usepackage{pdfpages}
\makeatletter
\AtBeginDocument{\let\LS@rot\@undefined}
\makeatother
\newcommand{\one}{\mathds{1}}
\newcommand{\Eq}[1]{Eq.~(\ref{#1})}

\newcommand{\Refe}[1]{Ref.~\cite{#1}}
\newcommand{\Refs}[1]{Refs.~\cite{#1}}

\newcommand{\Tab}[1]{Tab.~(\ref{#1})}
\newcommand{\BrackEq}[1]{[Eq.~(\ref{#1})]}

\newcommand{\Fig}[1]{Fig.~\ref{#1}}
\newcommand{\FigPart}[2]{Fig.~\ref{#1}(#2)}

\newcommand{\BrackFigPart}[2]{[Fig.~\ref{#1}(#2)]}

\newcommand{\BrackFigsPart}[2]{[Figs.~\ref{#1}(#2)]}
\newcommand{\Figs}[1]{Figs.~\ref{#1}}

\newcommand{\nbrack}[1]{\left(#1\right)}
\newcommand{\sqbrack}[1]{\left[#1\right]}
\newcommand{\lrsepa}{\quad,\quad}
\newcommand{\epsi}{\epsilon_i}
\newcommand{\epsSD}{\epsilon_{\text{S}}}
\newcommand{\eps}{\epsilon}
\newcommand{\Ecap}{E_{\text{cap}}}
\newcommand{\muL}{\mu_{\text{L}}}
\newcommand{\muR}{\mu_{\text{R}}}
\newcommand{\IL}{I_{\text{L}}}

\newcommand{\nd}{n}
\newcommand{\Ni}{n_i}
\newcommand{\nSD}{n_{\text{S}}}
\newcommand{\HdSC}{H_{\text{d,SC}}}
\newcommand{\ndSC}{n_{\text{d,SC}}}
\newcommand{\fpdSC}{(-\one)^{n_{\text{d,SC}}}}
\newcommand{\tQP}{t_{\text{QP}}}

\newcommand{\tbw}{t_{\text{b}}}
\newcommand{\Nosc}{N_{\text{osc}}}
\newcommand{\delt}{\delta t}
\newcommand{\delV}{\delta\mu}
\newcommand{\lameff}{\lambda_{\text{eff}}}
\newcommand{\lami}{\lambda_i}
\newcommand{\lamj}{\lambda_j}
\newcommand{\lamp}{\lambda^+}
\newcommand{\lamm}{\lambda^-}
\newcommand{\phii}{\phi_i}
\newcommand{\phij}{\phi_j}
\newcommand{\eiph}{e^{i\phi_i}}
\newcommand{\ui}{|u_i|}
\newcommand{\vi}{|v_i|}
\newcommand{\ali}{\alpha_i^{{}}}
\newcommand{\aldi}{\alpha^\dagger_i}
\newcommand{\al}{\alpha}
\newcommand{\ald}{\alpha^\dagger}
\newcommand{\fpdal}{(-\one)^{\nd + \ald\al}}
\newcommand{\dopdag}{d^\dagger}
\newcommand{\epssg}{\epsilon_{\text{sg}}}
\newcommand{\peff}{p_{\text{d}\alpha}}
\newcommand{\pdSC}{p_{\text{d,SC}}}

\newcommand{\Hcap}{H_{\text{cap}}}
\newcommand{\sig}{\mathcal{S}}
\newcommand{\dis}{\mathcal{D}}
\newcommand{\Ge}{G_{\text{e}}}
\newcommand{\Go}{G_{\text{o}}}
\newcommand{\Geo}{G_{\text{e/o}}}

\newcommand{\dGeo}{\delta G_{\text{e/o}}}
\newcommand{\tausp}{t_{\text{p}}}
\newcommand{\Qfac}{\mathcal{Q}}
\begin{document}
\title{Detecting Majorana modes by readout of poisoning-induced parity flips}

\author{Jens Schulenborg}
\affiliation{
  Center for Quantum Devices, Niels Bohr Institute, University of Copenhagen, 2100 Copenhagen, Denmark
}
\author{Svend Kr{\o}jer}
\affiliation{
  Center for Quantum Devices, Niels Bohr Institute, University of Copenhagen, 2100 Copenhagen, Denmark
}
\author{Michele Burrello}
\affiliation{
  Center for Quantum Devices, Niels Bohr Institute, University of Copenhagen, 2100 Copenhagen, Denmark
}
\affiliation{
 Niels Bohr International Academy, Niels Bohr Institute, University of Copenhagen, 2100 Copenhagen, Denmark
}
\author{Martin Leijnse}
\affiliation{
    Solid State Physics and NanoLund, Lund University, Box 118, S-221 00 Lund, Sweden
}
\affiliation{
  Center for Quantum Devices, Niels Bohr Institute, University of Copenhagen, 2100 Copenhagen, Denmark
}
\author{Karsten Flensberg}
\affiliation{
  Center for Quantum Devices, Niels Bohr Institute, University of Copenhagen, 2100 Copenhagen, Denmark
}

\begin{abstract}
Reading out the parity degree of freedom of Majorana bound states is key to demonstrating their non-Abelian exchange properties. Here, we present a low-energy model describing localized edge states in a two-arm device. We study parity-to-charge conversion based on coupling the superconductor bound states to a quantum dot whose charge is read out by a sensor. The dynamics of the system, including the readout device, is analyzed in full using a quantum-jump approach. We show how the resulting signal and signal-to-noise ratio differentiates between local Majorana and Andreev bound states.
\end{abstract}
\maketitle

Topological superconductors host Majorana zero-energy modes~\cite{Kitaev2001Oct,DasSarma2005Apr} that store quantum information nonlocally, and are thereby in principle protected against local perturbations. Many protocols to detect this nonlocal storage have been theorized in fractional quantum Hall systems \cite{Stern2008Jan} and, more recently, in superconducting wires \cite{Hassler2010Dec,Alicea2011Feb,Flensberg2011Mar,Hassler2011Sep,Hyart2013Jul,Aasen2016Aug,Plugge2017Jan,Karzig2017Jun,Manousakis2017Apr}. The latter are inspired by the idea of engineered topological superconductors, many of which have been realized experimentally.
Transport spectroscopy and interference~\cite{Fu2010Feb,Hell2018Apr,Drukier2018Oct,Whiticar2020Jun,Pikulin2021Mar,Cayao2021Jul,Aghaee2022Jul} indeed suggests the presence of Majoranas in such systems, but differentiating between topological and trivial states remains challenging~\cite{Prada2012Nov,Kells2012Sep,Cayao2015Jan,Moore2018Oct,Vuik2019Nov,Pan2020Mar,Hess2021Aug,FlensbergvonOppenStern2021}. Most importantly, however, transport via the states of interest themselves \emph{violates the conservation of fermionic parity} essential to most topological qubit proposals.

\begin{figure}[t!!]
\includegraphics[width=\linewidth]{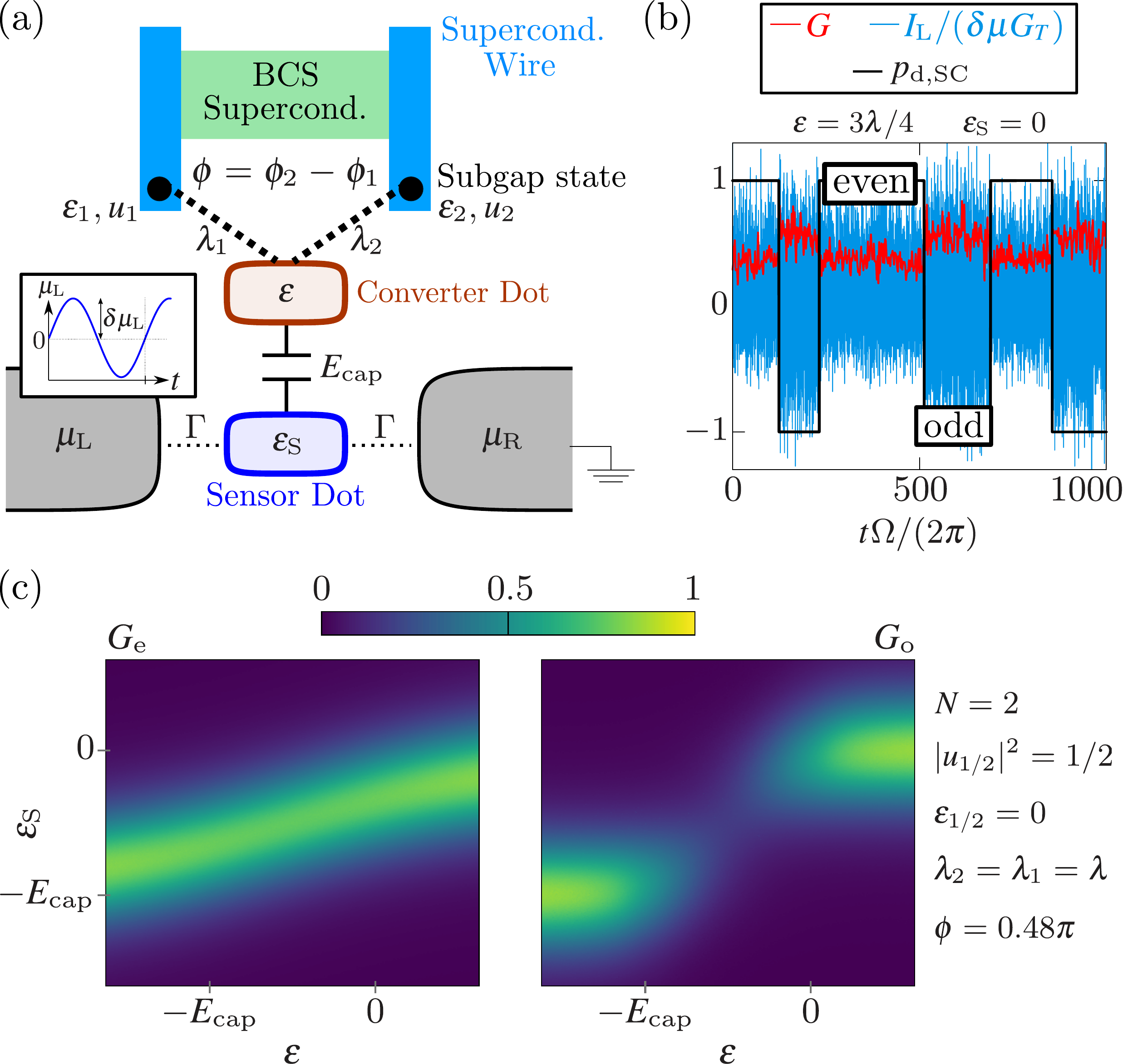}
\caption{(a) Subgap modes in two superconducting wires tunnel-coupled to a parity-to-charge converter dot \BrackEq{eq_hamiltonian_dSC} that capacitively couples to a charge sensor. (b) Sensor current $\IL(t)$ due to driven $\muL$, corresponding first harmonic $G(t)$ \BrackEq{eq_response_trajectory}, and dot-subgap parity $\pdSC$. (c) Sample-averaged conductances $\Geo = G_{\pdSC = \pm1}$ scaled by $G_T = \Gamma/(8T)$. (b) shares parameters with (c); other parameters are from \Fig{fig_zero}.\label{fig_setup}}
\end{figure}

The key question addressed in this Letter is thus how to test for Majoranas directly via the conserved parity of nonlocal Majorana pairs, using the same readout device as in the intended quantum information application. Typical parity measurement schemes are theorized to rely on a conversion to a charge or magnetic flux when the Majorana modes overlap~\cite{Hassler2010Dec,Flensberg2011Mar,Hyart2013Jul,Ohm2015Feb,Aasen2016Aug,Gharavi2016Oct,Plugge2017Jan,Karzig2017Jun,Malciu2018Oct,Li2018Nov,Schrade2018Dec,Grimsmo2019Jun,Szechenyi2020Jun,Munk2020Aug,Steiner2020Aug,Smith2020Nov}.
Here, we consider parity-to-charge conversion with a quantum dot coupled to the subgap end states of superconducting wires, such that the dot charge measures the combined dot-subgap parity~\cite{Flensberg2011Mar,Plugge2017Jan,Karzig2017Jun,Munk2020Aug,Steiner2020Aug,Schulenborg2021Jun}.
Integrating dots into semiconductor-superconductor structures is experimentally well developed~\cite{Hofstetter2009Oct,DeFranceschi2010Oct,Deng2014Dec,Deng2016Dec,Szombati2016Jun,Deng2018Aug,vanVeen2019Nov,Razmadze2020Sep}:
Couplings are accurately tuned via gate voltages, and charge readout is performed via electromagnetic resonators~\cite{Yoshie2004Nov,Reithmaier2004Nov,Delbecq2011Dec,Frey2012Jan,Petersson2012Oct,Xiang2013Apr,Stockklauser2017Mar,Burkard2020Mar,Deng2020Apr}, sensor dots~\cite{Schoelkopf1998May,Lu2003May,Fujisawa2004Mar,Bylander2005Mar,Buehler2005Apr,Barthel2010Apr,Maisi2011May}, or quantum point contacts ~\cite{Field1993Mar,Elzerman2003Apr,Elzerman2004Jul,Ihn2009Sep,Barthel2009Oct}.

However, while the parity-to-charge conversion principle is well established, its implementation raises many fundamental questions which we answer in this Letter. First, can the charge detection, even in principle, differentiate Majorana modes from Andreev bound states? Second, do readout fluctuations provide additional information about these modes? And third, how do the coupling strengths affect the detection scheme?

We concretely study the system in Fig.~\ref{fig_setup}(a): Two superconducting wires with one or several subgap states at their ends (black dots) are tunnel coupled to a parity-to-charge converter dot (CD). The stable charge of this dot changes with subgap parity flips~\cite{Flensberg2011Mar}, which we assume to be simply due to rare, but inevitable quasi-particle (QP) poisoning~\cite{Zgirski2011Jun,Janvier2015Sep,Higginbotham2015Dec,Hays2018Jul,Karzig2021Feb,Wesdorp2021Dec} on a time scale $\tQP$. As the occasional CD charge jumps affect a capacitively coupled sensor dot (SD), the parity flips are measurable via telegraph noise in the zero-bias SD conductance $G$ across two tunnel coupled leads \BrackFigPart{fig_setup}{b}.
We calculate this conductance by explicitly mimicking the experimental lock-in charge readout technique~\cite{Schoelkopf1998May,Razmadze2019Jun}: Applying a small ac-voltage at zero dc bias, the equal-frequency first harmonic current response provides a measure for $G$. The strength and unique feature of our approach is to quantify the conductance as directly sampled from the experimentally accessible readout signal $G(t)$, including SD back-action via a consistent quantum master equation ~\cite{Kirsanskas2018Jan,Nathan2020Sep,Munk2020Aug,Schulenborg2021Jun} in a quantum-jump Monte-Carlo simulation ~\cite{Molmer1993Mar,Plenio1998Jan,Daley2014Mar,Schulenborg2022Suppmat}.
We thereby capture the effect of measurement induced relaxation, thermal- and non-equilibrium noise, and of the lock-in drive as well as the readout integration time. This includes possible signal loss due to state projections away from the CD-subgap ground state~\cite{Flensberg2011Mar,Munk2020Aug,Steiner2020Aug,DerakhshanMaman2020Dec}.

The low-energy Hamiltonian for the single-level CD with energy $\eps$, $N$ superconducting subgap modes $i$ at energies $\epsi$ in either wire, and the CD-subgap tunnel couplings $\lambda_{i=1,\dotsc,N}$ reads $(|e| = \hbar = k_{\text{B}} = 1)$
\begin{equation}
\HdSC = \eps \nd + \sum_{i=1}^N\epsi \Ni + \sum_{i=1}^N\lami \eiph\dopdag\sqbrack{\ui\ali + \vi\aldi} + \text{H.c.}\label{eq_hamiltonian_dSC}
\end{equation}
Here, $\nd = \dopdag d$ is the CD occupation with fermionic creation/annihilation operator $\dopdag,d$; the subgap state occupations $n_i =\aldi\ali$ are likewise associated with the creation/annihilation operators $\aldi,\ali$.
To justify the single-level dot picture, we assume both the CD single-particle level spacing and onsite Coulomb interaction to be large compared to the tunneling amplitudes $\lami$~\Refe{Schulenborg2021Jun}. This implies sufficient CD-spin polarization, motivated by the large magnetic field needed for Majorana modes~\cite{Lutchyn2010Aug,Oreg2010Oct}. We, however, allow for subgap modes with $\epsi \neq 0$, with spin-axis orientations different from the CD~\cite{Sticlet2012Mar,Kjaergaard2012Jan}, and with (normalized, $\ui^2 + \vi^2 = 1$) couplings featuring unequal particle- and hole components $\ui \neq \vi$ and mode-dependent phases $\phii \neq \phij$.
Unlike previous works~\cite{Flensberg2011Mar,Plugge2017Jan,Munk2020Aug,Steiner2020Aug,Schulenborg2021Jun} which assumed a dot coupling to only one Majorana per wire, we thus represent $\HdSC$ in terms of fermionic fields $\aldi,\ali$ instead of Majorana operators. These fermions are gauged to obtain $\lami \geq 0$, so phases enter \Eq{eq_hamiltonian_dSC} exclusively via $\eiph$. These depend on uncontrollable CD/subgap wave function details, but in the important case of one fermion per wire, the only relevant phase difference $\phi_2 - \phi_1 = \phi$ is flux($\phi$)-tunable.

The model \eqref{eq_hamiltonian_dSC} implies quantum fluctuations in all occupations $\nd,\Ni$ and their sum $\ndSC = \nd + \sum_{i=1}^N\Ni$, but leaves the combined parity $\pdSC  = \fpdSC$ a good quantum number. Our central question is how this parity $\pdSC$  --- converted to the CD charge ($\langle n\rangle$) affecting the SD conductance \BrackFigsPart{fig_setup}{a,b} --- distinguishes between finite- and zero-energy Andreev- and Majorana wire modes. The capacitive coupling $\Hcap = \Ecap\nSD\nd$ to the SD charge $\nSD$ is quantified by $\Ecap$.
The SD tunnel-couples to metallic non-interacting leads, assuming symmetric, energy-independent tunneling rates $\Gamma_{\text{L/R}} = \Gamma$ ~\cite{Jauho1994Aug,Schulenborg2022Suppmat}. The Supplemental Material~\cite{Schulenborg2022Suppmat} details how we obtain the time-resolved conductance from the quantum jump method~\cite{Molmer1993Mar,Plenio1998Jan,Daley2014Mar}, with universal Lindblad operators~\cite{Kirsanskas2018Jan,Nathan2020Sep,Munk2020Aug,Schulenborg2021Jun} applicable even for the here relevant near-degenerate states. We assume weak coupling compared to the lead temperature, $\Gamma \ll T$, and lead-internal relaxation as the shortest time scale in the problem.

In brief, our method yields current-time series admitting sample averaging. After each numerical time step $\delt \ll \Gamma^{-1}$ of each series, we record the accumulated number of electron jumps $J_{r\eta}(t)$ to ($\eta = +$) and from ($\eta = -$) lead $r$, and calculate the current $\IL(t) = \sum_{\eta = \pm}(\eta/\tbw)\sqbrack{J_{\text{L}\eta}(t) - J_{\text{L}\eta}(t-\tbw)}$ with the bandwidth $(1/\tbw) < \Gamma$ reflecting detector-internal time averaging. Mimicking the experimental lock-in technique, the zero-bias conductance signal $G(t)$ is extracted from $\IL(t)$ by applying a low-amplitude voltage oscillation $\muL(t) = \muR + \delV\sin\left(\Omega t\right)$ with frequency $(1/\tQP) \ll \Omega \ll (1/\tbw)$, and by taking the first harmonic response divided by $\delV \lesssim T$. Explicitly, each $G(t)$-sample averages over $\Nosc$ voltage oscillations,
\begin{equation}
 G(t) = \frac{\Omega\delt/G_T}{\pi\Nosc\delV}\left|\sum_{n=0}^{\lceil\frac{2\pi\Nosc}{\Omega\delt}\rceil - 1}\IL(t + n\delt)e^{-in\Omega\delt}\right|\label{eq_response_trajectory},
\end{equation}
where $G_T = \Gamma/(8T)$ is the spinless on-resonance conductance~\cite{Jauho1994Aug}.
Figure~\ref{fig_setup}(b) shows the difference between the raw conductance time series $\IL(t)/(\delV G_T)$ and the first harmonic response. The sudden jumps causing the telegraph noise stem from randomly inserted $\pdSC$ flips~\cite{Zgirski2011Jun,Janvier2015Sep,Higginbotham2015Dec,Hays2018Jul,Karzig2021Feb,Wesdorp2021Dec}. Just as in experiments, $G(t)$ filters out high frequencies via the Fourier transform, bounding the noise spectrum not by the inaccessible bandwidth $1/\tbw$, but by the well-controlled lock-in frequency $\Omega$.

The even/odd-parity conductances $\Geo = \langle G \rangle_{M,\pdSC = \pm 1}$ in \FigPart{fig_setup}{c} are averages over $M = N_{\text{T}}M_{\text{T}}$ samples from $N_{\text{T}}$ trajectories $G(t)$ \emph{at fixed $\pdSC = \pm1$}; each trajectory consists of $M_{\text{T}}$ subsequent samples in a time $\Delta t = M_{\text{T}}\Nosc\frac{2\pi}{\Omega} \leq \tQP$ expected to be within two QP poisonings~\cite{Schulenborg2022Suppmat}, as exemplified by the typical $G$-plateau length in \FigPart{fig_setup}{b}. The $\Geo$ and their fluctuations $\dGeo = \sqrt{\langle G^2 \rangle_{M,\pdSC = \pm 1} - G^2_{\text{e/o}}}$ define the signal $\sig$ and signal-to-noise ratio (SNR) $\dis$:
\begin{equation}
 \sig = |\Ge - \Go| \lrsepa \dis = 2\sig/(\delta \Ge + \delta \Go).\label{eq_parity_signal}
\end{equation}
The ratio $\dis$ as a function of the tunable CD level $\epsilon$ and flux $\phi$ is our key observable to characterize the subgap states, as it sets the number of $G$-samples required for statistically significant parity distinguishability. We, however, also refer to $\sig$, mostly based on data in the Supplemental Material~\cite{Schulenborg2022Suppmat}, to rule out or identify any nontrivial scaling between noise and signal. This furthermore allows us to estimate how the distinguishability would diminish with additional noise unaccounted for here.

The detector setup and SD level $\epsSD$ for optimal $\dis$ depends on both $\Ecap$ and $\lami$~\cite{Schulenborg2022Suppmat}, since $\sig,\dis > 0$ are due to $\pdSC$-dependent CD-subgap hybridization inducing a $\pdSC$-dependent conductance peak shift away from the classical resonances $\epsSD = 0,-\Ecap$. We here focus on $\Ecap > \lami$, yielding nearly classical $\Go$-peaks and sizably deviating $\Ge$ \BrackFigPart{fig_setup}{c}.
The largest $\sig$ then appears between Coulomb peaks, $\epsSD = -\Ecap/2$, but this point is often susceptible to the here neglected higher-order-$\Gamma$ effects such as Kondo resonances. We hence instead fix $\epsSD = 0$ and reduce lead-induced broadening by demanding $\Gamma,T \ll \Ecap$. Larger $\frac{\Gamma}{T},\frac{\delV}{T}$ tend to improve $\dis$~\cite{Schulenborg2022Suppmat}, but we must keep $\Gamma$, $\delV$, and especially $T$ smaller than $\lami$; otherwise, capacitive back-action may drive the system too far away from the near-ground state in \Fig{fig_setup}(c) to obtain a signal~\cite{Steiner2020Aug,Schulenborg2021Jun}.

\begin{figure}[t!!]
\includegraphics[width=\linewidth]{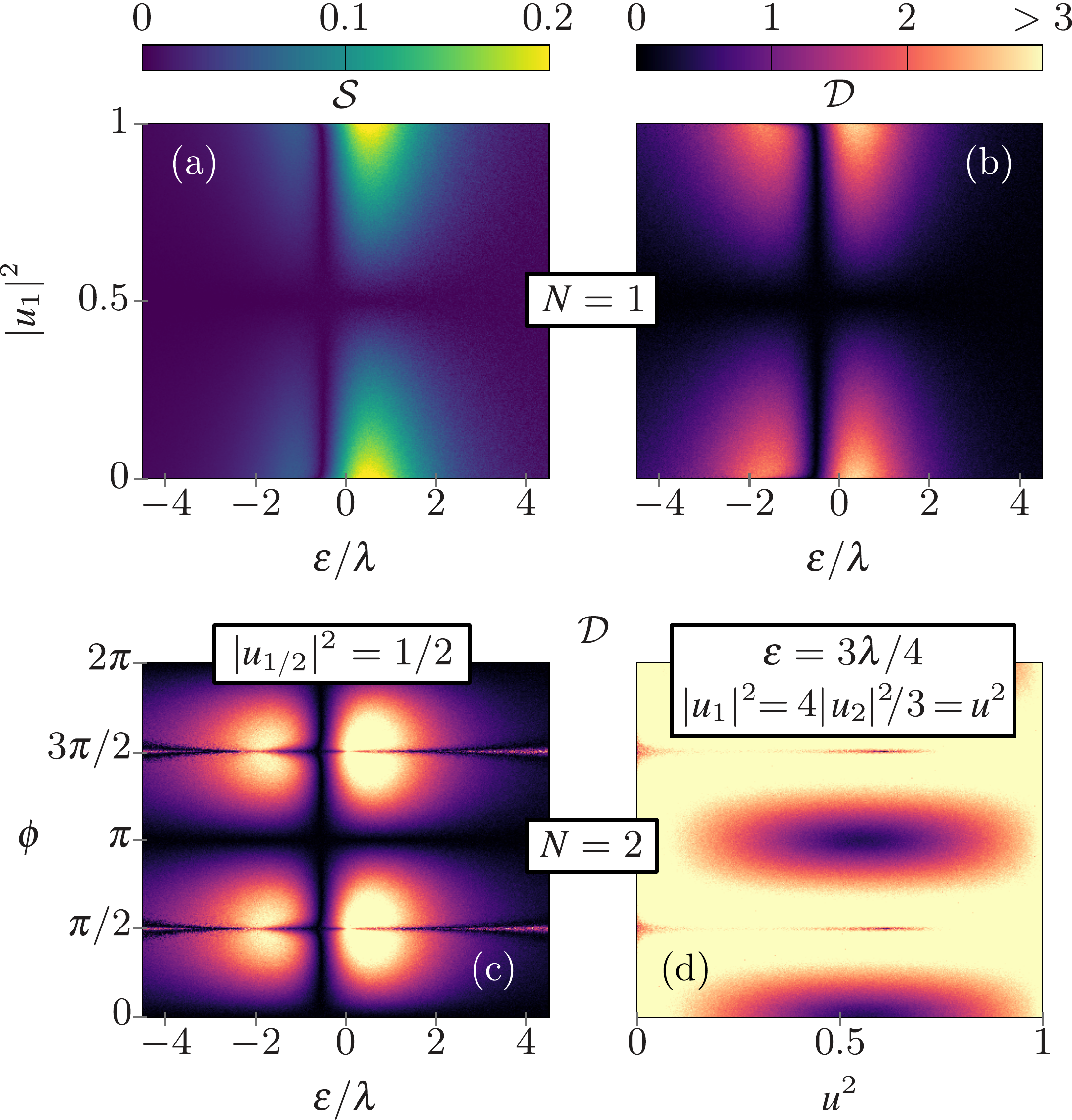}
\caption{Signal $\sig$ and signal-to-noise ratio $\dis$ \BrackEq{eq_parity_signal} as a function of various parameters for $N = 1$ (a,b), and $N = 2$ (c,d) zero-energy subgap modes. All panels use $\epsi = 0$, $\lambda_i = \lambda$, $\epsSD = 0$, $\Ecap = 2\lambda = 6T = 6\delV = 20\Gamma$, $2\pi/\Omega = 4\tbw = 2\pi\times 2400/\lambda =  10^4\delt = 1\mu\text{s}$, $\Nosc = 2$, $N_{\text{T}} = 10$, $M_{\text{T}} = 50$.\label{fig_zero}}
\end{figure}
Let us now first discuss the ideal limit in \Fig{fig_zero}, with all subgap energies $\epsi/\lami \rightarrow 0$. The CD then effectively couples to only one subgap fermion~\cite{Schulenborg2022Suppmat}, being a linear combination of all $N$ subgap modes in \Eq{eq_hamiltonian_dSC}:
\begin{equation}
 \HdSC = \eps n + \left[\dopdag\nbrack{\lamp\ald + \lamm\alpha} + \text{H.c.}\right],\label{eq_hamiltonian_dSC_single_mode}
\end{equation}
where $\alpha$ is a fermionic annihilation operator and
\begin{equation}
 \lambda^\pm = \lameff\sqrt{1 \pm \sqrt{1 - |\Qfac|^2}} \lrsepa \Qfac = \sum_{i=1}^N\frac{\lami^2}{\lameff^2}e^{i2\phii}\ui\vi \label{eq_single_mode_coupling}
\end{equation}
with $0 \leq |\Qfac| \leq 1$ and effective coupling $\lameff = \sqrt{\sum_{i=1}^N\lami^2/2}$. The capacitive readout likewise only senses the subparity of this one $\alpha$-mode and the CD, $\peff = \fpdal$; the other $N-1$ orthogonal subgap modes are invisible. For this single remaining mode, $\Qfac$ generalizes the Majorana quality factor of \Refs{Clarke2017Nov,Prada2017Aug}: Interpolating between a pure particle- or hole-like ($|\Qfac| = 0$), and an equally particle- and hole like coupling ($|\Qfac| = 1$), the latter crucially maps to the single-Majorana case:
\begin{equation}
\HdSC \overset{|\Qfac| = 1}{\rightarrow} \eps\nd + \lameff\nbrack{\dopdag - d}\gamma,\label{eq_single_majorana}
\end{equation}
with $\gamma = \ald + \al$. The sensor is then insensitive to $\peff$~\cite{Flensberg2011Mar}, and yields $\sig = \dis = 0$ for any $\eps,\lami$ within the single-level CD approximation. We emphasize that for this $(\eps,\lami)$-independent parity insensitivity, $\epsi/\lami \rightarrow 0$ and $|\Qfac| = 1$ are not only sufficient, but necessary~\cite{Schulenborg2022Suppmat}. Following \Eq{eq_single_mode_coupling}, this means that all $N$ subgap modes $i$ in \Eq{eq_hamiltonian_dSC} must couple as zero-energy Majoranas to the CD, $\ui = \vi$, and with equal $\phii$ up to a $\pi$-shift.

The key Majorana signature deriving from \Eq{eq_single_majorana} are constant $\sig,\dis = 0$ with varying individual conductances $\Geo$ in sweeps of the dot level $\eps$ and, if possible, the coupling strengths $\lami$: Constant $\Geo$ may merely indicate insufficient sensor coupling, and if $\sig,\dis = 0$ only for specific $\eps,\lami$, one can neither rule out fine-tuning unrelated to Majoranas, nor the quasi-Majorana case~\cite{Vuik2019Nov} with a coincidentally uncoupled zero mode. Importantly, such a sweep test is inherently robust to fluctuating $\eps,\epsSD,\lami$ due to, e.g., $1/f$ noise~\cite{Knapp2018Mar,Mishmash2020Feb,Khindanov2021Jun}, and towards unavoidable coupling asymmetries $\lami \neq \lamj$~\cite{Schulenborg2022Suppmat}.

Table \ref{tab_summary} lists the subgap mode setups for which the suggested prescription can or cannot yield $(\eps,\lami)$-independent $\sig,\dis = 0$. The simplest case involves only one wire with only one $(N = 1)$ coupled mode, see \cite{Leijnse2012Oct,Prada2017Aug,Clarke2017Nov,Dvir2023Feb}. For this $\phi$-independent situation, \Figs{fig_zero}(a,b) show that all $\eps$-traces approach $\sig,\dis = 0$ close to the particle-hole symmetry point $\eps = 0$, but only a Majorana ($|u_1|^2 = 0.5 \Rightarrow$ \Eq{eq_single_majorana}) yields $\sig = \dis = 0$ for all $\eps$.
This robustness is equivalent to the single-site protection in the minimal Kitaev chain~\cite{Leijnse2012Oct,Dvir2023Feb}, but using it to identify a Majorana in a single, \emph{actual} wire is difficult. First, it relies on the SNR in the Andreev case $|u_1|^2 \neq 0.5$: The parameters in \Fig{fig_zero}(b) yield $\dis > 1$ if $||u_1|^2 - 0.5| \gtrsim 0.1$, which may improve if
$\epsSD$, $\Ecap$, $\Gamma$ are optimizable within the stated constraints. But more importantly, convincing evidence would also include tunability towards a control case with $\sig,\dis > 0$ for some $\eps$ \emph{at fixed} $\epsilon_1/\lambda_1 \rightarrow 0$.

The latter is provided when both wires couple, making $\sig,\dis$ flux($\phi$)-tunable as shown for one mode per arm $(N = 2)$ in \Figs{fig_zero}(c,d). For Majoranas ($|u_{1/2}|^2 = 0.5$), $\eps$-independent $\sig,\dis = 0$ are seen exclusively at $\phi = 0,\pi$ fulfilling \Eq{eq_single_majorana}, and $\dis \gtrsim 1$ already if $\phi$ deviates by $\sim\pi/20$ with the detector parameters in \FigPart{fig_zero}{c}. Given instead at least one zero-energy Andreev mode $(\ui^2 \neq 0.5)$, $\eps$-independent traces $\sig,\dis = 0$ cannot be observed for any $\phi$, with \FigPart{fig_zero}{d} exhibiting $\dis \geq 1$ for $|\ui^2 - 0.5| \gtrsim 0.1$ close to $\phi = 0,\pi$. A $(\eps,\phi)$-profile as in \FigPart{fig_zero}{c} thus strongly indicates Majorana modes in both wires.

Moreover, while the steep $\phi$-profile increases the susceptibility to flux noise, it also enhances the sensitivity towards additional, typically unwanted zero-energy modes, $N > 2$. This is because even for Majoranas, $\ui^2 = 0.5$, the flux cannot generally fix the $\phii$ required for \Eq{eq_single_majorana} simultaneously for all $N > 2$ modes. The absence of a line $\sig(\eps) = \dis(\eps) = 0$  at a specific flux $\phi$ then no longer rules out Majoranas $[$\Tab{tab_summary}, last two lines$]$, but it remains conclusive in ruling out the Majorana \emph{pair interference} desired for applications in, e.g., qubits.

\begin{table}[t]
\begin{tabular}{|c||cc|c|}
\hline
\multirow{2}{*}{Subgap mode setup}                                                     & \multicolumn{2}{c|}{\vphantom{$\sum^{N^N}_N$}$\epsi = 0$ for all $i$}                 & at least one $\epsi \neq 0$ \\ \cline{2-4}
                                                                        & \multicolumn{1}{c|}{\vphantom{$\sum^{N^N}_N$}Maj.}              & And.                & Any $\ui$               \\ \cline{2-4}\hline\hline
\multicolumn{1}{|c|| }{$N = 1$}                                           & \multicolumn{1}{c|}{\vphantom{$\sum^{N^N}_{N}$}Yes}               & \multirow{4}{*}{\vphantom{$\sum^{N^N}_{N_N}$}No}                 & No                              \\\cline{1-2}\cline{4-4}
\multicolumn{1}{|c|| }{$N = 2$ , $\phi_2 - \phi_1 = \phi$}                & \multicolumn{1}{c|}{\vphantom{$\sum^{N^N}_{N_N}$}Iff $\phi = 0,\pi$} &                    & \multirow{3}{*}{No*}            \\ \cline{1-2}
\multicolumn{1}{|c|| }{$N \geq 2$ , $\phii \neq \phij \text{ mod } \pi$} & \multicolumn{1}{c|}{\vphantom{$\sum^{N^N}_{N_N}$}No}                &                     &                                 \\ \cline{1-2}
\multicolumn{1}{|c|| }{$N \geq 2$ , $\phii = \phij \text{ mod } \pi$} & \multicolumn{1}{c|}{\vphantom{$\sum^{N^N}_{N_N}$}Yes}                &                     &                                 \\ \hline
\end{tabular}
\caption{Setups that do (``Yes'')/do not (``No'') admit $(\eps,\lami)$-independent $\sig,\dis = 0$. ``Maj.'': All $N$ modes are Majoranas, $\ui = 0.5$. ``And.'': At least one Andreev mode, $\ui \neq 0.5$. ``No*'': $\dis \ll 1 \,\,\forall\eps,\phi$ possible if $\Delta t \gtrsim \tausp \sim (\lambda/\epsi)^2$ \BrackFigPart{fig_finite}{b}.\label{tab_summary}}
\end{table}

To finish the analysis in the limit $\epsi/\lami \rightarrow 0$, we also highlight the \emph{$(\eps,\lami)$-sensitive} signature at $\phi = \frac{\pi}{2},\frac{3\pi}{2}$ \BrackFigPart{fig_zero}{c}, showing reduced $\dis$ without $\sig$ loss~\cite{Schulenborg2022Suppmat} for $|\eps| \lesssim \lami$ , but enhanced $\dis$ at $|\eps| \gg \lami$. Here, the particle-hole-mixed subgap modes superpose to a fully electron/hole-like effective mode $(\Qfac = 0)$. The CD-subgap tunneling is then nearly blocked for one parity only, $\HdSC \rightarrow \eps n + \sqrt{2}\lameff\sqbrack{\dopdag\ald + \al d}$. The much longer tunneling time for this parity ($\peff = -1$ in our gauge) results in $\dis$-lowering noise at fixed $\sig$ if the time exceeds the sample time $\sim\Omega^{-1}$, but again raises $\sig,\dis$ if the hopping time even surpasses the poisoning time $\tQP$. Previous studies of this \emph{parity blockade} focused on one Majorana per wire end~\cite{Steiner2020Aug,Schulenborg2021Jun,Nitsch2022Nov}. Our analysis shows blockade for any number $N \geq 2$ of Majorana- or Andreev modes with $\epsi/\lami \rightarrow 0$~\cite{Schulenborg2022Suppmat}, but ---unlike for Majoranas at $\phi = 0,\pi$--- only for specific $\lami$ permitting $\Qfac = 0$ in \Eq{eq_single_mode_coupling}.

\begin{figure}[t!!]
\includegraphics[width=\linewidth]{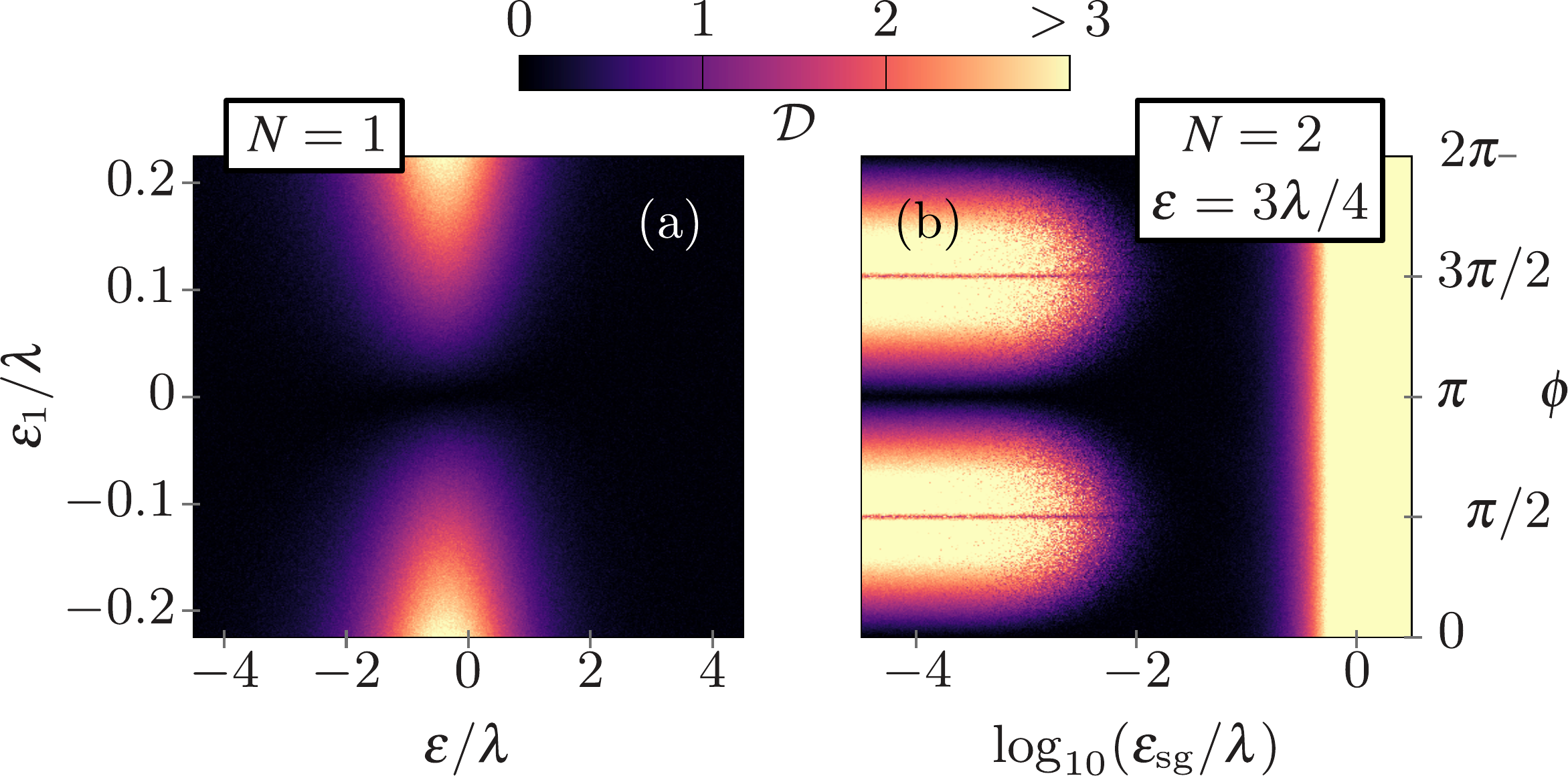}
\caption{Signal-to-noise ratio $\dis$ \BrackEq{eq_parity_signal} as a function of the subgap energies $\epsi$ for (a) one coupled wire with $N = 1, |u_1|^2 = 0.5$, and (b) one mode per wire, $N = 2,\ui^2 = 0.5$. In (b), we set $\epsilon_2 = 2\epsilon_1 = 2\epssg > 0$. Other parameters are as in \Fig{fig_zero}.\label{fig_finite}}
\end{figure}

The sensor's ability to discern finite subgap energies $\epsi \neq 0$ at specific particle-hole mixing is illustrated in \Fig{fig_finite} $[$\Tab{tab_summary}, rightmost column$]$. For a single wire $(N = 1)$, \FigPart{fig_finite}{a} shows an $\eps$-regime with $\dis \gtrsim 1$ already for small $|\epsilon_1|/\lambda_1 \gtrsim 0.1$ and the given detector parameters, even at $u_1^2 = 0.5$.
With $N \geq 2$ states in both wires giving rise to $\phi$-tunable interference, any $\epsi \neq 0$ now couples the effective mode to the other $N - 1$ formerly invisible subgap modes.
The no longer conserved parity $\peff$ then no longer protects against relaxation to an energetically favorable steady state that is independent of $\peff$ right after a QP poisoning. Given a small typical energy difference $\epssg$ between effective and orthogonal modes, $0 < \epssg/\lambda \ll 1$ for $\lami \sim \lambda$, and a sensor-dominated dissipation rate $\sim\Gamma$, this relaxation occurs on a time scale $\tausp \sim \frac{1}{\Gamma}(\frac{\lambda}{\epssg})^2$. If the trajectory time $\Delta t$ exceeds $\tausp$, the signal no longer represents the $\peff = \pm1$-difference. The signal and noise profiles $\sig(\eps,\phi),\dis(\eps,\phi)$ may then be suppressed and lose any conclusive feature in a broad parameter range.
We exemplify this for one mode per wire in \FigPart{fig_finite}{b}, showing the Majorana-specific $\phi$-profile of \FigPart{fig_zero}{c} to vanish if $\tausp(\epssg) \ll \Delta t$. Note that even for weak $\Gamma = \lambda/10$ with typical $\lambda \sim 2.4\text{GHz}$, already $\epsi/\lambda \lesssim 10^{-2}$ leads to rather small $\tausp \sim 10\,\mu\text{s}$. We furthermore stress that the $\phi$-variations are significantly attenuated for any $\eps$ and $\ui$ \cite{Schulenborg2022Suppmat}, also if $\epssg \gtrsim \lambda$ where $\sig,\dis \neq 0$. In the latter case, the CD effectively decouples from the wire with the higher-lying level, reducing the problem to the single-wire case in \FigPart{fig_finite}{a}.

In conclusion, identifying Majorana modes with a subgap parity readout via a capacitively sensed dot is both a major opportunity and challenge. Table~\ref{tab_summary} and \FigPart{fig_zero}{c} show the telltale signature for two topological wires providing a single, nonlocal pair of interfering Majoranas --- a level-independent sensor parity signal and signal-to-noise-ratio $\sig(\eps) = \dis(\eps) = 0$ exclusively at fluxes $\phi = 0,\pi$. This signature is conclusive in that it disappears whenever any Andreev mode, or any additional Majorana orthogonal to the other two Majoranas couples to the converter dot. Furthermore, our protocol is inherently robust to experimentally unavoidable $1/f$ charge noise and asymmetric wire couplings.
A clear inference may, however, still be impeded by flux noise. Moreover, already small subgap mode energies can suppress the $\phi$-dependence after a perhaps challengingly short decay time $\tausp$. On the flip side, \emph{extracting} $\tausp$ by varying measurement times may be an interferometric method to resolve deviations $\epsi \neq 0$ more precisely than dc transport spectroscopy.

\begin{acknowledgments}
We thank Serwan Asaad, Magnus Lykkegaard, Maximilian Nitsch, Felix Passmann, and Charles Marcus for very helpful discussions. The research was supported by the Danish National Research Foundation, the Danish Council for Independent Research Natural Sciences, and the Swedish Research Council (VR). M.B. is supported by the Villum Foundation (Research Grant No. 25310). This work was also funded by the Deutsche Forschungsgemeinschaft (DFG, German Research Foundation) - Project Number 277101999 - CRC 183  and the European Research Council (ERC) under the European Union’s Horizon 2020 research and innovation program under Grant agreement No. 856526.
\end{acknowledgments}

\onecolumngrid
\clearpage


\section*{Supplementary material}

# Supplementary material for
# Detecting Majorana modes by readout of poisoning-induced parity flips

Jens Schulenborg,[1] Svend Krøjer,[1] Michele Burrello,[1,2] Martin Leijnse,[3,1] and Karsten Flensberg[1]

[1]*Center for Quantum Devices, Niels Bohr Institute,
University of Copenhagen, 2100 Copenhagen, Denmark*
[2]*Niels Bohr International Academy, Niels Bohr Institute,
University of Copenhagen, 2100 Copenhagen, Denmark*
[3]*Solid State Physics and NanoLund, Lund University, Box 118, S-221 00 Lund, Sweden*

## CONTENTS

This supplementary material details the parity-to-charge readout description based on the Lindblad quantum master equation as well as the quantum jump method, and furthermore derives and supports various claims and relations from the main paper. Equations and figures in this supplementary are labeled with an additional S to distinguish them from labels referring to the main text. For example, Eq. 1 refers to the main text, whereas Eq. S1 refers to this supplementary material. Note that we set $k_\text{B} = |e| = \hbar = 1$ as in the main text. Moreover, symbols that have already been defined in the main text are not always explicitly redefined in this document.

## I. FULL HAMILTONIAN AND QUANTUM MASTER EQUATION

To describe the sensor-dot readout, let us start by giving the full Hamiltonian including the sensor dot and the coupled leads:

$$H_\text{tot} = H + H_\text{lead} + H_\text{tun} \tag{S1}$$

The subsystem consisting of sensor dot, main dot, and subgap states is described by

$$H = H_\text{d,SC} + \epsilon_\text{S} n_\text{S} + H_\text{cap} \quad , \quad H_\text{cap} = E_\text{cap} n_\text{S} n, \tag{S2}$$

with $H_\text{d,SC}$ given in Eq. 1 of the main text. The sensor dot excess charge $n_\text{S} = d_\text{S}^\dagger d_\text{S}$ with potential $\epsilon_\text{S}$ is created(annihilated) by $d_\text{S}^\dagger(d_\text{S})$, and capacitively coupled with strength $E_\text{cap}$ to the main dot. The Hamiltonian of the two non-interacting fermionic leads $r = \text{L,R}$ is given by

$$H_\text{lead} = \sum_{r=\text{L,R}} H_r \quad , \quad H_r = \sum_{k\nu} \epsilon_{rk\nu} c_{rk\nu}^\dagger c_{rk\nu}, \tag{S3}$$

where $c^\dagger_{rk\nu}$ and $c_{rk\nu}$, respectively, create and annihilate electrons in lead $r$ with energy $\epsilon_{rk\nu}$, wave number $k$, and all further discrete quantum numbers $\nu$ necessary to fully characterize the single-particle states. The tunnel coupling

$$H_{\text{tun}} = \sum_{r=\text{L,R}} H_{\text{tun},r} \quad , \quad H_{\text{tun},r} = \sum_{k\nu} \tau_{rk\nu} d^\dagger_{\text{S}} c_{rk\nu} + \text{H.c.} \tag{S4}$$

is characterized by the typical coupling strengths

$$\Gamma_r(E) = 2\pi \sum_{k\nu} \delta(E - \epsilon_{rk\nu}) |\tau_{rk\nu}|^2 \to \Gamma, \tag{S5}$$

where we assume the wideband limit and symmetric lead couplings, i.e., an $(r,E)$-independent coupling $\Gamma$.

To describe the dynamics of the subsystem $H$ in the presence of the leads, we first consider the reduced density operator

$$\rho(t) = \underset{\text{lead}}{\text{Tr}} [\rho_{\text{tot}}(t)] \tag{S6}$$

with the electronic leads $r =$ L,R traced out from the density operator $\rho_{\text{tot}}(t)$ of the total system. We assume the leads to act as Markovian, weakly coupled baths. Their states $\rho_r$ are initially, i.e., at time $t_0 = 0$ in the distant past relative to the steady-state on the scale of the typical frequencies in $H$, given by the grand-canonical ensemble

$$\rho_r = \exp\left(-\frac{H_r - \mu_r N_r}{T}\right) / \underset{\text{lead}}{\text{Tr}} \left[\exp\left(-\frac{H_r - \mu_r N_r}{T}\right)\right] \tag{S7}$$

with respect to the individual electro-chemical potentials $\mu_r$, particle numbers $N_r = \sum_{k\nu} c^\dagger_{rk\nu} c_{rk\nu}$, and common temperature $T$. The lead states (S7) are hence assumed to be initially uncorrelated both from each other, and from the initial main-system state,

$$\rho_{\text{tot}}(t_0 = 0) = \rho_0 \prod_{r=\text{L,R}} \rho_r \quad , \quad \rho_0 = \rho(t_0 = 0). \tag{S8}$$

This means that $\rho(t)$ evolves from $\rho_0$ approximately according to the Lindblad master equation as in Refs. 1–4:

$$\partial_t \rho(t) = -i[H + \Lambda, \rho(t)] + \sum_{r=\text{L,R},\eta=\pm} \mathcal{L}_{r\eta} \rho(t), \tag{S9}$$

including the Lamb shift $\Lambda$ and the dissipators

$$\mathcal{L}_{r\eta} \bullet = L_{r\eta} \bullet L^\dagger_{r\eta} - \frac{1}{2}\{L^\dagger_{r\eta} L_{r\eta}, \bullet\}. \tag{S10}$$

The jump operators $L_{r\eta}$ represent the dynamics due to electrons jumping to($\eta = +1$) or from($\eta = -1$) the sensor dot via the tunnel coupled lead $r$. With the tunneling $H_{\text{tun}}$ as described by Eqs. (S4)-(S5) in the wideband limit, these jumps are given by [1, 2]

$$L_{r\eta} = \sqrt{\Gamma} \sum_{ij} \sqrt{f^\eta_r(\eta E_{ij})} \langle i|d_{\text{S},\eta}|j\rangle \times |i\rangle\langle j|, \tag{S11}$$

where $f^\eta_r(E) = \left[\exp\left(\eta \frac{E - \mu_r}{T}\right) + 1\right]^{-1}$ are the lead Fermi functions, $d_{\text{S},\eta=+} = d^\dagger_{\text{S}}$, and $d_{\text{S},\eta=-} = d_{\text{S}}$. The symbols $E_{ij} = E_i - E_j$ denote energy differences between the eigenstates $|i\rangle, |j\rangle$ of the local Hamiltonian $H$ [Eq. (S2)]. The Hermitian Lamb shift $\Lambda = \Lambda^\dagger$ further modifies the local coherent dynamics due to this Hamiltonian. Following Ref. 2 and using a finite energy bandwidth $D$ for the frequency integral,

$$\Lambda = \Gamma \sum_{\substack{r\eta \\ ii'j}} F\left(\frac{E_{i'i} - \eta\mu_r}{T}, \frac{E_{i'j} - \eta\mu_r}{T}\right) (\langle i'|d_{\text{S},\eta}|i\rangle)^* \langle i'|d_{\text{S},\eta}|j\rangle \times |i\rangle\langle j|$$

$$F(a,b) = -\frac{1}{2\pi} \mathcal{P} \int_{-D/T}^{D/T} dx \frac{1}{x\sqrt{e^{x+a}+1}\sqrt{e^{x+b}+1}}, \tag{S12}$$

where $\mathcal{P}$ indicates the Cauchy principal value integral avoiding $x = 0$. Crucially, the precise choice of the bandwidth $D/T$ is not critical as long as it is much larger than any possible energy difference $(E_{i'i} \pm \mu_r)/T$ occurring in our simulations, in accordance with the wideband limit. Namely, while the integral in Eq. (S12) diverges logarithmically in $D$ on the negative $x$ branch, the $(a,b)$ dependence of this divergence is suppressed by the exponential factors $e^x$. For large enough $D$, we may thus always write $F(a,b) = F_0(a,b) + F_D$ with an $(a,b)$-dependent, but bandwidth-independent part $F_0(a,b)$, and a practically $(a,b)$-independent part $F_D \sim \ln(D)$ that diverges with bandwidth $D$. In the Lamb shift $\Lambda$, the latter only generates a constant shift $\Lambda_D \sim \ln(D)\mathbb{1}$ which drops out in the commutator of the master equation (S9). Note, however, that since $[H, n] \neq 0$, the non-divergent part of $\Lambda$ may generally depend on the subparity $p_{d\alpha} = (-\mathbb{1})^{n+\alpha^\dagger \alpha}$ of dot and effective subgap mode [Sec. (IV)]. We hence always account for $\Lambda$ by calculating $F(a,b)$ numerically as in Ref. 4, with $D/T = 10^4$. More precisely, Simpson's 1/3 rule with 8192 intervals of sufficiently small size is used to perform the integral once for a fixed set of points $(a, b)$ on a grid. This grid is chosen dense ($512 \times 512$ points) and large enough so that bilinear interpolation between those fixed points yields $F$ with sufficient accuracy for any energy pair. This works because the integrand in Eq. (S12) behaves well for $x, a, b \in \mathbb{R}$.

To obtain $\rho(t)$, we solve the master equation (S9) with an initial state $\rho_0$ of fixed subparity in the subspace of the dot and the effective mode described below in Sec. IV, i.e., $[p_{d\alpha}, \rho_0] = 0$ and $p_{d\alpha}\rho_0 = \pm\rho_0$, where the effective mode is defined with respect to the initial system parameters [Eq. (S24)]. Note that the parity $p_{d\alpha}$ is not generally conserved for finite-energy subgap modes, $[H_{tot}, p_{d\alpha}] = [H_{d,SC}, p_{d\alpha}] \neq 0$, and hence does not generally allow us to reduce the Hilbert space to blocks of fixed $p_{d\alpha}$. Conversely, Eq. (S9) conserves the parity $p_{d,SC}$ and does not generate any quantum coherences between states of different sensor dot occupations $n_S$. This reduces the effectively required size of the $2^{N+2} \times 2^{N+2}$ matrix for $\rho(t)$ describing $N$ subgap states, main dot, and sensor dot to four blocks of size $(2^N \times 2^N)$. The dissipators (S10) are accordingly represented by two blocks of size $4(2^N \times 2^N)^2$, namely one for $p_{d,SC} = +1$ and one for $p_{d,SC} = -1$.

With $\rho(t)$ at hand, we use the fact that the tunneling $H_{\text{tun},r} = \sum_{k\nu} \left[ \tau_{rk\nu} d_S^\dagger c_{rk\nu} + \text{H.c.} \right]$ to only one lead $r$ conserves the sum of charges in lead $r$ and in the sensor dot, $[n_S + N_r, H_{\text{tun},r}] = 0$. Namely, with the *total* Hamiltonian (S1), the time-dependent, ensemble-averaged tunnel current out of lead $r = L$ is given by

$$\bar{I}_L(t) = -\partial_t \underset{\text{tot}}{\text{Tr}} [N_L \rho_{tot}] = -\underset{\text{tot}}{\text{Tr}} [N_L \partial_t \rho_{tot}] = i\underset{\text{tot}}{\text{Tr}} [N_L [H_{tot}, \rho_{tot}]] = i\underset{\text{tot}}{\text{Tr}} [[N_L, H_{tot}] \rho_{tot}] = i\underset{\text{tot}}{\text{Tr}} [[N_L, H_{\text{tun},L}] \rho_{tot}]$$

$$\overset{[n_S+N_r, H_{\text{tun},r}]=0}{=} -i\underset{\text{tot}}{\text{Tr}} [[n_S, H_{\text{tun},L}] \rho_{tot}] \overset{[n_S, H+H_{\text{lead}}]=0}{=} -i\underset{\text{tot}}{\text{Tr}} [[n_S, H_{\text{tun},L} + H + H_{\text{lead}}] \rho_{tot}]$$

$$= \underset{\text{tot}}{\text{Tr}} [n_S(-i[H_{\text{tun},L} + H + H_{\text{lead}}, \rho_{tot}])] = \underset{\text{tot}}{\text{Tr}} \left[ n_S(-i[H_{tot}, \rho_{tot}])_{\tau_{Rk\nu}=0} \right]$$

$$\overset{\underset{\text{tot}}{\text{Tr}} \bullet = \underset{\text{lead}}{\text{Tr}} \text{Tr} \bullet}{\underset{\underset{\text{lead}}{\text{Tr}} n_S \bullet = n_S \underset{\text{lead}}{\text{Tr}} \bullet}{=}} \text{Tr} \left[ n_S \underset{\text{lead}}{\text{Tr}} (-i[H_{tot}, \rho_{tot}])_{\tau_{Rk\nu}=0} \right] \overset{(S6)}{=} \text{Tr} \left[ n_S [\partial_t \rho(t)]_{\tau_{Rk\nu}=0} \right]. \tag{S13}$$

Using the master equation (S9),

$$\bar{I}_L(t) \overset{(S9)}{\underset{(S13)}{=}} -i \text{Tr} [n_S [H + \Lambda, \rho(t)]]_{\tau_{Rk\nu}=0} + \text{Tr} \left[ n_S \left( \sum_{\eta=\pm} \mathcal{L}_{L\eta} \rho(t) \right) \right] \overset{[n_S, H]=0}{\underset{[n_S, \Lambda]=0}{=}} \text{Tr} \left[ n_S \left( \sum_{\eta=\pm} \mathcal{L}_{L\eta} \rho(t) \right) \right]$$

$$\overset{(S10),(S11)}{\underset{n_S L_{L\eta} = \delta_{\eta+} L_{L\eta}}{=}} \text{Tr} \left[ L_{L+} \rho(t) L_{L+}^\dagger \right] - \frac{1}{2} \sum_{\eta=\pm} \text{Tr} \left[ n_S L_{L\eta}^\dagger L_{L\eta} \rho(t) \right] - \frac{1}{2} \sum_{\eta=\pm} \text{Tr} \left[ L_{L\eta}^\dagger L_{L\eta} n_S \rho(t) \right]$$

$$\overset{n_S L_{L\eta}^\dagger = L_{L\eta}^\dagger \delta_{\eta-}}{\underset{L_{L\eta} n_S = L_{L\eta} \delta_{\eta-}}{=}} \sum_{\eta=\pm} \eta \, \text{Tr} \left[ L_{L\eta} \rho(t) L_{L\eta}^\dagger \right], \tag{S14}$$

where the overbar $\bar{\bullet}$ indicates the ensemble-average. Equation (S14) gives the current response

$$\bar{G}(t) = \frac{\bar{I}_L(t)}{\delta\mu} = \sum_{\eta=\pm} \frac{\eta}{\delta\mu} \text{Tr} \left[ L_{L\eta} \rho(t) L_{L\eta}^\dagger \right] \tag{S15}$$

to a small potential bias $\delta\mu = \mu_L - \mu_R$ applied to the left lead while keeping $\mu_R \equiv 0$ fixed. We use Eq. (S15) to confirm that Eq. 2 from the main text —as further detailed in the next Sec. II— approximates the absolute value of (S15) when averaged over many samples. In this comparison, we do not assume the steady-state limit $t \to \infty$ in $\bar{G}(t)$, but rather choose the sampled trajectory time $\Delta t = M_T N_{osc}(2\pi/\Omega)$, as detailed in the main text and below. This matters since for nearly degenerate subgap states, the dot-SC Hamiltonian $H_{d,SC}$ may contain small splittings $\Delta E < 1/\Delta t$ which result in a metastable state $\rho(t)$ decaying on times comparable to $\Delta t$.

## II. QUANTUM JUMP METHOD FOR FIRST HARMONIC CURRENT RESPONSE

This section details how to use the quantum jump method to evaluate the first harmonic current response in Eq. 2 of the main paper. The general approach is to unravel the master equation (S9) into quantum jumps [5–7], amounting to a Monte Carlo simulation yielding possible individual trajectories of the quantum state $|\Psi(t)\rangle$ describing the subsystem of sensor dot, main dot, and subgap states. The procedure can be divided into the following steps:

1. We generate the jump operators $L_{r\eta}(t) = L_{r\eta}(\mu_L(t))$, the Lamb shift $\Lambda(t) = \Lambda(\mu_L(t))$, and the no-jump evolution

$$\mathcal{U}(t) = \exp\left(-i\delta t \left[H + \Lambda(t) - \frac{i}{2}\sum_{r\eta} L^\dagger_{r\eta}(t) L_{r\eta}(t)\right]\right) \tag{S16}$$

   as a function of time for one driving period of the chemical potential $\mu_L(t) = \mu_R + \delta\mu \sin(\Omega t)$ by evaluating Eq. (S11) and Eq. (S12) with respect to the instantaneous $\mu_L(t)$, where $\delta\mu = T$ and $\Omega/2\pi = 1\,\text{GHz}$ as given in the main text. Note that for the here chosen, slow driving frequency compared to the lead coupling, $\Omega \ll \Gamma$, the driven dissipative evolution is well approximated by such instantaneous updates, see Ref. 2. Since we repeat the simulation for a large set of different subsystem Hamiltonian($H$) parameters ($400 \times 400$ points for each density plot), we reduce the numerical cost by evaluating Eqs. (S11) and the exponential (S16) including the Lamb shift (S12) only for a fixed subset of $10^3$ times points among the full set of $(2\pi/\Omega)/\delta t = 10^4$ time points per $\mu_L(t)$-oscillation; the remaining $L_{r\eta}(t)$ and $\mathcal{U}(t)$ are obtained by linear interpolation between those time points.

2. For each trajectory, the initial state is defined to be a pure state $|\Psi(t=0)\rangle$ with fixed $n_S$ and fixed effective-mode parity $p_{d\alpha} = (-1)^{n+\alpha^\dagger\alpha}$ [Sec. (IV)] with respect to the system parameters entering $H_{d,SC}$. This means that for $p_{d\alpha} = +1$, we randomly fix $n = \alpha^\dagger\alpha = 0$ or $n = \alpha^\dagger\alpha = 1$, whereas for $p_{d\alpha} = -1$, we randomly set either $n = 1 - \alpha^\dagger\alpha = 0$ or $n = 1 - \alpha^\dagger\alpha = 1$. Since $n_S = 0$ or $n_S = 1$ during the entire evolution, we track the sensor dot state with a simple boolean index, thereby reducing the dimension of $|\Psi\rangle$ from $2^{N+1}$ to $2^N$ for the state of the main dot and $N$ subgap modes at fixed parity $p_{d,SC}$. We also initialize the four jump numbers $J_{r\eta}(t_0 = 0) = 0$ counting the total number of jump events $J_{r\eta}(t)$ to $(\eta = +)$/from$(\eta = -)$ lead $r$ up to any time $t$.

3. To advance the state $|\Psi(t)\rangle$ at time $t$ by a step $\delta t$, we first perform the no-jump evolution

$$|\Psi(t+\delta t)\rangle = \mathcal{U}(t)|\Psi(t)\rangle \tag{S17}$$

   with $\mathcal{U}(t)$ chosen from the set generated in step 1 according to the time within the $\mu_L(t)$-driving period. Since the evolution in Eq. (S16) is not unitary but dissipative, the state needs to be renormalized afterwards, $|\Psi(t+\delta t)\rangle \to |\Psi(t+\delta t)\rangle/\sqrt{\langle\Psi(t+\delta t)|\Psi(t+\delta t)\rangle}$. Note that even though we have chosen a time step $\delta t$ similar to the typical oscillation frequencies in $H$, meaning $(1/\delta t) \sim \lambda$, the numerical precision suffices since we apply the full exponential in Eq. (S16), and since both the driving and the dissipative jump dynamics proceed on the much longer time scales $2\pi/\Omega$ and $1/\Gamma$.

4. We set the lead index $r$ randomly to either $r = L$ or $r = R$.

5. For the chosen lead index $r$ and time $t' = t + \delta t$, we pick the corresponding $L_{r\eta}(t')$ from step 1 to calculate the probability $P_{r\eta}(t') = \langle\Psi(t')|L^\dagger_{r\eta}(t')L_{r\eta}(t')|\Psi(t')\rangle\delta t$, where $\eta = +1$ if $n_S|\Psi(t')\rangle = 0$ and $\eta = -1$ if $n_S|\Psi(t')\rangle = |\Psi(t')\rangle$. We draw a uniformly distributed, random number $P \in [0,1]$, and proceed as follows:

   - If $P < P_{r\eta}(t')$, we perform the jump

$$|\Psi(t')\rangle \to \frac{L_{r\eta}(t')|\Psi(t')\rangle}{\sqrt{\langle\Psi(t')|L^\dagger_{r\eta}(t')L_{r\eta}(t')|\Psi(t')\rangle}} = \frac{L_{r\eta}(t')|\Psi(t')\rangle}{\sqrt{P_{r\eta}(t')/\delta t}}, \tag{S18}$$

     flip the sensor dot index, $n_S \to 1 - n_S$, and increment the jump number $J_{r\eta}(t') = J_{r\eta}(t) + 1$. We then repeat starting from step 3 with the new state $|\Psi(t')\rangle$.

   - If $P \geq P_{r\eta}(t')$, we repeat step 5 with the opposite lead index $r' \neq r$, i.e., $r' = R$ if $r = L$ and vice versa. If the newly generated random number $P' < P_{r'\eta}(t')$, we perform the jump as described in Eq. (S18) and below. If $P' \geq P_{r'\eta}(t')$, we repeat from step 3 with the new state $|\Psi(t')\rangle$, without performing a jump.

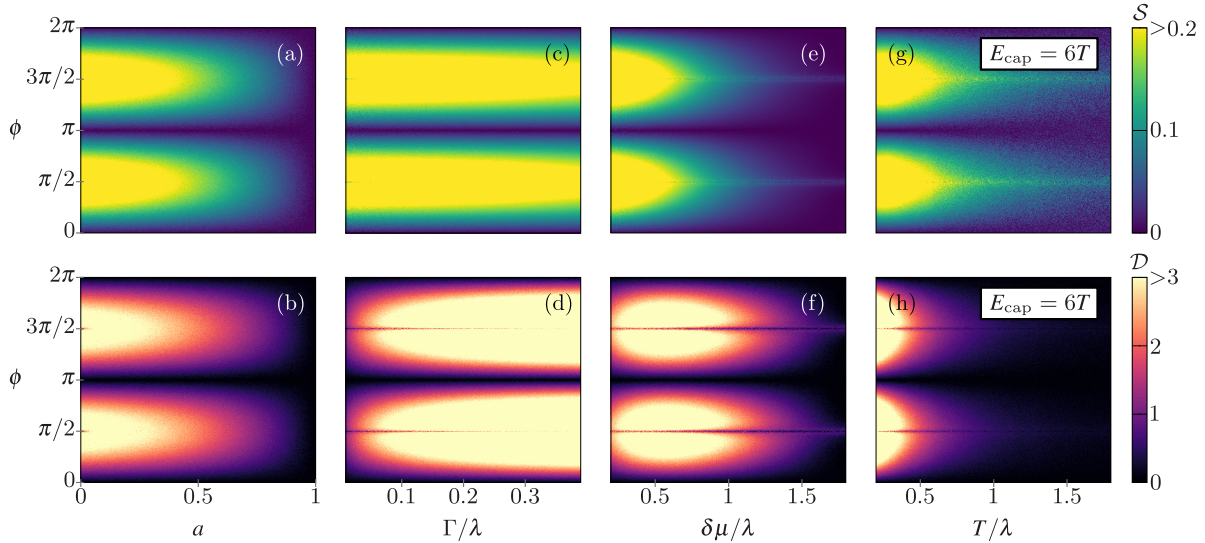

FIG. S1. Parity signal $\mathcal{S}$ and signal-to-noise ratio $\mathcal{D}$ for one mode per wire ($N = 2$) as a function of various parameters. In (a), we set $\lambda_2 = (1-a)\lambda_1 = \lambda$. If not specified in the plots, we set $\epsilon = 3\lambda/4$, $|u_i|^2 = 1/2$, $\epsilon_i = 0$, $a = 0$, $\epsilon_S = 0$, $E_{\text{cap}} = 2\lambda$, $T = \lambda/3$, $\delta\mu = \lambda/3$, $\Gamma = \lambda/10$, $2\pi/\Omega = 4t_b = 2\pi \times 2400/\lambda = 10^4 \delta t = 1\mu s$, $N_{\text{osc}} = 2$, $N_T = 10$, $M_T = 50$.

6. We take a current record

$$I_L(t) = \sum_{\eta=\pm} \frac{\eta}{t_b} \left[ J_{L\eta}(t) - J_{L\eta}(t - t_b) \right]. \tag{S19}$$

with detector bandwidth $t_b\Omega/2\pi = 1/4$ for a duration of $\Delta t + \Delta t_0$. This includes the measurement time $\Delta t = M_T N_{\text{osc}}(2\pi/\Omega)$ consisting of $M_T$ samples, each integrating over $N_{\text{osc}}$ oscillations with frequency $\Omega/2\pi$, and a one-oscillation offset $\Delta t_0 = 2\pi/\Omega \gg 1/\Gamma$ after which transients on a time scale $\sim \Gamma^{-1}$ have decayed. Equation (S19) then enters Eq. 2 from the main text, where we generate $N_T$ trajectories of $M_T$ samples. The expectation values in Eq. 3 are taken with respect to the combined number of $M = N_T M_T$ samples.

All random numbers are obtained from a 64-bit Xorshift* pseudo-random number generator that, per trajectory, is seeded by another, regular 64-bit Xorshift pseudo-random number generator.

## III. DEPENDENCE ON SENSOR LEVEL, COUPLINGS, DRIVING AMPLITUDE AND TEMPERATURE

In Fig. S1, we provide further results supporting our statements from the main text about how the flux($\phi$)-variations of the parity signal $\mathcal{S}$ and signal-to-noise ratio $\mathcal{D}$ depend on

- the subgap-converter coupling asymmetry $a$ for one mode per wire, with $\lambda_2 = (1-a)\lambda_1 = \lambda$ [Figs. S1(a,b)]
- the sensor-lead tunnel coupling $\Gamma$ [Figs. S1(c,d)]
- the driving amplitude $\delta\mu$ [Figs. S1(e,f)]
- and the lead temperature $T$ at constant ratio $E_{\text{cap}}/T = 6$ [Figs. S1(g,h)].

The constant $\mathcal{S}, \mathcal{D} = 0$ at $\phi = 0, \pi$ confirm that the Majorana signature is independent of any tunnel coupling strengths, sensor driving amplitude and lead temperature. However, more symmetric CD-subgap state couplings and larger $\Gamma$ give higher $\mathcal{D}$-contrast between $\phi = 0, \pi$ and any $\phi \neq 0, \pi$. The temperature in Figs. S1(g,h) also scales the capacitive coupling $E_{\text{cap}} = 6T$. This rules out any trivial effect due to increasing $T$ blurring the two sensor dot conductance peaks, and thereby links the $\mathcal{S}, \mathcal{D}$-suppression for $T/\lambda \gtrsim 1$ to sensor-induced thermal excitations of the converter-subgap system. Note also that the upper $\Gamma$ range in Figs. S1(c,d) needs to be viewed carefully, as $\Gamma > T = \lambda/3$ reaches beyond the point at which neglecting higher-order-$\Gamma$ effects as in our theory is justified.

6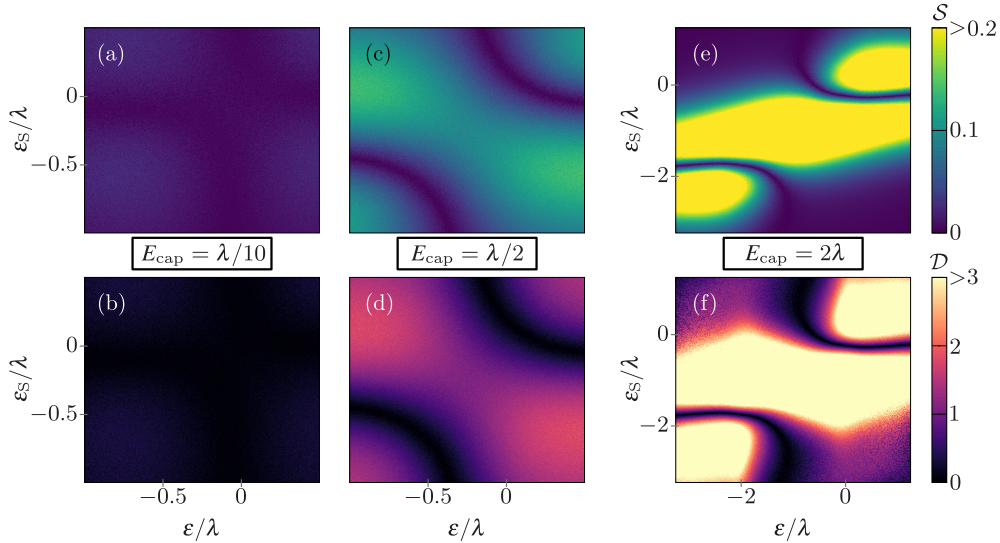

FIG. S2. Parity signal $\mathcal{S}$ and signal-to-noise ratio $\mathcal{D}$ as a function of CD level $\epsilon$ and SD level $\epsilon_S$ with one mode per wire ($N = 2$). Parameters not specified in the plots are $\phi = 0.48\pi$, $|u_i|^2 = 1/2$, $\epsilon_i = 0$, $\lambda_i = \lambda$, $T = \lambda/3$, $\delta\mu = \lambda/3$, $\Gamma = \lambda/10$, $2\pi/\Omega = 4t_b = 2\pi \times 2400/\lambda = 10^4 \delta t = 1\mu s$, $N_{\rm osc} = 2$, $N_{\rm T} = 10$, $M_{\rm T} = 50$.

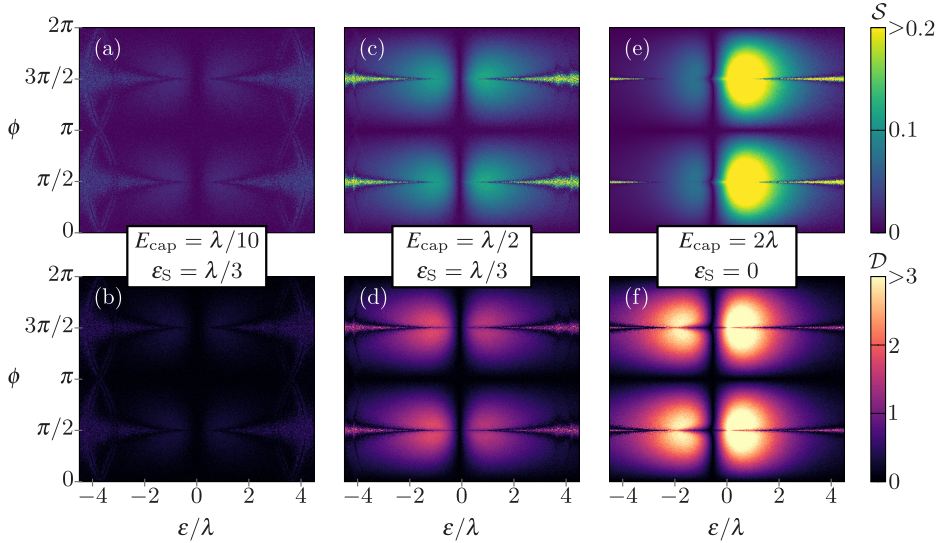

FIG. S3. Parity signal $\mathcal{S}$ and signal-to-noise ratio $\mathcal{D}$ as a function of CD level $\epsilon$ and flux $\phi$ with one mode per wire ($N = 2$). Parameters not specified in the plots are the same as in Fig. S2.

Figures S2 and S3 show how the optimal sensor dot level $\epsilon_S$ and the key Majorana signature from Fig. 2(c) of the main text depend on the capacitive coupling $E_{\rm cap}$ in relation to the CD-subgap tunnel couplings $\lambda_i$. More precisely, Fig. S2 shows generally very low $\mathcal{S}, \mathcal{D}$ for $E_{\rm cap} < T$ [Figs. S2(a,b)], and moderate $\mathcal{S}, \mathcal{D}$ across the considered $\epsilon$-range for $T < E_{\rm cap} < \lambda_i$ at $\epsilon_S$ slightly tuned away from resonance $\epsilon_S = 0$ [Figs. S2(c,d)]; the result from the main text for $E_{\rm cap} > \lambda_i > T$ at $\epsilon_S = 0$ corresponds to Figs. S2(e,f). Note that even though $\epsilon_S = -E_{\rm cap}/2$ features generally large $\mathcal{D}$ across the $\epsilon$-range, we do not choose this $\epsilon_S$ in the main text, as this off-resonant point in the Coulomb blockade regime may be more susceptible to higher-order-$\Gamma$ effects (such as a Kondo resonance) unaccounted for in our theory.

Given the above suggested $\epsilon_S$ choices, the $\mathcal{S}(\epsilon, \phi)$- and $\mathcal{D}(\epsilon, \phi)$ maps indicative of Majorana modes are displayed in Fig. S3 for the three different ratios $E_{\rm cap}/\lambda$ studied in Fig. S2. A capacitive coupling $E_{\rm cap} > \lambda_i$ is chosen in the main text, as it gives the highest $\mathcal{D}$ contrast between $\phi = 0, \pi$ and $\phi \neq 0, \pi$, and hence the clearest Majorana signature.

## IV. EFFECTIVE MODE

The expression of the creation and annihilation operators $\alpha^\dagger, \alpha$ for the effective mode entering Eq. 4 from the main text in terms of the fermionic operators $\alpha_i^\dagger, \alpha_i$ in $H_{\text{d,SC}}$ as given Eq. 1 is derived by *demanding*

$$\{\alpha, \alpha\} = \{\alpha^\dagger, \alpha^\dagger\} = 0 \quad, \quad \{\alpha^\dagger, \alpha\} = 1 \tag{S20}$$

$$d^\dagger \left[\lambda^+ \alpha^\dagger + \lambda^- \alpha\right] = \sum_{i=1}^N \lambda_i e^{i\phi_i} d^\dagger \left[|v_i|\alpha_i^\dagger + |u_i|\alpha_i\right]. \tag{S21}$$

Let us first construct fields $\alpha^\dagger, \alpha$ solving Eq. (S21) for any $\lambda^\pm$, and then constrain these coefficients with the anti-commutation relations (S20). To find $\alpha^\dagger$, we insert the ansatz

$$\alpha^\dagger = \sum_{i=1}^N \left(\xi_i^+ \alpha_i^\dagger + \xi_i^- \alpha_i\right) \quad, \quad \alpha = (\alpha^\dagger)^\dagger \tag{S22}$$

into the left-hand side of Eq. (S21), and then apply $\{[\bullet, d], \alpha_i\}$ as well as $\{[\bullet, d], \alpha_i^\dagger\}$ on both sides, yielding

$$\begin{pmatrix} \lambda^+ & \lambda^- \\ (\lambda^-)^* & (\lambda^+)^* \end{pmatrix} \cdot \begin{pmatrix} \xi_i^+ \\ (\xi_i^-)^* \end{pmatrix} = \lambda_i \begin{pmatrix} |v_i|e^{i\phi_i} \\ |u_i|e^{-i\phi_i} \end{pmatrix}. \tag{S23}$$

If $|\lambda^+| \neq |\lambda^-|$, Eq. (S23) becomes invertible:

$$\alpha^\dagger \stackrel{|\lambda^+| \neq |\lambda^-|}{=} \frac{1}{|\lambda^+|^2 - |\lambda^-|^2} \sum_{i=1}^N \lambda_i \left[\left((\lambda^+)^*|v_i|e^{i\phi_i} - \lambda^-|u_i|e^{-i\phi_i}\right)\alpha_i^\dagger + \left((\lambda^+)^*|u_i|e^{i\phi_i} - \lambda^-|v_i|e^{-i\phi_i}\right)\alpha_i\right] \quad, \quad \alpha = (\alpha^\dagger)^\dagger. \tag{S24}$$

In the special case $|\lambda^+| = |\lambda^-|$, we can always insert $\lambda^+ = |\lambda^+|e^{-i\phi^+}$, $\lambda^- = |\lambda^+|e^{i(\phi^+ + \phi^-)}$ into Eq. (S23), and get

$$|\lambda^+| \left(e^{-i\phi^+} \xi_i^+ + e^{i(\phi^+ + \phi^-)} (\xi_i^-)^*\right) = \lambda_i |v_i| e^{i\phi_i} = \lambda_i |u_i| e^{-i(\phi_i - \phi^-)}. \tag{S25}$$

Thus, $|\lambda^+| = |\lambda^-|$ is only possible if $|u_i| = |v_i| = 1/\sqrt{2}$ and if $e^{2i\phi_i} = e^{i\phi^-}$ is independent of $i$ for all coupled modes $i$ with $\lambda_i > 0$. This implies in particular that the phases are equal up to a $\pi$-shift, $\phi_i = \phi_{i'} + z\pi$ with $z \in \mathbb{Z}$. If this is guaranteed, we are free to set $\xi_i^- = 0$ and solve Eq. (S23) by

$$\alpha^\dagger \stackrel{|\lambda^+| = |\lambda^-|}{=} e^{i\phi^+} \sum_{i=1}^N \frac{\lambda_i}{\sqrt{2}|\lambda^+|} \alpha_i^\dagger \quad, \quad \alpha = (\alpha^\dagger)^\dagger. \tag{S26}$$

Having constructed solutions $\alpha^\dagger, \alpha$ to Eq. (S21) as a function of two yet-to-be-determined coefficients $\lambda^\pm$, we now fix the latter by imposing the anti-commutation relations (S20). Namely, we apply the commutator $[\bullet, d]$ to Eq. (S21) to factor out $d^\dagger$, apply the anti-commutator of both sides of the resulting equation with either themselves or their Hermitian conjugate, assume Eq. (S20) to be true to obtain $\lambda^\pm$ as functions of all $\lambda_i, |u_i|, |v_i|, \phi_i$, and finally substitute $\lambda^\pm$ by these functions in Eqs. (S24),(S26) to explicitly verify Eq. (S20). Using the known anti-commutation relations $\{\alpha_i, \alpha_j\} = \{\alpha_i^\dagger, \alpha_j^\dagger\} = 0$ and $\{\alpha_i^\dagger, \alpha_j\} = \delta_{ij}$ as well as the normalization $|u_i|^2 + |v_i|^2 = 1$ in Eq. (S21), this gives

$$\left|\lambda^+\right|^2 + \left|\lambda^-\right|^2 = \sum_{i=1}^N \lambda_i^2 = 2\lambda_{\text{eff}}^2 \quad, \quad \lambda^+ \lambda^- = \sum_{i=1}^N \lambda_i^2 e^{i2\phi_i} |u_i||v_i| = \mathcal{Q}\lambda_{\text{eff}}^2 \tag{S27}$$

with $\lambda_{\text{eff}}$ and $\mathcal{Q}$ as defined in the main text. We here assume $\lambda^\pm \geq 0$, since any two phase factors for the coefficients $\lambda^\pm$ appearing in Eq. (S21) can be gauged into the fields $\alpha, \alpha^\dagger, d^\dagger$ without modifying any other term of $H_{\text{d,SC}}$ as written in Eq. 4 of the main text. Equation (S27) then yields Eq. 5,

$$\lambda^\pm = \lambda_{\text{eff}} \sqrt{1 \pm \sqrt{1 - |\mathcal{Q}|^2}} \quad, \quad \mathcal{Q} = |\mathcal{Q}| = \frac{\lambda^+ \lambda^-}{\lambda_{\text{eff}}^2} = \sum_{i=1}^N \frac{\lambda_i^2}{\lambda_{\text{eff}}^2} e^{i2\phi_i} |u_i||v_i|, \tag{S28}$$

where $0 \leq |u_i||v_i| \leq 1/2$ ensures $0 \leq \mathcal{Q} \leq 1$. The upper bound $\mathcal{Q} = 1$ can, according to Eqs. (S27)-(S28) and due to $0 \leq |u_i||v_i| \leq 1/2$, only be saturated for the above highlighted special case $|u_i|^2 = |v_i|^2 = 1/2$ and $\phi_i = \phi_{i'} + z\pi$ with $z \in \mathbb{Z}$ for all coupled modes $i$ with $\lambda_i > 0$, leading to $\lambda^+ = \lambda^- = \lambda_{\text{eff}}$. We finish by inserting Eq. (S28) into Eqs. (S24),(S26) and using $\{\alpha_i, \alpha_j\} = 0$, $\{\alpha_i^\dagger, \alpha_j\} = \delta_{ij}$ to explicitly verify the anti-commutation relations (S20).

## V. PARITY-INDEPENDENT DYNAMICS IN THE SINGLE-MAJORANA COUPLING LIMIT

This section establishes the connection between $(-\mathbb{1})^{n+\alpha^\dagger \alpha}$-independent dot dynamics and the presence of Majorana modes. We have already argued below Eqs. (S25) and (S28) that $\mathcal{Q} = 1$ can only be fulfilled if all modes couple equally particle- and hole like $|u_i| = |v_i| = 1/\sqrt{2}$, and with equal phase $\phi_i = \phi$ up to a $\pi$ shift. The general dot-subgap Hamiltonian in Eq. 1 then simplifies to

$$H_{\text{d,SC}} \overset{\mathcal{Q}=1}{\to} \epsilon n + \lambda_{\text{eff}} (d^\dagger - d)\gamma + \sum_i \epsilon_i n_i \quad , \quad \gamma = \alpha + \alpha^\dagger. \tag{S29}$$

For Majorana modes with $\epsilon_i = 0$, this Hamiltonian evidently conserves $p_{\text{d}\alpha} = (-\mathbb{1})^{n+\alpha^\dagger \alpha}$ and admits a representation with two identical blocks for the two parities $p_{\text{d}\alpha} = \pm 1$, thus leading to complete independence of the steady-state dot dynamics from any $(-\mathbb{1})^{n_i}$ *for any choice of* $\epsilon, \lambda_i$. Such independence more generally requires the existence of at least one excitation $c_i = \zeta_i^+ \alpha_i^\dagger + \zeta_i^- \alpha_i$ *for each independent mode $i$* that commutes with the *total* Hamiltonian $H_{\text{tot}}$ as given Eq. (S1). Equating $[H_{\text{tot}}, c_i] = [H_{\text{d,SC}}, c_i] = 0$ with $H_{\text{d,SC}}$ given in Eq. 1, and subsequently applying $\left\{\bullet, \alpha_i^\dagger\right\}, \left\{\bullet, \alpha_i\right\}, \left\{\bullet, d^\dagger\right\}, \left\{\bullet, d\right\}$, we obtain

$$\begin{pmatrix} \epsilon_i & 0 \\ 0 & -\epsilon_i \end{pmatrix} \cdot \begin{pmatrix} \zeta_i^+ \\ \zeta_i^- \end{pmatrix} = \begin{pmatrix} 0 \\ 0 \end{pmatrix} \quad , \quad \begin{pmatrix} e^{i\phi_i}|u_i| & e^{i\phi_i}|v_i| \\ -e^{-i\phi_i}|v_i| & -e^{-i\phi_i}|u_i| \end{pmatrix} \cdot \begin{pmatrix} \zeta_i^+ \\ \zeta_i^- \end{pmatrix} = \begin{pmatrix} 0 \\ 0 \end{pmatrix}. \tag{S30}$$

A solution different from the trivial one $\zeta_i^\pm = 0$ only exists for $\epsilon_i = 0$ and $|u_i| = |v_i|$. For zero-energy modes $\epsilon_i = 0$, we, however, already know that the dot dynamics are only completely parity independent for $\mathcal{Q} = 1$, additionally requiring $\phi_i = \phi_{i'} + z\pi$ with $z \in \mathbb{Z}$. This confirms that the Majorana limit $\epsilon_i = 0, |u_i| = |v_i|, \phi_i = \phi_{i'} + z\pi$ is indeed necessary and sufficient for parity-independent dynamics with arbitrary converter dot level $\epsilon$ and couplings $\lambda_i$.

## VI. PARITY BLOCKADE

We show in this section that a state with *permanently* blocked dynamics despite finite couplings $\lambda_i \neq 0$ for at least one mode $i$ *only* exists if all $\epsilon_i = 0$ and $\mathcal{Q} = 0$. First, a blocked dot at *any* time $t$ implies that the system resides in a normalized state $|\Phi\rangle$ for which $[H_{\text{tot}}, n]|\Phi\rangle = [H_{\text{d,SC}}, n]|\Phi\rangle = 0$. This is only possible if $\det([H_{\text{d,SC}}, n]) = 0$, which by expressing $H_{\text{d,SC}}$ as in Eq. 4 reduces to either $\lambda^+ = 0$ or $\lambda^- = 0$. Since only $\lambda^- = 0$ is possible in our gauge [Eq. (S28)], we require $\mathcal{Q} = 0$ and a prepared blocked state with odd subparity in the subspace of dot and effective mode, $p_{\text{d}\alpha}|\Phi\rangle = -|\Phi\rangle$. A permanent blocking then requires that this specific subparity $p_{\text{d}\alpha}$ remains constant, $[H_{\text{tot}}, p_{\text{d}\alpha}] = [H_{\text{d,SC}}, p_{\text{d}\alpha}] = 0$, which by Eq. 4 simplifies to

$$\left[\sum_{i=1}^N \epsilon_i n_i, \alpha^\dagger \alpha\right] = \sum_{i=1}^N \epsilon_i \left(\alpha^\dagger \alpha_i^\dagger \{\alpha_i, \alpha\} - \alpha^\dagger \{\alpha_i^\dagger, \alpha\} \alpha_i + \alpha_i^\dagger \{\alpha_i, \alpha^\dagger\} \alpha - \{\alpha_i^\dagger, \alpha^\dagger\} \alpha_i \alpha\right) \overset{!}{=} 0. \tag{S31}$$

We evaluate this using Eq. (S24) for the effective-mode creation operator $\alpha^\dagger$ with $\mathcal{Q} = 0 \Rightarrow \lambda^- = 0$, and exclude both the single-mode case $N = 1$ and uncoupled modes $\lambda_i = 0$, for which Eq. (S31) is trivially fulfilled. Comparing coefficients for linearly independent operators, this eventually yields the two conditions

$$(\epsilon_i + \epsilon_j)\left(|u_i||v_j| - e^{-i2(\phi_i - \phi_j)}|u_j||v_i|\right) = 0 \quad , \quad (\epsilon_i - \epsilon_j)\left(|u_i||u_j| + e^{-i2(\phi_i - \phi_j)}|v_i||v_j|\right) = 0 \tag{S32}$$

for any pair of modes $i \neq j$. These conditions cannot be fulfilled simultaneously with proper normalization $|u_i|^2 + |v_i|^2 = 1$ if there is any two modes $i \neq j$ for which $|\epsilon_i| \neq |\epsilon_j|$. Hence, we either have only zero modes, $\epsilon_i = 0$ for all $0 \leq i \leq N$, or two groups of modes, one with energy $\epsilon_i = \epsilon_{\text{sg}} \neq 0$ and another group with $\epsilon_j = -\epsilon_{\text{sg}} \neq 0$. The sign of any finite $\epsilon_i$ in $H_{\text{d,SC}}$ can be freely changed by substituting particles with holes, $\alpha_i^\dagger \leftrightarrow \alpha_i$, since this only adds an irrelevant constant to $H_{\text{d,SC}}$ that drops out of the commutators $[\bullet, H_{\text{d,SC}}]$ leading to Eq. (S31). We can therefore assume that Eq. (S32) can only be fulfilled if all modes have equal energy $\epsilon_i = \epsilon_{\text{sg}}$. For $\epsilon_{\text{sg}} \neq 0$, the first relation in Eq. (S32) dictates $|u_i| = |u_j|$ and $\phi_i = \phi_j + z\pi$ with $z \in \mathbb{Z}$ for any pair of modes $i \neq j$. Equation (S28) shows that this is only compatible with $\mathcal{Q} = 0$ if the modes couple either all perfectly particle-like, $|u_i| = 1$, or all perfectly hole-like, $|v_i| = 1$. This case reduces to an ordinary Pauli blockade scenario between the dot and a perfectly particle- or hole-like effective mode that attains an overall energy shift from $\sum_{i=1}^N \epsilon_i n_i \to \epsilon_{\text{sg}} \sum_{i=1}^N n_i$. Nontrivial parity blockade with particle-hole interference, however, requires all modes to be zero-energy modes.

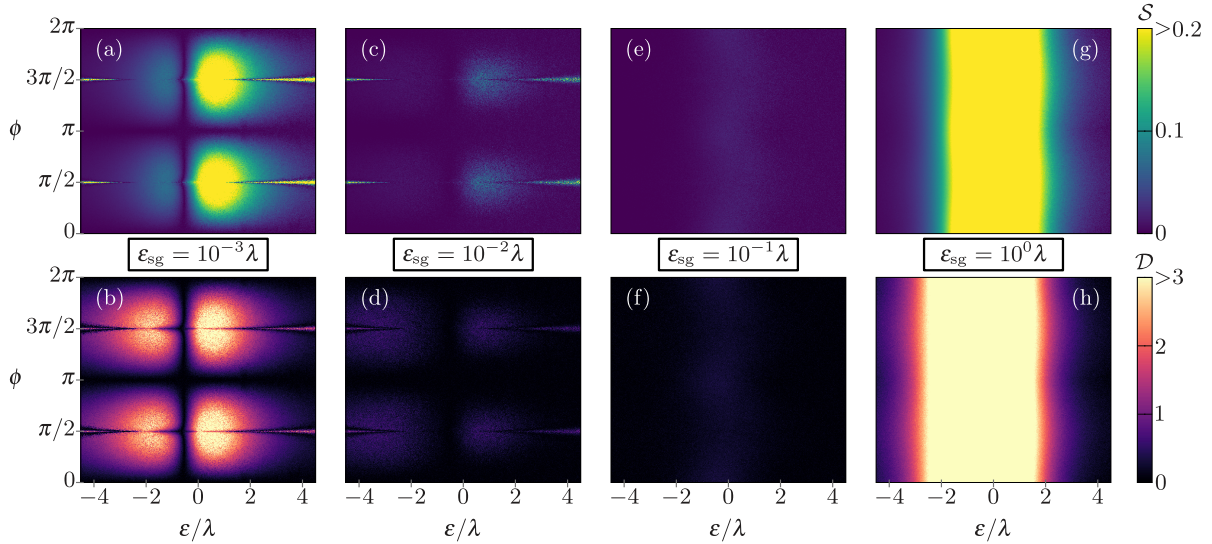

FIG. S4. Parity signals $\mathcal{S}$ and signal-to-noise ratios $\mathcal{D}$ for one coupled mode per wire ($N = 2$) with equal particle-hole mixing ($|u_i|^2 = 0.5$) but different subgap energies $\epsilon_2 = 2\epsilon_1 = 2\epsilon_{\rm sg} > 0$. Parameters: $\lambda_i = \lambda$, $\epsilon_{\rm S} = 0$, $E_{\rm cap} = 2\lambda = 6T = 6\delta\mu = 20\Gamma$, $2\pi/\Omega = 4t_{\rm b} = 2\pi \times 2400/\lambda = 10^4 \delta t = 1\mu{\rm s}$, $N_{\rm osc} = 2$, $N_{\rm T} = 10$, $M_{\rm T} = 50$.

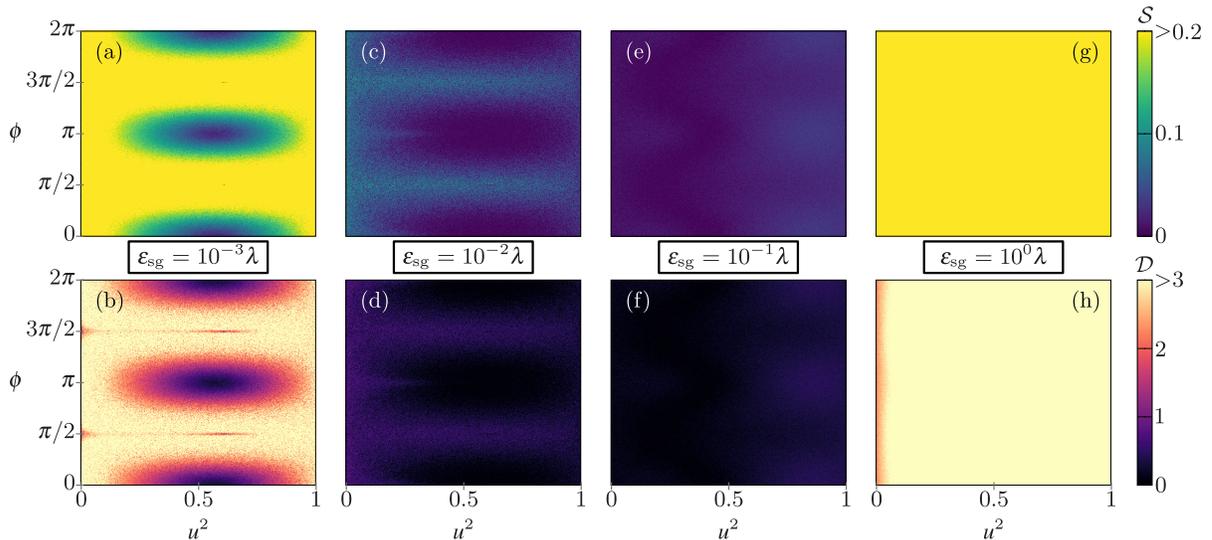

FIG. S5. Parity signals $\mathcal{S}$ and signal-to-noise ratios $\mathcal{D}$ for one coupled mode per wire ($N = 2$) and different subgap energies $\epsilon_2 = 2\epsilon_1 = 2\epsilon_{\rm sg} > 0$, where $|u_1|^2 = 4|u_2|^2/3 = u^2$ and $\epsilon = 3\lambda/4$. Parameters: $\lambda_i = \lambda$, $\epsilon_{\rm S} = 0$, $E_{\rm cap} = 2\lambda = 6T = 6\delta\mu = 20\Gamma$, $2\pi/\Omega = 4t_{\rm b} = 2\pi \times 2400/\lambda = 10^4 \delta t = 1\mu{\rm s}$, $N_{\rm osc} = 2$, $N_{\rm T} = 10$, $M_{\rm T} = 50$.

## VII. ADDITIONAL RESULTS IN SUPPORT OF MAIN TEXT

This appendix provides some further data supporting the main statements of the paper. In Fig. S4 and Fig. S5, we explicitly demonstrate that the suppressed flux($\phi$)-dependence at finite subgap energies $\epsilon_i \neq 0$ shown in Fig. 3 of the main text does not only occur at specific converter levels $\epsilon$ and equal particle hole mixing $|u_i|^2 = 0.5$, but affects nearly the entire $\mathcal{S}(\epsilon, \phi), \mathcal{D}(\epsilon, \phi)$ profile and a broad range of $|u_i|^2$.

In Fig. S6, we confirm that our results do not critically depend on any lead-SD-coupling asymmetry, $\Gamma_{\rm R} \neq \Gamma_{\rm L}$. In agreement with the well-known theory of electronic transport through quantum dots and metallic islands [8], the conductance and hence the signal $\mathcal{S}$ reduces with any finite coupling asymmetry $|\Gamma_{\rm L} - \Gamma_{\rm R}|/|\Gamma_{\rm L} + \Gamma_{\rm R}| > 0$. However, for not too small coupling, the noise is also reduced, and hence partly compensates the effect on the signal-to-noise




\begin{thebibliography}{94}%
\makeatletter
\providecommand \@ifxundefined [1]{%
 \@ifx{#1\undefined}
}%
\providecommand \@ifnum [1]{%
 \ifnum #1\expandafter \@firstoftwo
 \else \expandafter \@secondoftwo
 \fi
}%
\providecommand \@ifx [1]{%
 \ifx #1\expandafter \@firstoftwo
 \else \expandafter \@secondoftwo
 \fi
}%
\providecommand \natexlab [1]{#1}%
\providecommand \enquote  [1]{``#1''}%
\providecommand \bibnamefont  [1]{#1}%
\providecommand \bibfnamefont [1]{#1}%
\providecommand \citenamefont [1]{#1}%
\providecommand \href@noop [0]{\@secondoftwo}%
\providecommand \href [0]{\begingroup \@sanitize@url \@href}%
\providecommand \@href[1]{\@@startlink{#1}\@@href}%
\providecommand \@@href[1]{\endgroup#1\@@endlink}%
\providecommand \@sanitize@url [0]{\catcode `\\12\catcode `\$12\catcode
  `\&12\catcode `\#12\catcode `\^12\catcode `\_12\catcode `\%12\relax}%
\providecommand \@@startlink[1]{}%
\providecommand \@@endlink[0]{}%
\providecommand \url  [0]{\begingroup\@sanitize@url \@url }%
\providecommand \@url [1]{\endgroup\@href {#1}{\urlprefix }}%
\providecommand \urlprefix  [0]{URL }%
\providecommand \Eprint [0]{\href }%
\providecommand \doibase [0]{http://dx.doi.org/}%
\providecommand \selectlanguage [0]{\@gobble}%
\providecommand \bibinfo  [0]{\@secondoftwo}%
\providecommand \bibfield  [0]{\@secondoftwo}%
\providecommand \translation [1]{[#1]}%
\providecommand \BibitemOpen [0]{}%
\providecommand \bibitemStop [0]{}%
\providecommand \bibitemNoStop [0]{.\EOS\space}%
\providecommand \EOS [0]{\spacefactor3000\relax}%
\providecommand \BibitemShut  [1]{\csname bibitem#1\endcsname}%
\let\auto@bib@innerbib\@empty
\bibitem [{\citenamefont {Kitaev}(2001)}]{Kitaev2001Oct}%
  \BibitemOpen
  \bibfield  {author} {\bibinfo {author} {\bibfnamefont {A.~Y.}\ \bibnamefont
  {Kitaev}},\ }\href {\doibase 10.1070/1063-7869/44/10s/s29} {\bibfield
  {journal} {\bibinfo  {journal} {Phys.-Usp.}\ }\textbf {\bibinfo {volume}
  {44}},\ \bibinfo {pages} {131} (\bibinfo {year} {2001})}\BibitemShut
  {NoStop}%
\bibitem [{\citenamefont {Das~Sarma}\ \emph {et~al.}(2005)\citenamefont
  {Das~Sarma}, \citenamefont {Freedman},\ and\ \citenamefont
  {Nayak}}]{DasSarma2005Apr}%
  \BibitemOpen
  \bibfield  {author} {\bibinfo {author} {\bibfnamefont {S.}~\bibnamefont
  {Das~Sarma}}, \bibinfo {author} {\bibfnamefont {M.}~\bibnamefont {Freedman}},
  \ and\ \bibinfo {author} {\bibfnamefont {C.}~\bibnamefont {Nayak}},\ }\href
  {\doibase 10.1103/PhysRevLett.94.166802} {\bibfield  {journal} {\bibinfo
  {journal} {Phys. Rev. Lett.}\ }\textbf {\bibinfo {volume} {94}},\ \bibinfo
  {pages} {166802} (\bibinfo {year} {2005})}\BibitemShut {NoStop}%
\bibitem [{\citenamefont {Stern}(2008)}]{Stern2008Jan}%
  \BibitemOpen
  \bibfield  {author} {\bibinfo {author} {\bibfnamefont {A.}~\bibnamefont
  {Stern}},\ }\href {\doibase 10.1016/j.aop.2007.10.008} {\bibfield  {journal}
  {\bibinfo  {journal} {Ann. Phys.}\ }\textbf {\bibinfo {volume} {323}},\
  \bibinfo {pages} {204} (\bibinfo {year} {2008})}\BibitemShut {NoStop}%
\bibitem [{\citenamefont {Hassler}\ \emph {et~al.}(2010)\citenamefont
  {Hassler}, \citenamefont {Akhmerov}, \citenamefont {Hou},\ and\ \citenamefont
  {Beenakker}}]{Hassler2010Dec}%
  \BibitemOpen
  \bibfield  {author} {\bibinfo {author} {\bibfnamefont {F.}~\bibnamefont
  {Hassler}}, \bibinfo {author} {\bibfnamefont {A.~R.}\ \bibnamefont
  {Akhmerov}}, \bibinfo {author} {\bibfnamefont {C.-Y.}\ \bibnamefont {Hou}}, \
  and\ \bibinfo {author} {\bibfnamefont {C.~W.~J.}\ \bibnamefont {Beenakker}},\
  }\href {\doibase 10.1088/1367-2630/12/12/125002} {\bibfield  {journal}
  {\bibinfo  {journal} {New J. Phys.}\ }\textbf {\bibinfo {volume} {12}},\
  \bibinfo {pages} {125002} (\bibinfo {year} {2010})}\BibitemShut {NoStop}%
\bibitem [{\citenamefont {Alicea}\ \emph {et~al.}(2011)\citenamefont {Alicea},
  \citenamefont {Oreg}, \citenamefont {Refael}, \citenamefont {von Oppen},\
  and\ \citenamefont {Fisher}}]{Alicea2011Feb}%
  \BibitemOpen
  \bibfield  {author} {\bibinfo {author} {\bibfnamefont {J.}~\bibnamefont
  {Alicea}}, \bibinfo {author} {\bibfnamefont {Y.}~\bibnamefont {Oreg}},
  \bibinfo {author} {\bibfnamefont {G.}~\bibnamefont {Refael}}, \bibinfo
  {author} {\bibfnamefont {F.}~\bibnamefont {von Oppen}}, \ and\ \bibinfo
  {author} {\bibfnamefont {M.~P.~A.}\ \bibnamefont {Fisher}},\ }\href {\doibase
  10.1038/nphys1915} {\bibfield  {journal} {\bibinfo  {journal} {Nat. Phys.}\
  }\textbf {\bibinfo {volume} {7}},\ \bibinfo {pages} {412} (\bibinfo {year}
  {2011})}\BibitemShut {NoStop}%
\bibitem [{\citenamefont {Flensberg}(2011)}]{Flensberg2011Mar}%
  \BibitemOpen
  \bibfield  {author} {\bibinfo {author} {\bibfnamefont {K.}~\bibnamefont
  {Flensberg}},\ }\href {\doibase 10.1103/PhysRevLett.106.090503} {\bibfield
  {journal} {\bibinfo  {journal} {Phys. Rev. Lett.}\ }\textbf {\bibinfo
  {volume} {106}},\ \bibinfo {pages} {090503} (\bibinfo {year}
  {2011})}\BibitemShut {NoStop}%
\bibitem [{\citenamefont {Hassler}\ \emph {et~al.}(2011)\citenamefont
  {Hassler}, \citenamefont {Akhmerov},\ and\ \citenamefont
  {Beenakker}}]{Hassler2011Sep}%
  \BibitemOpen
  \bibfield  {author} {\bibinfo {author} {\bibfnamefont {F.}~\bibnamefont
  {Hassler}}, \bibinfo {author} {\bibfnamefont {A.~R.}\ \bibnamefont
  {Akhmerov}}, \ and\ \bibinfo {author} {\bibfnamefont {C.~W.~J.}\ \bibnamefont
  {Beenakker}},\ }\href {\doibase 10.1088/1367-2630/13/9/095004} {\bibfield
  {journal} {\bibinfo  {journal} {New J. Phys.}\ }\textbf {\bibinfo {volume}
  {13}},\ \bibinfo {pages} {095004} (\bibinfo {year} {2011})}\BibitemShut
  {NoStop}%
\bibitem [{\citenamefont {Hyart}\ \emph {et~al.}(2013)\citenamefont {Hyart},
  \citenamefont {van Heck}, \citenamefont {Fulga}, \citenamefont {Burrello},
  \citenamefont {Akhmerov},\ and\ \citenamefont {Beenakker}}]{Hyart2013Jul}%
  \BibitemOpen
  \bibfield  {author} {\bibinfo {author} {\bibfnamefont {T.}~\bibnamefont
  {Hyart}}, \bibinfo {author} {\bibfnamefont {B.}~\bibnamefont {van Heck}},
  \bibinfo {author} {\bibfnamefont {I.~C.}\ \bibnamefont {Fulga}}, \bibinfo
  {author} {\bibfnamefont {M.}~\bibnamefont {Burrello}}, \bibinfo {author}
  {\bibfnamefont {A.~R.}\ \bibnamefont {Akhmerov}}, \ and\ \bibinfo {author}
  {\bibfnamefont {C.~W.~J.}\ \bibnamefont {Beenakker}},\ }\href {\doibase
  10.1103/PhysRevB.88.035121} {\bibfield  {journal} {\bibinfo  {journal} {Phys.
  Rev. B}\ }\textbf {\bibinfo {volume} {88}},\ \bibinfo {pages} {035121}
  (\bibinfo {year} {2013})}\BibitemShut {NoStop}%
\bibitem [{\citenamefont {Aasen}\ \emph {et~al.}(2016)\citenamefont {Aasen},
  \citenamefont {Hell}, \citenamefont {Mishmash}, \citenamefont {Higginbotham},
  \citenamefont {Danon}, \citenamefont {Leijnse}, \citenamefont {Jespersen},
  \citenamefont {Folk}, \citenamefont {Marcus}, \citenamefont {Flensberg},\
  and\ \citenamefont {Alicea}}]{Aasen2016Aug}%
  \BibitemOpen
  \bibfield  {author} {\bibinfo {author} {\bibfnamefont {D.}~\bibnamefont
  {Aasen}}, \bibinfo {author} {\bibfnamefont {M.}~\bibnamefont {Hell}},
  \bibinfo {author} {\bibfnamefont {R.~V.}\ \bibnamefont {Mishmash}}, \bibinfo
  {author} {\bibfnamefont {A.}~\bibnamefont {Higginbotham}}, \bibinfo {author}
  {\bibfnamefont {J.}~\bibnamefont {Danon}}, \bibinfo {author} {\bibfnamefont
  {M.}~\bibnamefont {Leijnse}}, \bibinfo {author} {\bibfnamefont {T.~S.}\
  \bibnamefont {Jespersen}}, \bibinfo {author} {\bibfnamefont {J.~A.}\
  \bibnamefont {Folk}}, \bibinfo {author} {\bibfnamefont {C.~M.}\ \bibnamefont
  {Marcus}}, \bibinfo {author} {\bibfnamefont {K.}~\bibnamefont {Flensberg}}, \
  and\ \bibinfo {author} {\bibfnamefont {J.}~\bibnamefont {Alicea}},\ }\href
  {\doibase 10.1103/PhysRevX.6.031016} {\bibfield  {journal} {\bibinfo
  {journal} {Phys. Rev. X}\ }\textbf {\bibinfo {volume} {6}},\ \bibinfo {pages}
  {031016} (\bibinfo {year} {2016})}\BibitemShut {NoStop}%
\bibitem [{\citenamefont {Plugge}\ \emph {et~al.}(2017)\citenamefont {Plugge},
  \citenamefont {Rasmussen}, \citenamefont {Egger},\ and\ \citenamefont
  {Flensberg}}]{Plugge2017Jan}%
  \BibitemOpen
  \bibfield  {author} {\bibinfo {author} {\bibfnamefont {S.}~\bibnamefont
  {Plugge}}, \bibinfo {author} {\bibfnamefont {A.}~\bibnamefont {Rasmussen}},
  \bibinfo {author} {\bibfnamefont {R.}~\bibnamefont {Egger}}, \ and\ \bibinfo
  {author} {\bibfnamefont {K.}~\bibnamefont {Flensberg}},\ }\href {\doibase
  10.1088/1367-2630/aa54e1} {\bibfield  {journal} {\bibinfo  {journal} {New J.
  Phys.}\ }\textbf {\bibinfo {volume} {19}},\ \bibinfo {pages} {012001}
  (\bibinfo {year} {2017})}\BibitemShut {NoStop}%
\bibitem [{\citenamefont {Karzig}\ \emph {et~al.}(2017)\citenamefont {Karzig},
  \citenamefont {Knapp}, \citenamefont {Lutchyn}, \citenamefont {Bonderson},
  \citenamefont {Hastings}, \citenamefont {Nayak}, \citenamefont {Alicea},
  \citenamefont {Flensberg}, \citenamefont {Plugge}, \citenamefont {Oreg},
  \citenamefont {Marcus},\ and\ \citenamefont {Freedman}}]{Karzig2017Jun}%
  \BibitemOpen
  \bibfield  {author} {\bibinfo {author} {\bibfnamefont {T.}~\bibnamefont
  {Karzig}}, \bibinfo {author} {\bibfnamefont {C.}~\bibnamefont {Knapp}},
  \bibinfo {author} {\bibfnamefont {R.~M.}\ \bibnamefont {Lutchyn}}, \bibinfo
  {author} {\bibfnamefont {P.}~\bibnamefont {Bonderson}}, \bibinfo {author}
  {\bibfnamefont {M.~B.}\ \bibnamefont {Hastings}}, \bibinfo {author}
  {\bibfnamefont {C.}~\bibnamefont {Nayak}}, \bibinfo {author} {\bibfnamefont
  {J.}~\bibnamefont {Alicea}}, \bibinfo {author} {\bibfnamefont
  {K.}~\bibnamefont {Flensberg}}, \bibinfo {author} {\bibfnamefont
  {S.}~\bibnamefont {Plugge}}, \bibinfo {author} {\bibfnamefont
  {Y.}~\bibnamefont {Oreg}}, \bibinfo {author} {\bibfnamefont {C.~M.}\
  \bibnamefont {Marcus}}, \ and\ \bibinfo {author} {\bibfnamefont {M.~H.}\
  \bibnamefont {Freedman}},\ }\href {\doibase 10.1103/PhysRevB.95.235305}
  {\bibfield  {journal} {\bibinfo  {journal} {Phys. Rev. B}\ }\textbf {\bibinfo
  {volume} {95}},\ \bibinfo {pages} {235305} (\bibinfo {year}
  {2017})}\BibitemShut {NoStop}%
\bibitem [{\citenamefont {Manousakis}\ \emph {et~al.}(2017)\citenamefont
  {Manousakis}, \citenamefont {Altland}, \citenamefont {Bagrets}, \citenamefont
  {Egger},\ and\ \citenamefont {Ando}}]{Manousakis2017Apr}%
  \BibitemOpen
  \bibfield  {author} {\bibinfo {author} {\bibfnamefont {J.}~\bibnamefont
  {Manousakis}}, \bibinfo {author} {\bibfnamefont {A.}~\bibnamefont {Altland}},
  \bibinfo {author} {\bibfnamefont {D.}~\bibnamefont {Bagrets}}, \bibinfo
  {author} {\bibfnamefont {R.}~\bibnamefont {Egger}}, \ and\ \bibinfo {author}
  {\bibfnamefont {Y.}~\bibnamefont {Ando}},\ }\href {\doibase
  10.1103/PhysRevB.95.165424} {\bibfield  {journal} {\bibinfo  {journal} {Phys.
  Rev. B}\ }\textbf {\bibinfo {volume} {95}},\ \bibinfo {pages} {165424}
  (\bibinfo {year} {2017})}\BibitemShut {NoStop}%
\bibitem [{\citenamefont {Fu}(2010)}]{Fu2010Feb}%
  \BibitemOpen
  \bibfield  {author} {\bibinfo {author} {\bibfnamefont {L.}~\bibnamefont
  {Fu}},\ }\href {\doibase 10.1103/PhysRevLett.104.056402} {\bibfield
  {journal} {\bibinfo  {journal} {Phys. Rev. Lett.}\ }\textbf {\bibinfo
  {volume} {104}},\ \bibinfo {pages} {056402} (\bibinfo {year}
  {2010})}\BibitemShut {NoStop}%
\bibitem [{\citenamefont {Hell}\ \emph {et~al.}(2018)\citenamefont {Hell},
  \citenamefont {Flensberg},\ and\ \citenamefont {Leijnse}}]{Hell2018Apr}%
  \BibitemOpen
  \bibfield  {author} {\bibinfo {author} {\bibfnamefont {M.}~\bibnamefont
  {Hell}}, \bibinfo {author} {\bibfnamefont {K.}~\bibnamefont {Flensberg}}, \
  and\ \bibinfo {author} {\bibfnamefont {M.}~\bibnamefont {Leijnse}},\ }\href
  {\doibase 10.1103/PhysRevB.97.161401} {\bibfield  {journal} {\bibinfo
  {journal} {Phys. Rev. B}\ }\textbf {\bibinfo {volume} {97}},\ \bibinfo
  {pages} {161401(R)} (\bibinfo {year} {2018})}\BibitemShut {NoStop}%
\bibitem [{\citenamefont {Drukier}\ \emph {et~al.}(2018)\citenamefont
  {Drukier}, \citenamefont {Zirnstein}, \citenamefont {Rosenow}, \citenamefont
  {Stern},\ and\ \citenamefont {Oreg}}]{Drukier2018Oct}%
  \BibitemOpen
  \bibfield  {author} {\bibinfo {author} {\bibfnamefont {C.}~\bibnamefont
  {Drukier}}, \bibinfo {author} {\bibfnamefont {H.-G.}\ \bibnamefont
  {Zirnstein}}, \bibinfo {author} {\bibfnamefont {B.}~\bibnamefont {Rosenow}},
  \bibinfo {author} {\bibfnamefont {A.}~\bibnamefont {Stern}}, \ and\ \bibinfo
  {author} {\bibfnamefont {Y.}~\bibnamefont {Oreg}},\ }\href {\doibase
  10.1103/PhysRevB.98.161401} {\bibfield  {journal} {\bibinfo  {journal} {Phys.
  Rev. B}\ }\textbf {\bibinfo {volume} {98}},\ \bibinfo {pages} {161401}
  (\bibinfo {year} {2018})}\BibitemShut {NoStop}%
\bibitem [{\citenamefont {Whiticar}\ \emph {et~al.}(2020)\citenamefont
  {Whiticar}, \citenamefont {Fornieri}, \citenamefont {O{'}Farrell},
  \citenamefont {Drachmann}, \citenamefont {Wang}, \citenamefont {Thomas},
  \citenamefont {Gronin}, \citenamefont {Kallaher}, \citenamefont {Gardner},
  \citenamefont {Manfra}, \citenamefont {Marcus},\ and\ \citenamefont
  {Nichele}}]{Whiticar2020Jun}%
  \BibitemOpen
  \bibfield  {author} {\bibinfo {author} {\bibfnamefont {A.~M.}\ \bibnamefont
  {Whiticar}}, \bibinfo {author} {\bibfnamefont {A.}~\bibnamefont {Fornieri}},
  \bibinfo {author} {\bibfnamefont {E.~C.~T.}\ \bibnamefont {O{'}Farrell}},
  \bibinfo {author} {\bibfnamefont {A.~C.~C.}\ \bibnamefont {Drachmann}},
  \bibinfo {author} {\bibfnamefont {T.}~\bibnamefont {Wang}}, \bibinfo {author}
  {\bibfnamefont {C.}~\bibnamefont {Thomas}}, \bibinfo {author} {\bibfnamefont
  {S.}~\bibnamefont {Gronin}}, \bibinfo {author} {\bibfnamefont
  {R.}~\bibnamefont {Kallaher}}, \bibinfo {author} {\bibfnamefont {G.~C.}\
  \bibnamefont {Gardner}}, \bibinfo {author} {\bibfnamefont {M.~J.}\
  \bibnamefont {Manfra}}, \bibinfo {author} {\bibfnamefont {C.~M.}\
  \bibnamefont {Marcus}}, \ and\ \bibinfo {author} {\bibfnamefont
  {F.}~\bibnamefont {Nichele}},\ }\href {\doibase 10.1038/s41467-020-16988-x}
  {\bibfield  {journal} {\bibinfo  {journal} {Nat. Commun.}\ }\textbf {\bibinfo
  {volume} {11}},\ \bibinfo {pages} {1} (\bibinfo {year} {2020})}\BibitemShut
  {NoStop}%
\bibitem [{\citenamefont {Pikulin}\ \emph {et~al.}(2021)\citenamefont
  {Pikulin}, \citenamefont {van Heck}, \citenamefont {Karzig}, \citenamefont
  {Martinez}, \citenamefont {Nijholt}, \citenamefont {Laeven}, \citenamefont
  {Winkler}, \citenamefont {Watson}, \citenamefont {Heedt}, \citenamefont
  {Temurhan}, \citenamefont {Svidenko}, \citenamefont {Lutchyn}, \citenamefont
  {Thomas}, \citenamefont {de~Lange}, \citenamefont {Casparis},\ and\
  \citenamefont {Nayak}}]{Pikulin2021Mar}%
  \BibitemOpen
  \bibfield  {author} {\bibinfo {author} {\bibfnamefont {D.~I.}\ \bibnamefont
  {Pikulin}}, \bibinfo {author} {\bibfnamefont {B.}~\bibnamefont {van Heck}},
  \bibinfo {author} {\bibfnamefont {T.}~\bibnamefont {Karzig}}, \bibinfo
  {author} {\bibfnamefont {E.~A.}\ \bibnamefont {Martinez}}, \bibinfo {author}
  {\bibfnamefont {B.}~\bibnamefont {Nijholt}}, \bibinfo {author} {\bibfnamefont
  {T.}~\bibnamefont {Laeven}}, \bibinfo {author} {\bibfnamefont {G.~W.}\
  \bibnamefont {Winkler}}, \bibinfo {author} {\bibfnamefont {J.~D.}\
  \bibnamefont {Watson}}, \bibinfo {author} {\bibfnamefont {S.}~\bibnamefont
  {Heedt}}, \bibinfo {author} {\bibfnamefont {M.}~\bibnamefont {Temurhan}},
  \bibinfo {author} {\bibfnamefont {V.}~\bibnamefont {Svidenko}}, \bibinfo
  {author} {\bibfnamefont {R.~M.}\ \bibnamefont {Lutchyn}}, \bibinfo {author}
  {\bibfnamefont {M.}~\bibnamefont {Thomas}}, \bibinfo {author} {\bibfnamefont
  {G.}~\bibnamefont {de~Lange}}, \bibinfo {author} {\bibfnamefont
  {L.}~\bibnamefont {Casparis}}, \ and\ \bibinfo {author} {\bibfnamefont
  {C.}~\bibnamefont {Nayak}},\ }\href
  {https://doi.org/10.48550/arXiv.2103.12217} {\bibfield  {journal} {\bibinfo
  {journal} {arXiv}\ } (\bibinfo {year} {2021})},\ \Eprint
  {http://arxiv.org/abs/2103.12217} {2103.12217} \BibitemShut {NoStop}%
\bibitem [{\citenamefont {Cayao}\ and\ \citenamefont
  {Black-Schaffer}(2021)}]{Cayao2021Jul}%
  \BibitemOpen
  \bibfield  {author} {\bibinfo {author} {\bibfnamefont {J.}~\bibnamefont
  {Cayao}}\ and\ \bibinfo {author} {\bibfnamefont {A.~M.}\ \bibnamefont
  {Black-Schaffer}},\ }\href {\doibase 10.1103/PhysRevB.104.L020501} {\bibfield
   {journal} {\bibinfo  {journal} {Phys. Rev. B}\ }\textbf {\bibinfo {volume}
  {104}},\ \bibinfo {pages} {L020501} (\bibinfo {year} {2021})}\BibitemShut
  {NoStop}%
\bibitem [{\citenamefont {Aghaee}\ \emph {et~al.}(2022)\citenamefont {Aghaee},
  \citenamefont {Akkala}, \citenamefont {Alam}, \citenamefont {Ali},
  \citenamefont {Ramirez}, \citenamefont {Andrzejczuk}, \citenamefont
  {Antipov}, \citenamefont {Aseev}, \citenamefont {Astafev}, \citenamefont
  {Bauer}, \citenamefont {Becker}, \citenamefont {Boddapati}, \citenamefont
  {Boekhout}, \citenamefont {Bommer}, \citenamefont {Hansen}, \citenamefont
  {Bosma}, \citenamefont {Bourdet}, \citenamefont {Boutin}, \citenamefont
  {Caroff}, \citenamefont {Casparis}, \citenamefont {Cassidy}, \citenamefont
  {Christensen}, \citenamefont {Clay}, \citenamefont {Cole}, \citenamefont
  {Corsetti}, \citenamefont {Cui}, \citenamefont {Dalampiras}, \citenamefont
  {Dokania}, \citenamefont {de~Lange}, \citenamefont {de~Moor}, \citenamefont
  {Salda{\ifmmode\tilde{n}\else\~{n}\fi}a}, \citenamefont {Fallahi},
  \citenamefont {Fathabad}, \citenamefont {Gamble}, \citenamefont {Gardner},
  \citenamefont {Govender}, \citenamefont {Griggio}, \citenamefont {Grigoryan},
  \citenamefont {Gronin}, \citenamefont {Gukelberger}, \citenamefont {Heedt},
  \citenamefont {Zamorano}, \citenamefont {Ho}, \citenamefont {Holgaard},
  \citenamefont {Nielsen}, \citenamefont {Ingerslev}, \citenamefont
  {Krogstrup}, \citenamefont {Johansson}, \citenamefont {Jones}, \citenamefont
  {Kallaher}, \citenamefont {Karimi}, \citenamefont {Karzig}, \citenamefont
  {King}, \citenamefont {Kloster}, \citenamefont {Knapp}, \citenamefont
  {Kocon}, \citenamefont {Koski}, \citenamefont {Kostamo}, \citenamefont
  {Kumar}, \citenamefont {Laeven}, \citenamefont {Larsen}, \citenamefont {Li},
  \citenamefont {Lindemann}, \citenamefont {Love}, \citenamefont {Lutchyn},
  \citenamefont {Manfra}, \citenamefont {Memisevic}, \citenamefont {Nayak},
  \citenamefont {Nijholt}, \citenamefont {Madsen}, \citenamefont {Markussen},
  \citenamefont {Martinez}, \citenamefont {McNeil}, \citenamefont {Mullally},
  \citenamefont {Nielsen}, \citenamefont {Nurmohamed}, \citenamefont
  {O'Farrell}, \citenamefont {Otani}, \citenamefont {Pauka}, \citenamefont
  {Petersson}, \citenamefont {Petit}, \citenamefont {Pikulin}, \citenamefont
  {Preiss}, \citenamefont {Perez}, \citenamefont {Rasmussen}, \citenamefont
  {Rajpalke}, \citenamefont {Razmadze}, \citenamefont {Reentila}, \citenamefont
  {Reilly}, \citenamefont {Rouse}, \citenamefont {Sadovskyy}, \citenamefont
  {Sainiemi}, \citenamefont {Schreppler}, \citenamefont {Sidorkin},
  \citenamefont {Singh}, \citenamefont {Singh}, \citenamefont {Sinha},
  \citenamefont {Sohr}, \citenamefont
  {Stankevi{\ifmmode\check{c}\else\v{c}\fi}}, \citenamefont {Stek},
  \citenamefont {Suominen}, \citenamefont {Suter}, \citenamefont {Svidenko},
  \citenamefont {Teicher}, \citenamefont {Temuerhan}, \citenamefont
  {Thiyagarajah}, \citenamefont {Tholapi}, \citenamefont {Thomas},
  \citenamefont {Toomey}, \citenamefont {Upadhyay}, \citenamefont {Urban},
  \citenamefont {Vaitiek{\ifmmode\dot{e}\else\.{e}\fi}nas}, \citenamefont
  {Van~Hoogdalem}, \citenamefont {Viazmitinov}, \citenamefont {Waddy},
  \citenamefont {Van~Woerkom}, \citenamefont {Vogel}, \citenamefont {Watson},
  \citenamefont {Weston}, \citenamefont {Winkler}, \citenamefont {Yang},
  \citenamefont {Yau}, \citenamefont {Yi}, \citenamefont {Yucelen},
  \citenamefont {Webster}, \citenamefont {Zeisel},\ and\ \citenamefont
  {Zhao}}]{Aghaee2022Jul}%
  \BibitemOpen
  \bibfield  {author} {\bibinfo {author} {\bibfnamefont {M.}~\bibnamefont
  {Aghaee}}, \bibinfo {author} {\bibfnamefont {A.}~\bibnamefont {Akkala}},
  \bibinfo {author} {\bibfnamefont {Z.}~\bibnamefont {Alam}}, \bibinfo {author}
  {\bibfnamefont {R.}~\bibnamefont {Ali}}, \bibinfo {author} {\bibfnamefont
  {A.~A.}\ \bibnamefont {Ramirez}}, \bibinfo {author} {\bibfnamefont
  {M.}~\bibnamefont {Andrzejczuk}}, \bibinfo {author} {\bibfnamefont {A.~E.}\
  \bibnamefont {Antipov}}, \bibinfo {author} {\bibfnamefont {P.}~\bibnamefont
  {Aseev}}, \bibinfo {author} {\bibfnamefont {M.}~\bibnamefont {Astafev}},
  \bibinfo {author} {\bibfnamefont {B.}~\bibnamefont {Bauer}}, \bibinfo
  {author} {\bibfnamefont {J.}~\bibnamefont {Becker}}, \bibinfo {author}
  {\bibfnamefont {S.}~\bibnamefont {Boddapati}}, \bibinfo {author}
  {\bibfnamefont {F.}~\bibnamefont {Boekhout}}, \bibinfo {author}
  {\bibfnamefont {J.}~\bibnamefont {Bommer}}, \bibinfo {author} {\bibfnamefont
  {E.~B.}\ \bibnamefont {Hansen}}, \bibinfo {author} {\bibfnamefont
  {T.}~\bibnamefont {Bosma}}, \bibinfo {author} {\bibfnamefont
  {L.}~\bibnamefont {Bourdet}}, \bibinfo {author} {\bibfnamefont
  {S.}~\bibnamefont {Boutin}}, \bibinfo {author} {\bibfnamefont
  {P.}~\bibnamefont {Caroff}}, \bibinfo {author} {\bibfnamefont
  {L.}~\bibnamefont {Casparis}}, \bibinfo {author} {\bibfnamefont
  {M.}~\bibnamefont {Cassidy}}, \bibinfo {author} {\bibfnamefont {A.~W.}\
  \bibnamefont {Christensen}}, \bibinfo {author} {\bibfnamefont
  {N.}~\bibnamefont {Clay}}, \bibinfo {author} {\bibfnamefont {W.~S.}\
  \bibnamefont {Cole}}, \bibinfo {author} {\bibfnamefont {F.}~\bibnamefont
  {Corsetti}}, \bibinfo {author} {\bibfnamefont {A.}~\bibnamefont {Cui}},
  \bibinfo {author} {\bibfnamefont {P.}~\bibnamefont {Dalampiras}}, \bibinfo
  {author} {\bibfnamefont {A.}~\bibnamefont {Dokania}}, \bibinfo {author}
  {\bibfnamefont {G.}~\bibnamefont {de~Lange}}, \bibinfo {author}
  {\bibfnamefont {M.}~\bibnamefont {de~Moor}}, \bibinfo {author} {\bibfnamefont
  {J.~C.~E.}\ \bibnamefont {Salda{\ifmmode\tilde{n}\else\~{n}\fi}a}}, \bibinfo
  {author} {\bibfnamefont {S.}~\bibnamefont {Fallahi}}, \bibinfo {author}
  {\bibfnamefont {Z.~H.}\ \bibnamefont {Fathabad}}, \bibinfo {author}
  {\bibfnamefont {J.}~\bibnamefont {Gamble}}, \bibinfo {author} {\bibfnamefont
  {G.}~\bibnamefont {Gardner}}, \bibinfo {author} {\bibfnamefont
  {D.}~\bibnamefont {Govender}}, \bibinfo {author} {\bibfnamefont
  {F.}~\bibnamefont {Griggio}}, \bibinfo {author} {\bibfnamefont
  {R.}~\bibnamefont {Grigoryan}}, \bibinfo {author} {\bibfnamefont
  {S.}~\bibnamefont {Gronin}}, \bibinfo {author} {\bibfnamefont
  {J.}~\bibnamefont {Gukelberger}}, \bibinfo {author} {\bibfnamefont
  {S.}~\bibnamefont {Heedt}}, \bibinfo {author} {\bibfnamefont {J.~H.}\
  \bibnamefont {Zamorano}}, \bibinfo {author} {\bibfnamefont {S.}~\bibnamefont
  {Ho}}, \bibinfo {author} {\bibfnamefont {U.~L.}\ \bibnamefont {Holgaard}},
  \bibinfo {author} {\bibfnamefont {W.~H.~P.}\ \bibnamefont {Nielsen}},
  \bibinfo {author} {\bibfnamefont {H.}~\bibnamefont {Ingerslev}}, \bibinfo
  {author} {\bibfnamefont {P.~J.}\ \bibnamefont {Krogstrup}}, \bibinfo {author}
  {\bibfnamefont {L.}~\bibnamefont {Johansson}}, \bibinfo {author}
  {\bibfnamefont {J.}~\bibnamefont {Jones}}, \bibinfo {author} {\bibfnamefont
  {R.}~\bibnamefont {Kallaher}}, \bibinfo {author} {\bibfnamefont
  {F.}~\bibnamefont {Karimi}}, \bibinfo {author} {\bibfnamefont
  {T.}~\bibnamefont {Karzig}}, \bibinfo {author} {\bibfnamefont
  {C.}~\bibnamefont {King}}, \bibinfo {author} {\bibfnamefont {M.~E.}\
  \bibnamefont {Kloster}}, \bibinfo {author} {\bibfnamefont {C.}~\bibnamefont
  {Knapp}}, \bibinfo {author} {\bibfnamefont {D.}~\bibnamefont {Kocon}},
  \bibinfo {author} {\bibfnamefont {J.}~\bibnamefont {Koski}}, \bibinfo
  {author} {\bibfnamefont {P.}~\bibnamefont {Kostamo}}, \bibinfo {author}
  {\bibfnamefont {M.}~\bibnamefont {Kumar}}, \bibinfo {author} {\bibfnamefont
  {T.}~\bibnamefont {Laeven}}, \bibinfo {author} {\bibfnamefont
  {T.}~\bibnamefont {Larsen}}, \bibinfo {author} {\bibfnamefont
  {K.}~\bibnamefont {Li}}, \bibinfo {author} {\bibfnamefont {T.}~\bibnamefont
  {Lindemann}}, \bibinfo {author} {\bibfnamefont {J.}~\bibnamefont {Love}},
  \bibinfo {author} {\bibfnamefont {R.}~\bibnamefont {Lutchyn}}, \bibinfo
  {author} {\bibfnamefont {M.}~\bibnamefont {Manfra}}, \bibinfo {author}
  {\bibfnamefont {E.}~\bibnamefont {Memisevic}}, \bibinfo {author}
  {\bibfnamefont {C.}~\bibnamefont {Nayak}}, \bibinfo {author} {\bibfnamefont
  {B.}~\bibnamefont {Nijholt}}, \bibinfo {author} {\bibfnamefont {M.~H.}\
  \bibnamefont {Madsen}}, \bibinfo {author} {\bibfnamefont {S.}~\bibnamefont
  {Markussen}}, \bibinfo {author} {\bibfnamefont {E.}~\bibnamefont {Martinez}},
  \bibinfo {author} {\bibfnamefont {R.}~\bibnamefont {McNeil}}, \bibinfo
  {author} {\bibfnamefont {A.}~\bibnamefont {Mullally}}, \bibinfo {author}
  {\bibfnamefont {J.}~\bibnamefont {Nielsen}}, \bibinfo {author} {\bibfnamefont
  {A.}~\bibnamefont {Nurmohamed}}, \bibinfo {author} {\bibfnamefont
  {E.}~\bibnamefont {O'Farrell}}, \bibinfo {author} {\bibfnamefont
  {K.}~\bibnamefont {Otani}}, \bibinfo {author} {\bibfnamefont
  {S.}~\bibnamefont {Pauka}}, \bibinfo {author} {\bibfnamefont
  {K.}~\bibnamefont {Petersson}}, \bibinfo {author} {\bibfnamefont
  {L.}~\bibnamefont {Petit}}, \bibinfo {author} {\bibfnamefont
  {D.}~\bibnamefont {Pikulin}}, \bibinfo {author} {\bibfnamefont
  {F.}~\bibnamefont {Preiss}}, \bibinfo {author} {\bibfnamefont {M.~Q.}\
  \bibnamefont {Perez}}, \bibinfo {author} {\bibfnamefont {K.}~\bibnamefont
  {Rasmussen}}, \bibinfo {author} {\bibfnamefont {M.}~\bibnamefont {Rajpalke}},
  \bibinfo {author} {\bibfnamefont {D.}~\bibnamefont {Razmadze}}, \bibinfo
  {author} {\bibfnamefont {O.}~\bibnamefont {Reentila}}, \bibinfo {author}
  {\bibfnamefont {D.}~\bibnamefont {Reilly}}, \bibinfo {author} {\bibfnamefont
  {R.}~\bibnamefont {Rouse}}, \bibinfo {author} {\bibfnamefont
  {I.}~\bibnamefont {Sadovskyy}}, \bibinfo {author} {\bibfnamefont
  {L.}~\bibnamefont {Sainiemi}}, \bibinfo {author} {\bibfnamefont
  {S.}~\bibnamefont {Schreppler}}, \bibinfo {author} {\bibfnamefont
  {V.}~\bibnamefont {Sidorkin}}, \bibinfo {author} {\bibfnamefont
  {A.}~\bibnamefont {Singh}}, \bibinfo {author} {\bibfnamefont
  {S.}~\bibnamefont {Singh}}, \bibinfo {author} {\bibfnamefont
  {S.}~\bibnamefont {Sinha}}, \bibinfo {author} {\bibfnamefont
  {P.}~\bibnamefont {Sohr}}, \bibinfo {author} {\bibfnamefont {T.}~\bibnamefont
  {Stankevi{\ifmmode\check{c}\else\v{c}\fi}}}, \bibinfo {author} {\bibfnamefont
  {L.}~\bibnamefont {Stek}}, \bibinfo {author} {\bibfnamefont {H.}~\bibnamefont
  {Suominen}}, \bibinfo {author} {\bibfnamefont {J.}~\bibnamefont {Suter}},
  \bibinfo {author} {\bibfnamefont {V.}~\bibnamefont {Svidenko}}, \bibinfo
  {author} {\bibfnamefont {S.}~\bibnamefont {Teicher}}, \bibinfo {author}
  {\bibfnamefont {M.}~\bibnamefont {Temuerhan}}, \bibinfo {author}
  {\bibfnamefont {N.}~\bibnamefont {Thiyagarajah}}, \bibinfo {author}
  {\bibfnamefont {R.}~\bibnamefont {Tholapi}}, \bibinfo {author} {\bibfnamefont
  {M.}~\bibnamefont {Thomas}}, \bibinfo {author} {\bibfnamefont
  {E.}~\bibnamefont {Toomey}}, \bibinfo {author} {\bibfnamefont
  {S.}~\bibnamefont {Upadhyay}}, \bibinfo {author} {\bibfnamefont
  {I.}~\bibnamefont {Urban}}, \bibinfo {author} {\bibfnamefont
  {S.}~\bibnamefont {Vaitiek{\ifmmode\dot{e}\else\.{e}\fi}nas}}, \bibinfo
  {author} {\bibfnamefont {K.}~\bibnamefont {Van~Hoogdalem}}, \bibinfo {author}
  {\bibfnamefont {D.~V.}\ \bibnamefont {Viazmitinov}}, \bibinfo {author}
  {\bibfnamefont {S.}~\bibnamefont {Waddy}}, \bibinfo {author} {\bibfnamefont
  {D.}~\bibnamefont {Van~Woerkom}}, \bibinfo {author} {\bibfnamefont
  {D.}~\bibnamefont {Vogel}}, \bibinfo {author} {\bibfnamefont
  {J.}~\bibnamefont {Watson}}, \bibinfo {author} {\bibfnamefont
  {J.}~\bibnamefont {Weston}}, \bibinfo {author} {\bibfnamefont {G.~W.}\
  \bibnamefont {Winkler}}, \bibinfo {author} {\bibfnamefont {C.~K.}\
  \bibnamefont {Yang}}, \bibinfo {author} {\bibfnamefont {S.}~\bibnamefont
  {Yau}}, \bibinfo {author} {\bibfnamefont {D.}~\bibnamefont {Yi}}, \bibinfo
  {author} {\bibfnamefont {E.}~\bibnamefont {Yucelen}}, \bibinfo {author}
  {\bibfnamefont {A.}~\bibnamefont {Webster}}, \bibinfo {author} {\bibfnamefont
  {R.}~\bibnamefont {Zeisel}}, \ and\ \bibinfo {author} {\bibfnamefont
  {R.}~\bibnamefont {Zhao}},\ }\href
  {https://doi.org/10.48550/arXiv.2207.02472} {\bibfield  {journal} {\bibinfo
  {journal} {arXiv}\ } (\bibinfo {year} {2022})},\ \Eprint
  {http://arxiv.org/abs/2207.02472} {2207.02472} \BibitemShut {NoStop}%
\bibitem [{\citenamefont {Prada}\ \emph {et~al.}(2012)\citenamefont {Prada},
  \citenamefont {San-Jose},\ and\ \citenamefont {Aguado}}]{Prada2012Nov}%
  \BibitemOpen
  \bibfield  {author} {\bibinfo {author} {\bibfnamefont {E.}~\bibnamefont
  {Prada}}, \bibinfo {author} {\bibfnamefont {P.}~\bibnamefont {San-Jose}}, \
  and\ \bibinfo {author} {\bibfnamefont {R.}~\bibnamefont {Aguado}},\ }\href
  {\doibase 10.1103/PhysRevB.86.180503} {\bibfield  {journal} {\bibinfo
  {journal} {Phys. Rev. B}\ }\textbf {\bibinfo {volume} {86}},\ \bibinfo
  {pages} {180503} (\bibinfo {year} {2012})}\BibitemShut {NoStop}%
\bibitem [{\citenamefont {Kells}\ \emph {et~al.}(2012)\citenamefont {Kells},
  \citenamefont {Meidan},\ and\ \citenamefont {Brouwer}}]{Kells2012Sep}%
  \BibitemOpen
  \bibfield  {author} {\bibinfo {author} {\bibfnamefont {G.}~\bibnamefont
  {Kells}}, \bibinfo {author} {\bibfnamefont {D.}~\bibnamefont {Meidan}}, \
  and\ \bibinfo {author} {\bibfnamefont {P.~W.}\ \bibnamefont {Brouwer}},\
  }\href {\doibase 10.1103/PhysRevB.86.100503} {\bibfield  {journal} {\bibinfo
  {journal} {Phys. Rev. B}\ }\textbf {\bibinfo {volume} {86}},\ \bibinfo
  {pages} {100503} (\bibinfo {year} {2012})}\BibitemShut {NoStop}%
\bibitem [{\citenamefont {Cayao}\ \emph {et~al.}(2015)\citenamefont {Cayao},
  \citenamefont {Prada}, \citenamefont {San-Jose},\ and\ \citenamefont
  {Aguado}}]{Cayao2015Jan}%
  \BibitemOpen
  \bibfield  {author} {\bibinfo {author} {\bibfnamefont {J.}~\bibnamefont
  {Cayao}}, \bibinfo {author} {\bibfnamefont {E.}~\bibnamefont {Prada}},
  \bibinfo {author} {\bibfnamefont {P.}~\bibnamefont {San-Jose}}, \ and\
  \bibinfo {author} {\bibfnamefont {R.}~\bibnamefont {Aguado}},\ }\href
  {\doibase 10.1103/PhysRevB.91.024514} {\bibfield  {journal} {\bibinfo
  {journal} {Phys. Rev. B}\ }\textbf {\bibinfo {volume} {91}},\ \bibinfo
  {pages} {024514} (\bibinfo {year} {2015})}\BibitemShut {NoStop}%
\bibitem [{\citenamefont {Moore}\ \emph {et~al.}(2018)\citenamefont {Moore},
  \citenamefont {Zeng}, \citenamefont {Stanescu},\ and\ \citenamefont
  {Tewari}}]{Moore2018Oct}%
  \BibitemOpen
  \bibfield  {author} {\bibinfo {author} {\bibfnamefont {C.}~\bibnamefont
  {Moore}}, \bibinfo {author} {\bibfnamefont {C.}~\bibnamefont {Zeng}},
  \bibinfo {author} {\bibfnamefont {T.~D.}\ \bibnamefont {Stanescu}}, \ and\
  \bibinfo {author} {\bibfnamefont {S.}~\bibnamefont {Tewari}},\ }\href
  {\doibase 10.1103/PhysRevB.98.155314} {\bibfield  {journal} {\bibinfo
  {journal} {Phys. Rev. B}\ }\textbf {\bibinfo {volume} {98}},\ \bibinfo
  {pages} {155314} (\bibinfo {year} {2018})}\BibitemShut {NoStop}%
\bibitem [{\citenamefont {Vuik}\ \emph {et~al.}(2019)\citenamefont {Vuik},
  \citenamefont {Nijholt}, \citenamefont {Akhmerov},\ and\ \citenamefont
  {Wimmer}}]{Vuik2019Nov}%
  \BibitemOpen
  \bibfield  {author} {\bibinfo {author} {\bibfnamefont {A.}~\bibnamefont
  {Vuik}}, \bibinfo {author} {\bibfnamefont {B.}~\bibnamefont {Nijholt}},
  \bibinfo {author} {\bibfnamefont {A.}~\bibnamefont {Akhmerov}}, \ and\
  \bibinfo {author} {\bibfnamefont {M.}~\bibnamefont {Wimmer}},\ }\href
  {\doibase 10.21468/SciPostPhys.7.5.061} {\bibfield  {journal} {\bibinfo
  {journal} {SciPost Phys.}\ }\textbf {\bibinfo {volume} {7}},\ \bibinfo
  {pages} {061} (\bibinfo {year} {2019})}\BibitemShut {NoStop}%
\bibitem [{\citenamefont {Pan}\ and\ \citenamefont
  {Das~Sarma}(2020)}]{Pan2020Mar}%
  \BibitemOpen
  \bibfield  {author} {\bibinfo {author} {\bibfnamefont {H.}~\bibnamefont
  {Pan}}\ and\ \bibinfo {author} {\bibfnamefont {S.}~\bibnamefont
  {Das~Sarma}},\ }\href {\doibase 10.1103/PhysRevResearch.2.013377} {\bibfield
  {journal} {\bibinfo  {journal} {Phys. Rev. Res.}\ }\textbf {\bibinfo {volume}
  {2}},\ \bibinfo {pages} {013377} (\bibinfo {year} {2020})}\BibitemShut
  {NoStop}%
\bibitem [{\citenamefont {Hess}\ \emph {et~al.}(2021)\citenamefont {Hess},
  \citenamefont {Legg}, \citenamefont {Loss},\ and\ \citenamefont
  {Klinovaja}}]{Hess2021Aug}%
  \BibitemOpen
  \bibfield  {author} {\bibinfo {author} {\bibfnamefont {R.}~\bibnamefont
  {Hess}}, \bibinfo {author} {\bibfnamefont {H.~F.}\ \bibnamefont {Legg}},
  \bibinfo {author} {\bibfnamefont {D.}~\bibnamefont {Loss}}, \ and\ \bibinfo
  {author} {\bibfnamefont {J.}~\bibnamefont {Klinovaja}},\ }\href {\doibase
  10.1103/PhysRevB.104.075405} {\bibfield  {journal} {\bibinfo  {journal}
  {Phys. Rev. B}\ }\textbf {\bibinfo {volume} {104}},\ \bibinfo {pages}
  {075405} (\bibinfo {year} {2021})}\BibitemShut {NoStop}%
\bibitem [{\citenamefont {Flensberg}\ \emph {et~al.}(2021)\citenamefont
  {Flensberg}, \citenamefont {von Oppen},\ and\ \citenamefont
  {Stern}}]{FlensbergvonOppenStern2021}%
  \BibitemOpen
  \bibfield  {author} {\bibinfo {author} {\bibfnamefont {K.}~\bibnamefont
  {Flensberg}}, \bibinfo {author} {\bibfnamefont {F.}~\bibnamefont {von
  Oppen}}, \ and\ \bibinfo {author} {\bibfnamefont {A.}~\bibnamefont {Stern}},\
  }\href {\doibase 10.1038/s41578-021-00336-6} {\bibfield  {journal} {\bibinfo
  {journal} {Nat. Rev. Mat.}\ }\textbf {\bibinfo {volume} {6}},\ \bibinfo
  {pages} {944} (\bibinfo {year} {2021})}\BibitemShut {NoStop}%
\bibitem [{\citenamefont {Ohm}\ and\ \citenamefont
  {Hassler}(2015)}]{Ohm2015Feb}%
  \BibitemOpen
  \bibfield  {author} {\bibinfo {author} {\bibfnamefont {C.}~\bibnamefont
  {Ohm}}\ and\ \bibinfo {author} {\bibfnamefont {F.}~\bibnamefont {Hassler}},\
  }\href {\doibase 10.1103/PhysRevB.91.085406} {\bibfield  {journal} {\bibinfo
  {journal} {Phys. Rev. B}\ }\textbf {\bibinfo {volume} {91}},\ \bibinfo
  {pages} {085406} (\bibinfo {year} {2015})}\BibitemShut {NoStop}%
\bibitem [{\citenamefont {Gharavi}\ \emph {et~al.}(2016)\citenamefont
  {Gharavi}, \citenamefont {Hoving},\ and\ \citenamefont
  {Baugh}}]{Gharavi2016Oct}%
  \BibitemOpen
  \bibfield  {author} {\bibinfo {author} {\bibfnamefont {K.}~\bibnamefont
  {Gharavi}}, \bibinfo {author} {\bibfnamefont {D.}~\bibnamefont {Hoving}}, \
  and\ \bibinfo {author} {\bibfnamefont {J.}~\bibnamefont {Baugh}},\ }\href
  {\doibase 10.1103/PhysRevB.94.155417} {\bibfield  {journal} {\bibinfo
  {journal} {Phys. Rev. B}\ }\textbf {\bibinfo {volume} {94}},\ \bibinfo
  {pages} {155417} (\bibinfo {year} {2016})}\BibitemShut {NoStop}%
\bibitem [{\citenamefont {Malciu}\ \emph {et~al.}(2018)\citenamefont {Malciu},
  \citenamefont {Mazza},\ and\ \citenamefont {Mora}}]{Malciu2018Oct}%
  \BibitemOpen
  \bibfield  {author} {\bibinfo {author} {\bibfnamefont {C.}~\bibnamefont
  {Malciu}}, \bibinfo {author} {\bibfnamefont {L.}~\bibnamefont {Mazza}}, \
  and\ \bibinfo {author} {\bibfnamefont {C.}~\bibnamefont {Mora}},\ }\href
  {\doibase 10.1103/PhysRevB.98.165426} {\bibfield  {journal} {\bibinfo
  {journal} {Phys. Rev. B}\ }\textbf {\bibinfo {volume} {98}},\ \bibinfo
  {pages} {165426} (\bibinfo {year} {2018})}\BibitemShut {NoStop}%
\bibitem [{\citenamefont {Li}\ \emph {et~al.}(2018)\citenamefont {Li},
  \citenamefont {Coish}, \citenamefont {Hell}, \citenamefont {Flensberg},\ and\
  \citenamefont {Leijnse}}]{Li2018Nov}%
  \BibitemOpen
  \bibfield  {author} {\bibinfo {author} {\bibfnamefont {T.}~\bibnamefont
  {Li}}, \bibinfo {author} {\bibfnamefont {W.~A.}\ \bibnamefont {Coish}},
  \bibinfo {author} {\bibfnamefont {M.}~\bibnamefont {Hell}}, \bibinfo {author}
  {\bibfnamefont {K.}~\bibnamefont {Flensberg}}, \ and\ \bibinfo {author}
  {\bibfnamefont {M.}~\bibnamefont {Leijnse}},\ }\href {\doibase
  10.1103/PhysRevB.98.205403} {\bibfield  {journal} {\bibinfo  {journal} {Phys.
  Rev. B}\ }\textbf {\bibinfo {volume} {98}},\ \bibinfo {pages} {205403}
  (\bibinfo {year} {2018})}\BibitemShut {NoStop}%
\bibitem [{\citenamefont {Schrade}\ and\ \citenamefont
  {Fu}(2018)}]{Schrade2018Dec}%
  \BibitemOpen
  \bibfield  {author} {\bibinfo {author} {\bibfnamefont {C.}~\bibnamefont
  {Schrade}}\ and\ \bibinfo {author} {\bibfnamefont {L.}~\bibnamefont {Fu}},\
  }\href {\doibase 10.1103/PhysRevLett.121.267002} {\bibfield  {journal}
  {\bibinfo  {journal} {Phys. Rev. Lett.}\ }\textbf {\bibinfo {volume} {121}},\
  \bibinfo {pages} {267002} (\bibinfo {year} {2018})}\BibitemShut {NoStop}%
\bibitem [{\citenamefont {Grimsmo}\ and\ \citenamefont
  {Smith}(2019)}]{Grimsmo2019Jun}%
  \BibitemOpen
  \bibfield  {author} {\bibinfo {author} {\bibfnamefont {A.~L.}\ \bibnamefont
  {Grimsmo}}\ and\ \bibinfo {author} {\bibfnamefont {T.~B.}\ \bibnamefont
  {Smith}},\ }\href {\doibase 10.1103/PhysRevB.99.235420} {\bibfield  {journal}
  {\bibinfo  {journal} {Phys. Rev. B}\ }\textbf {\bibinfo {volume} {99}},\
  \bibinfo {pages} {235420} (\bibinfo {year} {2019})}\BibitemShut {NoStop}%
\bibitem [{\citenamefont {Sz{\ifmmode\acute{e}\else\'{e}\fi}chenyi}\ and\
  \citenamefont
  {P{\ifmmode\acute{a}\else\'{a}\fi}lyi}(2020)}]{Szechenyi2020Jun}%
  \BibitemOpen
  \bibfield  {author} {\bibinfo {author} {\bibfnamefont {G.}~\bibnamefont
  {Sz{\ifmmode\acute{e}\else\'{e}\fi}chenyi}}\ and\ \bibinfo {author}
  {\bibfnamefont {A.}~\bibnamefont {P{\ifmmode\acute{a}\else\'{a}\fi}lyi}},\
  }\href {\doibase 10.1103/PhysRevB.101.235441} {\bibfield  {journal} {\bibinfo
   {journal} {Phys. Rev. B}\ }\textbf {\bibinfo {volume} {101}},\ \bibinfo
  {pages} {235441} (\bibinfo {year} {2020})}\BibitemShut {NoStop}%
\bibitem [{\citenamefont {Munk}\ \emph {et~al.}(2020)\citenamefont {Munk},
  \citenamefont {Schulenborg}, \citenamefont {Egger},\ and\ \citenamefont
  {Flensberg}}]{Munk2020Aug}%
  \BibitemOpen
  \bibfield  {author} {\bibinfo {author} {\bibfnamefont {M.~I.~K.}\
  \bibnamefont {Munk}}, \bibinfo {author} {\bibfnamefont {J.}~\bibnamefont
  {Schulenborg}}, \bibinfo {author} {\bibfnamefont {R.}~\bibnamefont {Egger}},
  \ and\ \bibinfo {author} {\bibfnamefont {K.}~\bibnamefont {Flensberg}},\
  }\href {\doibase 10.1103/PhysRevResearch.2.033254} {\bibfield  {journal}
  {\bibinfo  {journal} {Phys. Rev. Res.}\ }\textbf {\bibinfo {volume} {2}},\
  \bibinfo {pages} {033254} (\bibinfo {year} {2020})}\BibitemShut {NoStop}%
\bibitem [{\citenamefont {Steiner}\ and\ \citenamefont {von
  Oppen}(2020)}]{Steiner2020Aug}%
  \BibitemOpen
  \bibfield  {author} {\bibinfo {author} {\bibfnamefont {J.~F.}\ \bibnamefont
  {Steiner}}\ and\ \bibinfo {author} {\bibfnamefont {F.}~\bibnamefont {von
  Oppen}},\ }\href {\doibase 10.1103/PhysRevResearch.2.033255} {\bibfield
  {journal} {\bibinfo  {journal} {Phys. Rev. Res.}\ }\textbf {\bibinfo {volume}
  {2}},\ \bibinfo {pages} {033255} (\bibinfo {year} {2020})}\BibitemShut
  {NoStop}%
\bibitem [{\citenamefont {Smith}\ \emph {et~al.}(2020)\citenamefont {Smith},
  \citenamefont {Cassidy}, \citenamefont {Reilly}, \citenamefont {Bartlett},\
  and\ \citenamefont {Grimsmo}}]{Smith2020Nov}%
  \BibitemOpen
  \bibfield  {author} {\bibinfo {author} {\bibfnamefont {T.~B.}\ \bibnamefont
  {Smith}}, \bibinfo {author} {\bibfnamefont {M.~C.}\ \bibnamefont {Cassidy}},
  \bibinfo {author} {\bibfnamefont {D.~J.}\ \bibnamefont {Reilly}}, \bibinfo
  {author} {\bibfnamefont {S.~D.}\ \bibnamefont {Bartlett}}, \ and\ \bibinfo
  {author} {\bibfnamefont {A.~L.}\ \bibnamefont {Grimsmo}},\ }\href {\doibase
  10.1103/PRXQuantum.1.020313} {\bibfield  {journal} {\bibinfo  {journal} {PRX
  Quantum}\ }\textbf {\bibinfo {volume} {1}},\ \bibinfo {pages} {020313}
  (\bibinfo {year} {2020})}\BibitemShut {NoStop}%
\bibitem [{\citenamefont {Schulenborg}\ \emph {et~al.}(2021)\citenamefont
  {Schulenborg}, \citenamefont {Burrello}, \citenamefont {Leijnse},\ and\
  \citenamefont {Flensberg}}]{Schulenborg2021Jun}%
  \BibitemOpen
  \bibfield  {author} {\bibinfo {author} {\bibfnamefont {J.}~\bibnamefont
  {Schulenborg}}, \bibinfo {author} {\bibfnamefont {M.}~\bibnamefont
  {Burrello}}, \bibinfo {author} {\bibfnamefont {M.}~\bibnamefont {Leijnse}}, \
  and\ \bibinfo {author} {\bibfnamefont {K.}~\bibnamefont {Flensberg}},\ }\href
  {\doibase 10.1103/PhysRevB.103.245407} {\bibfield  {journal} {\bibinfo
  {journal} {Phys. Rev. B}\ }\textbf {\bibinfo {volume} {103}},\ \bibinfo
  {pages} {245407} (\bibinfo {year} {2021})}\BibitemShut {NoStop}%
\bibitem [{\citenamefont {Hofstetter}\ \emph {et~al.}(2009)\citenamefont
  {Hofstetter}, \citenamefont {Csonka}, \citenamefont {Nyg{\aa}rd},\ and\
  \citenamefont
  {Sch{\ifmmode\ddot{o}\else\"{o}\fi}nenberger}}]{Hofstetter2009Oct}%
  \BibitemOpen
  \bibfield  {author} {\bibinfo {author} {\bibfnamefont {L.}~\bibnamefont
  {Hofstetter}}, \bibinfo {author} {\bibfnamefont {S.}~\bibnamefont {Csonka}},
  \bibinfo {author} {\bibfnamefont {J.}~\bibnamefont {Nyg{\aa}rd}}, \ and\
  \bibinfo {author} {\bibfnamefont {C.}~\bibnamefont
  {Sch{\ifmmode\ddot{o}\else\"{o}\fi}nenberger}},\ }\href {\doibase
  10.1038/nature08432} {\bibfield  {journal} {\bibinfo  {journal} {Nature}\
  }\textbf {\bibinfo {volume} {461}},\ \bibinfo {pages} {960} (\bibinfo {year}
  {2009})}\BibitemShut {NoStop}%
\bibitem [{\citenamefont {De~Franceschi}\ \emph {et~al.}(2010)\citenamefont
  {De~Franceschi}, \citenamefont {Kouwenhoven}, \citenamefont
  {Sch{\ifmmode\ddot{o}\else\"{o}\fi}nenberger},\ and\ \citenamefont
  {Wernsdorfer}}]{DeFranceschi2010Oct}%
  \BibitemOpen
  \bibfield  {author} {\bibinfo {author} {\bibfnamefont {S.}~\bibnamefont
  {De~Franceschi}}, \bibinfo {author} {\bibfnamefont {L.}~\bibnamefont
  {Kouwenhoven}}, \bibinfo {author} {\bibfnamefont {C.}~\bibnamefont
  {Sch{\ifmmode\ddot{o}\else\"{o}\fi}nenberger}}, \ and\ \bibinfo {author}
  {\bibfnamefont {W.}~\bibnamefont {Wernsdorfer}},\ }\href {\doibase
  10.1038/nnano.2010.173} {\bibfield  {journal} {\bibinfo  {journal} {Nat.
  Nanotechnol.}\ }\textbf {\bibinfo {volume} {5}},\ \bibinfo {pages} {703}
  (\bibinfo {year} {2010})}\BibitemShut {NoStop}%
\bibitem [{\citenamefont {Deng}\ \emph {et~al.}(2014)\citenamefont {Deng},
  \citenamefont {Yu}, \citenamefont {Huang}, \citenamefont {Larsson},
  \citenamefont {Caroff},\ and\ \citenamefont {Xu}}]{Deng2014Dec}%
  \BibitemOpen
  \bibfield  {author} {\bibinfo {author} {\bibfnamefont {M.~T.}\ \bibnamefont
  {Deng}}, \bibinfo {author} {\bibfnamefont {C.~L.}\ \bibnamefont {Yu}},
  \bibinfo {author} {\bibfnamefont {G.~Y.}\ \bibnamefont {Huang}}, \bibinfo
  {author} {\bibfnamefont {M.}~\bibnamefont {Larsson}}, \bibinfo {author}
  {\bibfnamefont {P.}~\bibnamefont {Caroff}}, \ and\ \bibinfo {author}
  {\bibfnamefont {H.~Q.}\ \bibnamefont {Xu}},\ }\href {\doibase
  10.1038/srep07261} {\bibfield  {journal} {\bibinfo  {journal} {Sci. Rep.}\
  }\textbf {\bibinfo {volume} {4}},\ \bibinfo {pages} {1} (\bibinfo {year}
  {2014})}\BibitemShut {NoStop}%
\bibitem [{\citenamefont {Deng}\ \emph {et~al.}(2016)\citenamefont {Deng},
  \citenamefont {Vaitiek{\ifmmode\dot{e}\else\.{e}\fi}nas}, \citenamefont
  {Hansen}, \citenamefont {Danon}, \citenamefont {Leijnse}, \citenamefont
  {Flensberg}, \citenamefont {Nyg{\aa}rd}, \citenamefont {Krogstrup},\ and\
  \citenamefont {Marcus}}]{Deng2016Dec}%
  \BibitemOpen
  \bibfield  {author} {\bibinfo {author} {\bibfnamefont {M.~T.}\ \bibnamefont
  {Deng}}, \bibinfo {author} {\bibfnamefont {S.}~\bibnamefont
  {Vaitiek{\ifmmode\dot{e}\else\.{e}\fi}nas}}, \bibinfo {author} {\bibfnamefont
  {E.~B.}\ \bibnamefont {Hansen}}, \bibinfo {author} {\bibfnamefont
  {J.}~\bibnamefont {Danon}}, \bibinfo {author} {\bibfnamefont
  {M.}~\bibnamefont {Leijnse}}, \bibinfo {author} {\bibfnamefont
  {K.}~\bibnamefont {Flensberg}}, \bibinfo {author} {\bibfnamefont
  {J.}~\bibnamefont {Nyg{\aa}rd}}, \bibinfo {author} {\bibfnamefont
  {P.}~\bibnamefont {Krogstrup}}, \ and\ \bibinfo {author} {\bibfnamefont
  {C.~M.}\ \bibnamefont {Marcus}},\ }\href {\doibase 10.1126/science.aaf3961}
  {\bibfield  {journal} {\bibinfo  {journal} {Science}\ }\textbf {\bibinfo
  {volume} {354}},\ \bibinfo {pages} {1557} (\bibinfo {year}
  {2016})}\BibitemShut {NoStop}%
\bibitem [{\citenamefont {Szombati}\ \emph {et~al.}(2016)\citenamefont
  {Szombati}, \citenamefont {Nadj-Perge}, \citenamefont {Car}, \citenamefont
  {Plissard}, \citenamefont {Bakkers},\ and\ \citenamefont
  {Kouwenhoven}}]{Szombati2016Jun}%
  \BibitemOpen
  \bibfield  {author} {\bibinfo {author} {\bibfnamefont {D.~B.}\ \bibnamefont
  {Szombati}}, \bibinfo {author} {\bibfnamefont {S.}~\bibnamefont
  {Nadj-Perge}}, \bibinfo {author} {\bibfnamefont {D.}~\bibnamefont {Car}},
  \bibinfo {author} {\bibfnamefont {S.~R.}\ \bibnamefont {Plissard}}, \bibinfo
  {author} {\bibfnamefont {E.~P. A.~M.}\ \bibnamefont {Bakkers}}, \ and\
  \bibinfo {author} {\bibfnamefont {L.~P.}\ \bibnamefont {Kouwenhoven}},\
  }\href {\doibase 10.1038/nphys3742} {\bibfield  {journal} {\bibinfo
  {journal} {Nat. Phys.}\ }\textbf {\bibinfo {volume} {12}},\ \bibinfo {pages}
  {568} (\bibinfo {year} {2016})}\BibitemShut {NoStop}%
\bibitem [{\citenamefont {Deng}\ \emph {et~al.}(2018)\citenamefont {Deng},
  \citenamefont {Vaitiek{\ifmmode\dot{e}\else\.{e}\fi}nas}, \citenamefont
  {Prada}, \citenamefont {San-Jose}, \citenamefont {Nyg{\aa}rd}, \citenamefont
  {Krogstrup}, \citenamefont {Aguado},\ and\ \citenamefont
  {Marcus}}]{Deng2018Aug}%
  \BibitemOpen
  \bibfield  {author} {\bibinfo {author} {\bibfnamefont {M.-T.}\ \bibnamefont
  {Deng}}, \bibinfo {author} {\bibfnamefont {S.}~\bibnamefont
  {Vaitiek{\ifmmode\dot{e}\else\.{e}\fi}nas}}, \bibinfo {author} {\bibfnamefont
  {E.}~\bibnamefont {Prada}}, \bibinfo {author} {\bibfnamefont
  {P.}~\bibnamefont {San-Jose}}, \bibinfo {author} {\bibfnamefont
  {J.}~\bibnamefont {Nyg{\aa}rd}}, \bibinfo {author} {\bibfnamefont
  {P.}~\bibnamefont {Krogstrup}}, \bibinfo {author} {\bibfnamefont
  {R.}~\bibnamefont {Aguado}}, \ and\ \bibinfo {author} {\bibfnamefont {C.~M.}\
  \bibnamefont {Marcus}},\ }\href {\doibase 10.1103/PhysRevB.98.085125}
  {\bibfield  {journal} {\bibinfo  {journal} {Phys. Rev. B}\ }\textbf {\bibinfo
  {volume} {98}},\ \bibinfo {pages} {085125} (\bibinfo {year}
  {2018})}\BibitemShut {NoStop}%
\bibitem [{\citenamefont {van Veen}\ \emph {et~al.}(2019)\citenamefont {van
  Veen}, \citenamefont {de~Jong}, \citenamefont {Han}, \citenamefont {Prosko},
  \citenamefont {Krogstrup}, \citenamefont {Watson}, \citenamefont
  {Kouwenhoven},\ and\ \citenamefont {Pfaff}}]{vanVeen2019Nov}%
  \BibitemOpen
  \bibfield  {author} {\bibinfo {author} {\bibfnamefont {J.}~\bibnamefont {van
  Veen}}, \bibinfo {author} {\bibfnamefont {D.}~\bibnamefont {de~Jong}},
  \bibinfo {author} {\bibfnamefont {L.}~\bibnamefont {Han}}, \bibinfo {author}
  {\bibfnamefont {C.}~\bibnamefont {Prosko}}, \bibinfo {author} {\bibfnamefont
  {P.}~\bibnamefont {Krogstrup}}, \bibinfo {author} {\bibfnamefont {J.~D.}\
  \bibnamefont {Watson}}, \bibinfo {author} {\bibfnamefont {L.~P.}\
  \bibnamefont {Kouwenhoven}}, \ and\ \bibinfo {author} {\bibfnamefont
  {W.}~\bibnamefont {Pfaff}},\ }\href {\doibase 10.1103/PhysRevB.100.174508}
  {\bibfield  {journal} {\bibinfo  {journal} {Phys. Rev. B}\ }\textbf {\bibinfo
  {volume} {100}},\ \bibinfo {pages} {174508} (\bibinfo {year}
  {2019})}\BibitemShut {NoStop}%
\bibitem [{\citenamefont {Razmadze}\ \emph {et~al.}(2020)\citenamefont
  {Razmadze}, \citenamefont {O{'}Farrell}, \citenamefont {Krogstrup},\ and\
  \citenamefont {Marcus}}]{Razmadze2020Sep}%
  \BibitemOpen
  \bibfield  {author} {\bibinfo {author} {\bibfnamefont {D.}~\bibnamefont
  {Razmadze}}, \bibinfo {author} {\bibfnamefont {E.~C.~T.}\ \bibnamefont
  {O{'}Farrell}}, \bibinfo {author} {\bibfnamefont {P.}~\bibnamefont
  {Krogstrup}}, \ and\ \bibinfo {author} {\bibfnamefont {C.~M.}\ \bibnamefont
  {Marcus}},\ }\href {\doibase 10.1103/PhysRevLett.125.116803} {\bibfield
  {journal} {\bibinfo  {journal} {Phys. Rev. Lett.}\ }\textbf {\bibinfo
  {volume} {125}},\ \bibinfo {pages} {116803} (\bibinfo {year}
  {2020})}\BibitemShut {NoStop}%
\bibitem [{\citenamefont {Yoshie}\ \emph {et~al.}(2004)\citenamefont {Yoshie},
  \citenamefont {Scherer}, \citenamefont {Hendrickson}, \citenamefont
  {Khitrova}, \citenamefont {Gibbs}, \citenamefont {Rupper}, \citenamefont
  {Ell}, \citenamefont {Shchekin},\ and\ \citenamefont
  {Deppe}}]{Yoshie2004Nov}%
  \BibitemOpen
  \bibfield  {author} {\bibinfo {author} {\bibfnamefont {T.}~\bibnamefont
  {Yoshie}}, \bibinfo {author} {\bibfnamefont {A.}~\bibnamefont {Scherer}},
  \bibinfo {author} {\bibfnamefont {J.}~\bibnamefont {Hendrickson}}, \bibinfo
  {author} {\bibfnamefont {G.}~\bibnamefont {Khitrova}}, \bibinfo {author}
  {\bibfnamefont {H.~M.}\ \bibnamefont {Gibbs}}, \bibinfo {author}
  {\bibfnamefont {G.}~\bibnamefont {Rupper}}, \bibinfo {author} {\bibfnamefont
  {C.}~\bibnamefont {Ell}}, \bibinfo {author} {\bibfnamefont {O.~B.}\
  \bibnamefont {Shchekin}}, \ and\ \bibinfo {author} {\bibfnamefont {D.~G.}\
  \bibnamefont {Deppe}},\ }\href {\doibase 10.1038/nature03119} {\bibfield
  {journal} {\bibinfo  {journal} {Nature}\ }\textbf {\bibinfo {volume} {432}},\
  \bibinfo {pages} {200} (\bibinfo {year} {2004})}\BibitemShut {NoStop}%
\bibitem [{\citenamefont {Reithmaier}\ \emph {et~al.}(2004)\citenamefont
  {Reithmaier}, \citenamefont {Sek}, \citenamefont
  {L{\ifmmode\ddot{o}\else\"{o}\fi}ffler}, \citenamefont {Hofmann},
  \citenamefont {Kuhn}, \citenamefont {Reitzenstein}, \citenamefont {Keldysh},
  \citenamefont {Kulakovskii}, \citenamefont {Reinecke},\ and\ \citenamefont
  {Forchel}}]{Reithmaier2004Nov}%
  \BibitemOpen
  \bibfield  {author} {\bibinfo {author} {\bibfnamefont {J.~P.}\ \bibnamefont
  {Reithmaier}}, \bibinfo {author} {\bibfnamefont {G.}~\bibnamefont {Sek}},
  \bibinfo {author} {\bibfnamefont {A.}~\bibnamefont
  {L{\ifmmode\ddot{o}\else\"{o}\fi}ffler}}, \bibinfo {author} {\bibfnamefont
  {C.}~\bibnamefont {Hofmann}}, \bibinfo {author} {\bibfnamefont
  {S.}~\bibnamefont {Kuhn}}, \bibinfo {author} {\bibfnamefont {S.}~\bibnamefont
  {Reitzenstein}}, \bibinfo {author} {\bibfnamefont {L.~V.}\ \bibnamefont
  {Keldysh}}, \bibinfo {author} {\bibfnamefont {V.~D.}\ \bibnamefont
  {Kulakovskii}}, \bibinfo {author} {\bibfnamefont {T.~L.}\ \bibnamefont
  {Reinecke}}, \ and\ \bibinfo {author} {\bibfnamefont {A.}~\bibnamefont
  {Forchel}},\ }\href {\doibase 10.1038/nature02969} {\bibfield  {journal}
  {\bibinfo  {journal} {Nature}\ }\textbf {\bibinfo {volume} {432}},\ \bibinfo
  {pages} {197} (\bibinfo {year} {2004})}\BibitemShut {NoStop}%
\bibitem [{\citenamefont {Delbecq}\ \emph {et~al.}(2011)\citenamefont
  {Delbecq}, \citenamefont {Schmitt}, \citenamefont {Parmentier}, \citenamefont
  {Roch}, \citenamefont {Viennot}, \citenamefont
  {F{\ifmmode\grave{e}\else\`{e}\fi}ve}, \citenamefont {Huard}, \citenamefont
  {Mora}, \citenamefont {Cottet},\ and\ \citenamefont
  {Kontos}}]{Delbecq2011Dec}%
  \BibitemOpen
  \bibfield  {author} {\bibinfo {author} {\bibfnamefont {M.~R.}\ \bibnamefont
  {Delbecq}}, \bibinfo {author} {\bibfnamefont {V.}~\bibnamefont {Schmitt}},
  \bibinfo {author} {\bibfnamefont {F.~D.}\ \bibnamefont {Parmentier}},
  \bibinfo {author} {\bibfnamefont {N.}~\bibnamefont {Roch}}, \bibinfo {author}
  {\bibfnamefont {J.~J.}\ \bibnamefont {Viennot}}, \bibinfo {author}
  {\bibfnamefont {G.}~\bibnamefont {F{\ifmmode\grave{e}\else\`{e}\fi}ve}},
  \bibinfo {author} {\bibfnamefont {B.}~\bibnamefont {Huard}}, \bibinfo
  {author} {\bibfnamefont {C.}~\bibnamefont {Mora}}, \bibinfo {author}
  {\bibfnamefont {A.}~\bibnamefont {Cottet}}, \ and\ \bibinfo {author}
  {\bibfnamefont {T.}~\bibnamefont {Kontos}},\ }\href {\doibase
  10.1103/PhysRevLett.107.256804} {\bibfield  {journal} {\bibinfo  {journal}
  {Phys. Rev. Lett.}\ }\textbf {\bibinfo {volume} {107}},\ \bibinfo {pages}
  {256804} (\bibinfo {year} {2011})}\BibitemShut {NoStop}%
\bibitem [{\citenamefont {Frey}\ \emph {et~al.}(2012)\citenamefont {Frey},
  \citenamefont {Leek}, \citenamefont {Beck}, \citenamefont {Blais},
  \citenamefont {Ihn}, \citenamefont {Ensslin},\ and\ \citenamefont
  {Wallraff}}]{Frey2012Jan}%
  \BibitemOpen
  \bibfield  {author} {\bibinfo {author} {\bibfnamefont {T.}~\bibnamefont
  {Frey}}, \bibinfo {author} {\bibfnamefont {P.~J.}\ \bibnamefont {Leek}},
  \bibinfo {author} {\bibfnamefont {M.}~\bibnamefont {Beck}}, \bibinfo {author}
  {\bibfnamefont {A.}~\bibnamefont {Blais}}, \bibinfo {author} {\bibfnamefont
  {T.}~\bibnamefont {Ihn}}, \bibinfo {author} {\bibfnamefont {K.}~\bibnamefont
  {Ensslin}}, \ and\ \bibinfo {author} {\bibfnamefont {A.}~\bibnamefont
  {Wallraff}},\ }\href {\doibase 10.1103/PhysRevLett.108.046807} {\bibfield
  {journal} {\bibinfo  {journal} {Phys. Rev. Lett.}\ }\textbf {\bibinfo
  {volume} {108}},\ \bibinfo {pages} {046807} (\bibinfo {year}
  {2012})}\BibitemShut {NoStop}%
\bibitem [{\citenamefont {Petersson}\ \emph {et~al.}(2012)\citenamefont
  {Petersson}, \citenamefont {McFaul}, \citenamefont {Schroer}, \citenamefont
  {Jung}, \citenamefont {Taylor}, \citenamefont {Houck},\ and\ \citenamefont
  {Petta}}]{Petersson2012Oct}%
  \BibitemOpen
  \bibfield  {author} {\bibinfo {author} {\bibfnamefont {K.~D.}\ \bibnamefont
  {Petersson}}, \bibinfo {author} {\bibfnamefont {L.~W.}\ \bibnamefont
  {McFaul}}, \bibinfo {author} {\bibfnamefont {M.~D.}\ \bibnamefont {Schroer}},
  \bibinfo {author} {\bibfnamefont {M.}~\bibnamefont {Jung}}, \bibinfo {author}
  {\bibfnamefont {J.~M.}\ \bibnamefont {Taylor}}, \bibinfo {author}
  {\bibfnamefont {A.~A.}\ \bibnamefont {Houck}}, \ and\ \bibinfo {author}
  {\bibfnamefont {J.~R.}\ \bibnamefont {Petta}},\ }\href {\doibase
  10.1038/nature11559} {\bibfield  {journal} {\bibinfo  {journal} {Nature}\
  }\textbf {\bibinfo {volume} {490}},\ \bibinfo {pages} {380} (\bibinfo {year}
  {2012})}\BibitemShut {NoStop}%
\bibitem [{\citenamefont {Xiang}\ \emph {et~al.}(2013)\citenamefont {Xiang},
  \citenamefont {Ashhab}, \citenamefont {You},\ and\ \citenamefont
  {Nori}}]{Xiang2013Apr}%
  \BibitemOpen
  \bibfield  {author} {\bibinfo {author} {\bibfnamefont {Z.-L.}\ \bibnamefont
  {Xiang}}, \bibinfo {author} {\bibfnamefont {S.}~\bibnamefont {Ashhab}},
  \bibinfo {author} {\bibfnamefont {J.~Q.}\ \bibnamefont {You}}, \ and\
  \bibinfo {author} {\bibfnamefont {F.}~\bibnamefont {Nori}},\ }\href {\doibase
  10.1103/RevModPhys.85.623} {\bibfield  {journal} {\bibinfo  {journal} {Rev.
  Mod. Phys.}\ }\textbf {\bibinfo {volume} {85}},\ \bibinfo {pages} {623}
  (\bibinfo {year} {2013})}\BibitemShut {NoStop}%
\bibitem [{\citenamefont {Stockklauser}\ \emph {et~al.}(2017)\citenamefont
  {Stockklauser}, \citenamefont {Scarlino}, \citenamefont {Koski},
  \citenamefont {Gasparinetti}, \citenamefont {Andersen}, \citenamefont
  {Reichl}, \citenamefont {Wegscheider}, \citenamefont {Ihn}, \citenamefont
  {Ensslin},\ and\ \citenamefont {Wallraff}}]{Stockklauser2017Mar}%
  \BibitemOpen
  \bibfield  {author} {\bibinfo {author} {\bibfnamefont {A.}~\bibnamefont
  {Stockklauser}}, \bibinfo {author} {\bibfnamefont {P.}~\bibnamefont
  {Scarlino}}, \bibinfo {author} {\bibfnamefont {J.~V.}\ \bibnamefont {Koski}},
  \bibinfo {author} {\bibfnamefont {S.}~\bibnamefont {Gasparinetti}}, \bibinfo
  {author} {\bibfnamefont {C.~K.}\ \bibnamefont {Andersen}}, \bibinfo {author}
  {\bibfnamefont {C.}~\bibnamefont {Reichl}}, \bibinfo {author} {\bibfnamefont
  {W.}~\bibnamefont {Wegscheider}}, \bibinfo {author} {\bibfnamefont
  {T.}~\bibnamefont {Ihn}}, \bibinfo {author} {\bibfnamefont {K.}~\bibnamefont
  {Ensslin}}, \ and\ \bibinfo {author} {\bibfnamefont {A.}~\bibnamefont
  {Wallraff}},\ }\href {\doibase 10.1103/PhysRevX.7.011030} {\bibfield
  {journal} {\bibinfo  {journal} {Phys. Rev. X}\ }\textbf {\bibinfo {volume}
  {7}},\ \bibinfo {pages} {011030} (\bibinfo {year} {2017})}\BibitemShut
  {NoStop}%
\bibitem [{\citenamefont {Burkard}\ \emph {et~al.}(2020)\citenamefont
  {Burkard}, \citenamefont {Gullans}, \citenamefont {Mi},\ and\ \citenamefont
  {Petta}}]{Burkard2020Mar}%
  \BibitemOpen
  \bibfield  {author} {\bibinfo {author} {\bibfnamefont {G.}~\bibnamefont
  {Burkard}}, \bibinfo {author} {\bibfnamefont {M.~J.}\ \bibnamefont
  {Gullans}}, \bibinfo {author} {\bibfnamefont {X.}~\bibnamefont {Mi}}, \ and\
  \bibinfo {author} {\bibfnamefont {J.~R.}\ \bibnamefont {Petta}},\ }\href
  {\doibase 10.1038/s42254-019-0135-2} {\bibfield  {journal} {\bibinfo
  {journal} {Nat. Rev. Phys.}\ }\textbf {\bibinfo {volume} {2}},\ \bibinfo
  {pages} {129} (\bibinfo {year} {2020})}\BibitemShut {NoStop}%
\bibitem [{\citenamefont {Deng}\ \emph {et~al.}(2020)\citenamefont {Deng},
  \citenamefont {Xu},\ and\ \citenamefont {Li}}]{Deng2020Apr}%
  \BibitemOpen
  \bibfield  {author} {\bibinfo {author} {\bibfnamefont {G.-W.}\ \bibnamefont
  {Deng}}, \bibinfo {author} {\bibfnamefont {N.}~\bibnamefont {Xu}}, \ and\
  \bibinfo {author} {\bibfnamefont {W.-J.}\ \bibnamefont {Li}},\ }in\ \href
  {\doibase 10.1007/978-3-030-35813-6_4} {\emph {\bibinfo {booktitle} {{Quantum
  Dot Optoelectronic Devices}}}}\ (\bibinfo  {publisher} {Springer},\ \bibinfo
  {address} {Cham, Switzerland},\ \bibinfo {year} {2020})\ pp.\ \bibinfo
  {pages} {107--133}\BibitemShut {NoStop}%
\bibitem [{\citenamefont {Schoelkopf}\ \emph {et~al.}(1998)\citenamefont
  {Schoelkopf}, \citenamefont {Wahlgren}, \citenamefont {Kozhevnikov},
  \citenamefont {Delsing},\ and\ \citenamefont {Prober}}]{Schoelkopf1998May}%
  \BibitemOpen
  \bibfield  {author} {\bibinfo {author} {\bibfnamefont {R.~J.}\ \bibnamefont
  {Schoelkopf}}, \bibinfo {author} {\bibfnamefont {P.}~\bibnamefont
  {Wahlgren}}, \bibinfo {author} {\bibfnamefont {A.~A.}\ \bibnamefont
  {Kozhevnikov}}, \bibinfo {author} {\bibfnamefont {P.}~\bibnamefont
  {Delsing}}, \ and\ \bibinfo {author} {\bibfnamefont {D.~E.}\ \bibnamefont
  {Prober}},\ }\href {\doibase 10.1126/science.280.5367.1238} {\bibfield
  {journal} {\bibinfo  {journal} {Science}\ }\textbf {\bibinfo {volume}
  {280}},\ \bibinfo {pages} {1238} (\bibinfo {year} {1998})}\BibitemShut
  {NoStop}%
\bibitem [{\citenamefont {Lu}\ \emph {et~al.}(2003)\citenamefont {Lu},
  \citenamefont {Ji}, \citenamefont {Pfeiffer}, \citenamefont {West},\ and\
  \citenamefont {Rimberg}}]{Lu2003May}%
  \BibitemOpen
  \bibfield  {author} {\bibinfo {author} {\bibfnamefont {W.}~\bibnamefont
  {Lu}}, \bibinfo {author} {\bibfnamefont {Z.}~\bibnamefont {Ji}}, \bibinfo
  {author} {\bibfnamefont {L.}~\bibnamefont {Pfeiffer}}, \bibinfo {author}
  {\bibfnamefont {K.~W.}\ \bibnamefont {West}}, \ and\ \bibinfo {author}
  {\bibfnamefont {A.~J.}\ \bibnamefont {Rimberg}},\ }\href {\doibase
  10.1038/nature01642} {\bibfield  {journal} {\bibinfo  {journal} {Nature}\
  }\textbf {\bibinfo {volume} {423}},\ \bibinfo {pages} {422} (\bibinfo {year}
  {2003})}\BibitemShut {NoStop}%
\bibitem [{\citenamefont {Fujisawa}\ \emph {et~al.}(2004)\citenamefont
  {Fujisawa}, \citenamefont {Hayashi}, \citenamefont {Hirayama}, \citenamefont
  {Cheong},\ and\ \citenamefont {Jeong}}]{Fujisawa2004Mar}%
  \BibitemOpen
  \bibfield  {author} {\bibinfo {author} {\bibfnamefont {T.}~\bibnamefont
  {Fujisawa}}, \bibinfo {author} {\bibfnamefont {T.}~\bibnamefont {Hayashi}},
  \bibinfo {author} {\bibfnamefont {Y.}~\bibnamefont {Hirayama}}, \bibinfo
  {author} {\bibfnamefont {H.~D.}\ \bibnamefont {Cheong}}, \ and\ \bibinfo
  {author} {\bibfnamefont {Y.~H.}\ \bibnamefont {Jeong}},\ }\href {\doibase
  10.1063/1.1691491} {\bibfield  {journal} {\bibinfo  {journal} {Appl. Phys.
  Lett.}\ }\textbf {\bibinfo {volume} {84}},\ \bibinfo {pages} {2343} (\bibinfo
  {year} {2004})}\BibitemShut {NoStop}%
\bibitem [{\citenamefont {Bylander}\ \emph {et~al.}(2005)\citenamefont
  {Bylander}, \citenamefont {Duty},\ and\ \citenamefont
  {Delsing}}]{Bylander2005Mar}%
  \BibitemOpen
  \bibfield  {author} {\bibinfo {author} {\bibfnamefont {J.}~\bibnamefont
  {Bylander}}, \bibinfo {author} {\bibfnamefont {T.}~\bibnamefont {Duty}}, \
  and\ \bibinfo {author} {\bibfnamefont {P.}~\bibnamefont {Delsing}},\ }\href
  {\doibase 10.1038/nature03375} {\bibfield  {journal} {\bibinfo  {journal}
  {Nature}\ }\textbf {\bibinfo {volume} {434}},\ \bibinfo {pages} {361}
  (\bibinfo {year} {2005})}\BibitemShut {NoStop}%
\bibitem [{\citenamefont {Buehler}\ \emph {et~al.}(2005)\citenamefont
  {Buehler}, \citenamefont {Reilly}, \citenamefont {Starrett}, \citenamefont
  {Greentree}, \citenamefont {Hamilton}, \citenamefont {Dzurak},\ and\
  \citenamefont {Clark}}]{Buehler2005Apr}%
  \BibitemOpen
  \bibfield  {author} {\bibinfo {author} {\bibfnamefont {T.~M.}\ \bibnamefont
  {Buehler}}, \bibinfo {author} {\bibfnamefont {D.~J.}\ \bibnamefont {Reilly}},
  \bibinfo {author} {\bibfnamefont {R.~P.}\ \bibnamefont {Starrett}}, \bibinfo
  {author} {\bibfnamefont {A.~D.}\ \bibnamefont {Greentree}}, \bibinfo {author}
  {\bibfnamefont {A.~R.}\ \bibnamefont {Hamilton}}, \bibinfo {author}
  {\bibfnamefont {A.~S.}\ \bibnamefont {Dzurak}}, \ and\ \bibinfo {author}
  {\bibfnamefont {R.~G.}\ \bibnamefont {Clark}},\ }\href {\doibase
  10.1063/1.1897423} {\bibfield  {journal} {\bibinfo  {journal} {Appl. Phys.
  Lett.}\ }\textbf {\bibinfo {volume} {86}},\ \bibinfo {pages} {143117}
  (\bibinfo {year} {2005})}\BibitemShut {NoStop}%
\bibitem [{\citenamefont {Barthel}\ \emph {et~al.}(2010)\citenamefont
  {Barthel}, \citenamefont {Kj{\ae}rgaard}, \citenamefont {Medford},
  \citenamefont {Stopa}, \citenamefont {Marcus}, \citenamefont {Hanson},\ and\
  \citenamefont {Gossard}}]{Barthel2010Apr}%
  \BibitemOpen
  \bibfield  {author} {\bibinfo {author} {\bibfnamefont {C.}~\bibnamefont
  {Barthel}}, \bibinfo {author} {\bibfnamefont {M.}~\bibnamefont
  {Kj{\ae}rgaard}}, \bibinfo {author} {\bibfnamefont {J.}~\bibnamefont
  {Medford}}, \bibinfo {author} {\bibfnamefont {M.}~\bibnamefont {Stopa}},
  \bibinfo {author} {\bibfnamefont {C.~M.}\ \bibnamefont {Marcus}}, \bibinfo
  {author} {\bibfnamefont {M.~P.}\ \bibnamefont {Hanson}}, \ and\ \bibinfo
  {author} {\bibfnamefont {A.~C.}\ \bibnamefont {Gossard}},\ }\href {\doibase
  10.1103/PhysRevB.81.161308} {\bibfield  {journal} {\bibinfo  {journal} {Phys.
  Rev. B}\ }\textbf {\bibinfo {volume} {81}},\ \bibinfo {pages} {161308}
  (\bibinfo {year} {2010})}\BibitemShut {NoStop}%
\bibitem [{\citenamefont {Maisi}\ \emph {et~al.}(2011)\citenamefont {Maisi},
  \citenamefont {Saira}, \citenamefont {Pashkin}, \citenamefont {Tsai},
  \citenamefont {Averin},\ and\ \citenamefont {Pekola}}]{Maisi2011May}%
  \BibitemOpen
  \bibfield  {author} {\bibinfo {author} {\bibfnamefont {V.~F.}\ \bibnamefont
  {Maisi}}, \bibinfo {author} {\bibfnamefont {O.-P.}\ \bibnamefont {Saira}},
  \bibinfo {author} {\bibfnamefont {{\relax Yu}.~A.}\ \bibnamefont {Pashkin}},
  \bibinfo {author} {\bibfnamefont {J.~S.}\ \bibnamefont {Tsai}}, \bibinfo
  {author} {\bibfnamefont {D.~V.}\ \bibnamefont {Averin}}, \ and\ \bibinfo
  {author} {\bibfnamefont {J.~P.}\ \bibnamefont {Pekola}},\ }\href {\doibase
  10.1103/PhysRevLett.106.217003} {\bibfield  {journal} {\bibinfo  {journal}
  {Phys. Rev. Lett.}\ }\textbf {\bibinfo {volume} {106}},\ \bibinfo {pages}
  {217003} (\bibinfo {year} {2011})}\BibitemShut {NoStop}%
\bibitem [{\citenamefont {Field}\ \emph {et~al.}(1993)\citenamefont {Field},
  \citenamefont {Smith}, \citenamefont {Pepper}, \citenamefont {Ritchie},
  \citenamefont {Frost}, \citenamefont {Jones},\ and\ \citenamefont
  {Hasko}}]{Field1993Mar}%
  \BibitemOpen
  \bibfield  {author} {\bibinfo {author} {\bibfnamefont {M.}~\bibnamefont
  {Field}}, \bibinfo {author} {\bibfnamefont {C.~G.}\ \bibnamefont {Smith}},
  \bibinfo {author} {\bibfnamefont {M.}~\bibnamefont {Pepper}}, \bibinfo
  {author} {\bibfnamefont {D.~A.}\ \bibnamefont {Ritchie}}, \bibinfo {author}
  {\bibfnamefont {J.~E.~F.}\ \bibnamefont {Frost}}, \bibinfo {author}
  {\bibfnamefont {G.~A.~C.}\ \bibnamefont {Jones}}, \ and\ \bibinfo {author}
  {\bibfnamefont {D.~G.}\ \bibnamefont {Hasko}},\ }\href {\doibase
  10.1103/PhysRevLett.70.1311} {\bibfield  {journal} {\bibinfo  {journal}
  {Phys. Rev. Lett.}\ }\textbf {\bibinfo {volume} {70}},\ \bibinfo {pages}
  {1311} (\bibinfo {year} {1993})}\BibitemShut {NoStop}%
\bibitem [{\citenamefont {Elzerman}\ \emph {et~al.}(2003)\citenamefont
  {Elzerman}, \citenamefont {Hanson}, \citenamefont {Greidanus}, \citenamefont
  {Willems~van Beveren}, \citenamefont {De~Franceschi}, \citenamefont
  {Vandersypen}, \citenamefont {Tarucha},\ and\ \citenamefont
  {Kouwenhoven}}]{Elzerman2003Apr}%
  \BibitemOpen
  \bibfield  {author} {\bibinfo {author} {\bibfnamefont {J.~M.}\ \bibnamefont
  {Elzerman}}, \bibinfo {author} {\bibfnamefont {R.}~\bibnamefont {Hanson}},
  \bibinfo {author} {\bibfnamefont {J.~S.}\ \bibnamefont {Greidanus}}, \bibinfo
  {author} {\bibfnamefont {L.~H.}\ \bibnamefont {Willems~van Beveren}},
  \bibinfo {author} {\bibfnamefont {S.}~\bibnamefont {De~Franceschi}}, \bibinfo
  {author} {\bibfnamefont {L.~M.~K.}\ \bibnamefont {Vandersypen}}, \bibinfo
  {author} {\bibfnamefont {S.}~\bibnamefont {Tarucha}}, \ and\ \bibinfo
  {author} {\bibfnamefont {L.~P.}\ \bibnamefont {Kouwenhoven}},\ }\href
  {\doibase 10.1103/PhysRevB.67.161308} {\bibfield  {journal} {\bibinfo
  {journal} {Phys. Rev. B}\ }\textbf {\bibinfo {volume} {67}},\ \bibinfo
  {pages} {161308(R)} (\bibinfo {year} {2003})}\BibitemShut {NoStop}%
\bibitem [{\citenamefont {Elzerman}\ \emph {et~al.}(2004)\citenamefont
  {Elzerman}, \citenamefont {Hanson}, \citenamefont {Willems~van Beveren},
  \citenamefont {Witkamp}, \citenamefont {Vandersypen},\ and\ \citenamefont
  {Kouwenhoven}}]{Elzerman2004Jul}%
  \BibitemOpen
  \bibfield  {author} {\bibinfo {author} {\bibfnamefont {J.~M.}\ \bibnamefont
  {Elzerman}}, \bibinfo {author} {\bibfnamefont {R.}~\bibnamefont {Hanson}},
  \bibinfo {author} {\bibfnamefont {L.~H.}\ \bibnamefont {Willems~van
  Beveren}}, \bibinfo {author} {\bibfnamefont {B.}~\bibnamefont {Witkamp}},
  \bibinfo {author} {\bibfnamefont {L.~M.~K.}\ \bibnamefont {Vandersypen}}, \
  and\ \bibinfo {author} {\bibfnamefont {L.~P.}\ \bibnamefont {Kouwenhoven}},\
  }\href {\doibase 10.1038/nature02693} {\bibfield  {journal} {\bibinfo
  {journal} {Nature}\ }\textbf {\bibinfo {volume} {430}},\ \bibinfo {pages}
  {431} (\bibinfo {year} {2004})}\BibitemShut {NoStop}%
\bibitem [{\citenamefont {Ihn}\ \emph {et~al.}(2009)\citenamefont {Ihn},
  \citenamefont {Gustavsson}, \citenamefont {Gasser}, \citenamefont
  {K{\ifmmode\ddot{u}\else\"{u}\fi}ng}, \citenamefont
  {M{\ifmmode\ddot{u}\else\"{u}\fi}ller}, \citenamefont {Schleser},
  \citenamefont {Sigrist}, \citenamefont {Shorubalko}, \citenamefont
  {Leturcq},\ and\ \citenamefont {Ensslin}}]{Ihn2009Sep}%
  \BibitemOpen
  \bibfield  {author} {\bibinfo {author} {\bibfnamefont {T.}~\bibnamefont
  {Ihn}}, \bibinfo {author} {\bibfnamefont {S.}~\bibnamefont {Gustavsson}},
  \bibinfo {author} {\bibfnamefont {U.}~\bibnamefont {Gasser}}, \bibinfo
  {author} {\bibfnamefont {B.}~\bibnamefont
  {K{\ifmmode\ddot{u}\else\"{u}\fi}ng}}, \bibinfo {author} {\bibfnamefont
  {T.}~\bibnamefont {M{\ifmmode\ddot{u}\else\"{u}\fi}ller}}, \bibinfo {author}
  {\bibfnamefont {R.}~\bibnamefont {Schleser}}, \bibinfo {author}
  {\bibfnamefont {M.}~\bibnamefont {Sigrist}}, \bibinfo {author} {\bibfnamefont
  {I.}~\bibnamefont {Shorubalko}}, \bibinfo {author} {\bibfnamefont
  {R.}~\bibnamefont {Leturcq}}, \ and\ \bibinfo {author} {\bibfnamefont
  {K.}~\bibnamefont {Ensslin}},\ }\href {\doibase 10.1016/j.ssc.2009.04.040}
  {\bibfield  {journal} {\bibinfo  {journal} {Solid State Commun.}\ }\textbf
  {\bibinfo {volume} {149}},\ \bibinfo {pages} {1419} (\bibinfo {year}
  {2009})}\BibitemShut {NoStop}%
\bibitem [{\citenamefont {Barthel}\ \emph {et~al.}(2009)\citenamefont
  {Barthel}, \citenamefont {Reilly}, \citenamefont {Marcus}, \citenamefont
  {Hanson},\ and\ \citenamefont {Gossard}}]{Barthel2009Oct}%
  \BibitemOpen
  \bibfield  {author} {\bibinfo {author} {\bibfnamefont {C.}~\bibnamefont
  {Barthel}}, \bibinfo {author} {\bibfnamefont {D.~J.}\ \bibnamefont {Reilly}},
  \bibinfo {author} {\bibfnamefont {C.~M.}\ \bibnamefont {Marcus}}, \bibinfo
  {author} {\bibfnamefont {M.~P.}\ \bibnamefont {Hanson}}, \ and\ \bibinfo
  {author} {\bibfnamefont {A.~C.}\ \bibnamefont {Gossard}},\ }\href {\doibase
  10.1103/PhysRevLett.103.160503} {\bibfield  {journal} {\bibinfo  {journal}
  {Phys. Rev. Lett.}\ }\textbf {\bibinfo {volume} {103}},\ \bibinfo {pages}
  {160503} (\bibinfo {year} {2009})}\BibitemShut {NoStop}%
\bibitem [{\citenamefont {Zgirski}\ \emph {et~al.}(2011)\citenamefont
  {Zgirski}, \citenamefont {Bretheau}, \citenamefont {Le~Masne}, \citenamefont
  {Pothier}, \citenamefont {Esteve},\ and\ \citenamefont
  {Urbina}}]{Zgirski2011Jun}%
  \BibitemOpen
  \bibfield  {author} {\bibinfo {author} {\bibfnamefont {M.}~\bibnamefont
  {Zgirski}}, \bibinfo {author} {\bibfnamefont {L.}~\bibnamefont {Bretheau}},
  \bibinfo {author} {\bibfnamefont {Q.}~\bibnamefont {Le~Masne}}, \bibinfo
  {author} {\bibfnamefont {H.}~\bibnamefont {Pothier}}, \bibinfo {author}
  {\bibfnamefont {D.}~\bibnamefont {Esteve}}, \ and\ \bibinfo {author}
  {\bibfnamefont {C.}~\bibnamefont {Urbina}},\ }\href {\doibase
  10.1103/PhysRevLett.106.257003} {\bibfield  {journal} {\bibinfo  {journal}
  {Phys. Rev. Lett.}\ }\textbf {\bibinfo {volume} {106}},\ \bibinfo {pages}
  {257003} (\bibinfo {year} {2011})}\BibitemShut {NoStop}%
\bibitem [{\citenamefont {Janvier}\ \emph {et~al.}(2015)\citenamefont
  {Janvier}, \citenamefont {Tosi}, \citenamefont {Bretheau}, \citenamefont
  {Girit}, \citenamefont {Stern}, \citenamefont {Bertet}, \citenamefont
  {Joyez}, \citenamefont {Vion}, \citenamefont {Esteve}, \citenamefont
  {Goffman}, \citenamefont {Pothier},\ and\ \citenamefont
  {Urbina}}]{Janvier2015Sep}%
  \BibitemOpen
  \bibfield  {author} {\bibinfo {author} {\bibfnamefont {C.}~\bibnamefont
  {Janvier}}, \bibinfo {author} {\bibfnamefont {L.}~\bibnamefont {Tosi}},
  \bibinfo {author} {\bibfnamefont {L.}~\bibnamefont {Bretheau}}, \bibinfo
  {author} {\bibfnamefont
  {{\ifmmode\mbox{\c{C}}\else\c{C}\fi}.~{\ifmmode\ddot{O}\else\"{O}\fi}.}\
  \bibnamefont {Girit}}, \bibinfo {author} {\bibfnamefont {M.}~\bibnamefont
  {Stern}}, \bibinfo {author} {\bibfnamefont {P.}~\bibnamefont {Bertet}},
  \bibinfo {author} {\bibfnamefont {P.}~\bibnamefont {Joyez}}, \bibinfo
  {author} {\bibfnamefont {D.}~\bibnamefont {Vion}}, \bibinfo {author}
  {\bibfnamefont {D.}~\bibnamefont {Esteve}}, \bibinfo {author} {\bibfnamefont
  {M.~F.}\ \bibnamefont {Goffman}}, \bibinfo {author} {\bibfnamefont
  {H.}~\bibnamefont {Pothier}}, \ and\ \bibinfo {author} {\bibfnamefont
  {C.}~\bibnamefont {Urbina}},\ }\href {\doibase 10.1126/science.aab2179}
  {\bibfield  {journal} {\bibinfo  {journal} {Science}\ }\textbf {\bibinfo
  {volume} {349}},\ \bibinfo {pages} {1199} (\bibinfo {year}
  {2015})}\BibitemShut {NoStop}%
\bibitem [{\citenamefont {Higginbotham}\ \emph {et~al.}(2015)\citenamefont
  {Higginbotham}, \citenamefont {Albrecht}, \citenamefont
  {Kir{\ifmmode\check{s}\else\v{s}\fi}anskas}, \citenamefont {Chang},
  \citenamefont {Kuemmeth}, \citenamefont {Krogstrup}, \citenamefont
  {Jespersen}, \citenamefont {Nyg{\aa}rd}, \citenamefont {Flensberg},\ and\
  \citenamefont {Marcus}}]{Higginbotham2015Dec}%
  \BibitemOpen
  \bibfield  {author} {\bibinfo {author} {\bibfnamefont {A.~P.}\ \bibnamefont
  {Higginbotham}}, \bibinfo {author} {\bibfnamefont {S.~M.}\ \bibnamefont
  {Albrecht}}, \bibinfo {author} {\bibfnamefont {G.}~\bibnamefont
  {Kir{\ifmmode\check{s}\else\v{s}\fi}anskas}}, \bibinfo {author}
  {\bibfnamefont {W.}~\bibnamefont {Chang}}, \bibinfo {author} {\bibfnamefont
  {F.}~\bibnamefont {Kuemmeth}}, \bibinfo {author} {\bibfnamefont
  {P.}~\bibnamefont {Krogstrup}}, \bibinfo {author} {\bibfnamefont {T.~S.}\
  \bibnamefont {Jespersen}}, \bibinfo {author} {\bibfnamefont {J.}~\bibnamefont
  {Nyg{\aa}rd}}, \bibinfo {author} {\bibfnamefont {K.}~\bibnamefont
  {Flensberg}}, \ and\ \bibinfo {author} {\bibfnamefont {C.~M.}\ \bibnamefont
  {Marcus}},\ }\href {\doibase 10.1038/nphys3461} {\bibfield  {journal}
  {\bibinfo  {journal} {Nat. Phys.}\ }\textbf {\bibinfo {volume} {11}},\
  \bibinfo {pages} {1017} (\bibinfo {year} {2015})}\BibitemShut {NoStop}%
\bibitem [{\citenamefont {Hays}\ \emph {et~al.}(2018)\citenamefont {Hays},
  \citenamefont {de~Lange}, \citenamefont {Serniak}, \citenamefont {van
  Woerkom}, \citenamefont {Bouman}, \citenamefont {Krogstrup}, \citenamefont
  {Nyg{\aa}rd}, \citenamefont {Geresdi},\ and\ \citenamefont
  {Devoret}}]{Hays2018Jul}%
  \BibitemOpen
  \bibfield  {author} {\bibinfo {author} {\bibfnamefont {M.}~\bibnamefont
  {Hays}}, \bibinfo {author} {\bibfnamefont {G.}~\bibnamefont {de~Lange}},
  \bibinfo {author} {\bibfnamefont {K.}~\bibnamefont {Serniak}}, \bibinfo
  {author} {\bibfnamefont {D.~J.}\ \bibnamefont {van Woerkom}}, \bibinfo
  {author} {\bibfnamefont {D.}~\bibnamefont {Bouman}}, \bibinfo {author}
  {\bibfnamefont {P.}~\bibnamefont {Krogstrup}}, \bibinfo {author}
  {\bibfnamefont {J.}~\bibnamefont {Nyg{\aa}rd}}, \bibinfo {author}
  {\bibfnamefont {A.}~\bibnamefont {Geresdi}}, \ and\ \bibinfo {author}
  {\bibfnamefont {M.~H.}\ \bibnamefont {Devoret}},\ }\href {\doibase
  10.1103/PhysRevLett.121.047001} {\bibfield  {journal} {\bibinfo  {journal}
  {Phys. Rev. Lett.}\ }\textbf {\bibinfo {volume} {121}},\ \bibinfo {pages}
  {047001} (\bibinfo {year} {2018})}\BibitemShut {NoStop}%
\bibitem [{\citenamefont {Karzig}\ \emph {et~al.}(2021)\citenamefont {Karzig},
  \citenamefont {Cole},\ and\ \citenamefont {Pikulin}}]{Karzig2021Feb}%
  \BibitemOpen
  \bibfield  {author} {\bibinfo {author} {\bibfnamefont {T.}~\bibnamefont
  {Karzig}}, \bibinfo {author} {\bibfnamefont {W.~S.}\ \bibnamefont {Cole}}, \
  and\ \bibinfo {author} {\bibfnamefont {D.~I.}\ \bibnamefont {Pikulin}},\
  }\href {\doibase 10.1103/PhysRevLett.126.057702} {\bibfield  {journal}
  {\bibinfo  {journal} {Phys. Rev. Lett.}\ }\textbf {\bibinfo {volume} {126}},\
  \bibinfo {pages} {057702} (\bibinfo {year} {2021})}\BibitemShut {NoStop}%
\bibitem [{\citenamefont {Wesdorp}\ \emph {et~al.}(2021)\citenamefont
  {Wesdorp}, \citenamefont {Gr{\ifmmode\ddot{u}\else\"{u}\fi}nhaupt},
  \citenamefont {Vaartjes}, \citenamefont {Pita-Vidal}, \citenamefont
  {Bargerbos}, \citenamefont {Splitthoff}, \citenamefont {Krogstrup},
  \citenamefont {van Heck},\ and\ \citenamefont {de~Lange}}]{Wesdorp2021Dec}%
  \BibitemOpen
  \bibfield  {author} {\bibinfo {author} {\bibfnamefont {J.~J.}\ \bibnamefont
  {Wesdorp}}, \bibinfo {author} {\bibfnamefont {L.}~\bibnamefont
  {Gr{\ifmmode\ddot{u}\else\"{u}\fi}nhaupt}}, \bibinfo {author} {\bibfnamefont
  {A.}~\bibnamefont {Vaartjes}}, \bibinfo {author} {\bibfnamefont
  {M.}~\bibnamefont {Pita-Vidal}}, \bibinfo {author} {\bibfnamefont
  {A.}~\bibnamefont {Bargerbos}}, \bibinfo {author} {\bibfnamefont {L.~J.}\
  \bibnamefont {Splitthoff}}, \bibinfo {author} {\bibfnamefont
  {P.}~\bibnamefont {Krogstrup}}, \bibinfo {author} {\bibfnamefont
  {B.}~\bibnamefont {van Heck}}, \ and\ \bibinfo {author} {\bibfnamefont
  {G.}~\bibnamefont {de~Lange}},\ }\href
  {https://doi.org/10.48550/arXiv.2112.01936} {\bibfield  {journal} {\bibinfo
  {journal} {arXiv}\ } (\bibinfo {year} {2021})},\ \Eprint
  {http://arxiv.org/abs/2112.01936} {2112.01936} \BibitemShut {NoStop}%
\bibitem [{\citenamefont {Razmadze}\ \emph {et~al.}(2019)\citenamefont
  {Razmadze}, \citenamefont {Sabonis}, \citenamefont {Malinowski},
  \citenamefont {M{\ifmmode\acute{e}\else\'{e}\fi}nard}, \citenamefont {Pauka},
  \citenamefont {Nguyen}, \citenamefont {van Zanten}, \citenamefont
  {O'Farrell}, \citenamefont {Suter}, \citenamefont {Krogstrup}, \citenamefont
  {Kuemmeth},\ and\ \citenamefont {Marcus}}]{Razmadze2019Jun}%
  \BibitemOpen
  \bibfield  {author} {\bibinfo {author} {\bibfnamefont {D.}~\bibnamefont
  {Razmadze}}, \bibinfo {author} {\bibfnamefont {D.}~\bibnamefont {Sabonis}},
  \bibinfo {author} {\bibfnamefont {F.~K.}\ \bibnamefont {Malinowski}},
  \bibinfo {author} {\bibfnamefont {G.~C.}\ \bibnamefont
  {M{\ifmmode\acute{e}\else\'{e}\fi}nard}}, \bibinfo {author} {\bibfnamefont
  {S.}~\bibnamefont {Pauka}}, \bibinfo {author} {\bibfnamefont
  {H.}~\bibnamefont {Nguyen}}, \bibinfo {author} {\bibfnamefont {D.~M.~T.}\
  \bibnamefont {van Zanten}}, \bibinfo {author} {\bibfnamefont {E.~C.~T.}\
  \bibnamefont {O'Farrell}}, \bibinfo {author} {\bibfnamefont {J.}~\bibnamefont
  {Suter}}, \bibinfo {author} {\bibfnamefont {P.}~\bibnamefont {Krogstrup}},
  \bibinfo {author} {\bibfnamefont {F.}~\bibnamefont {Kuemmeth}}, \ and\
  \bibinfo {author} {\bibfnamefont {C.~M.}\ \bibnamefont {Marcus}},\ }\href
  {\doibase 10.1103/PhysRevApplied.11.064011} {\bibfield  {journal} {\bibinfo
  {journal} {Phys. Rev. Appl.}\ }\textbf {\bibinfo {volume} {11}},\ \bibinfo
  {pages} {064011} (\bibinfo {year} {2019})}\BibitemShut {NoStop}%
\bibitem [{\citenamefont {Kir{\ifmmode\check{s}\else\v{s}\fi}anskas}\ \emph
  {et~al.}(2018)\citenamefont {Kir{\ifmmode\check{s}\else\v{s}\fi}anskas},
  \citenamefont {Francki{\ifmmode\acute{e}\else\'{e}\fi}},\ and\ \citenamefont
  {Wacker}}]{Kirsanskas2018Jan}%
  \BibitemOpen
  \bibfield  {author} {\bibinfo {author} {\bibfnamefont {G.}~\bibnamefont
  {Kir{\ifmmode\check{s}\else\v{s}\fi}anskas}}, \bibinfo {author}
  {\bibfnamefont {M.}~\bibnamefont {Francki{\ifmmode\acute{e}\else\'{e}\fi}}},
  \ and\ \bibinfo {author} {\bibfnamefont {A.}~\bibnamefont {Wacker}},\ }\href
  {\doibase 10.1103/PhysRevB.97.035432} {\bibfield  {journal} {\bibinfo
  {journal} {Phys. Rev. B}\ }\textbf {\bibinfo {volume} {97}},\ \bibinfo
  {pages} {035432} (\bibinfo {year} {2018})}\BibitemShut {NoStop}%
\bibitem [{\citenamefont {Nathan}\ and\ \citenamefont
  {Rudner}(2020)}]{Nathan2020Sep}%
  \BibitemOpen
  \bibfield  {author} {\bibinfo {author} {\bibfnamefont {F.}~\bibnamefont
  {Nathan}}\ and\ \bibinfo {author} {\bibfnamefont {M.~S.}\ \bibnamefont
  {Rudner}},\ }\href {\doibase 10.1103/PhysRevB.102.115109} {\bibfield
  {journal} {\bibinfo  {journal} {Phys. Rev. B}\ }\textbf {\bibinfo {volume}
  {102}},\ \bibinfo {pages} {115109} (\bibinfo {year} {2020})}\BibitemShut
  {NoStop}%
\bibitem [{\citenamefont {M{\o}lmer}\ \emph {et~al.}(1993)\citenamefont
  {M{\o}lmer}, \citenamefont {Castin},\ and\ \citenamefont
  {Dalibard}}]{Molmer1993Mar}%
  \BibitemOpen
  \bibfield  {author} {\bibinfo {author} {\bibfnamefont {K.}~\bibnamefont
  {M{\o}lmer}}, \bibinfo {author} {\bibfnamefont {Y.}~\bibnamefont {Castin}}, \
  and\ \bibinfo {author} {\bibfnamefont {J.}~\bibnamefont {Dalibard}},\ }\href
  {\doibase 10.1364/JOSAB.10.000524} {\bibfield  {journal} {\bibinfo  {journal}
  {J. Opt. Soc. Am. B}\ }\textbf {\bibinfo {volume} {10}},\ \bibinfo {pages}
  {524} (\bibinfo {year} {1993})}\BibitemShut {NoStop}%
\bibitem [{\citenamefont {Plenio}\ and\ \citenamefont
  {Knight}(1998)}]{Plenio1998Jan}%
  \BibitemOpen
  \bibfield  {author} {\bibinfo {author} {\bibfnamefont {M.~B.}\ \bibnamefont
  {Plenio}}\ and\ \bibinfo {author} {\bibfnamefont {P.~L.}\ \bibnamefont
  {Knight}},\ }\href {\doibase 10.1103/RevModPhys.70.101} {\bibfield  {journal}
  {\bibinfo  {journal} {Rev. Mod. Phys.}\ }\textbf {\bibinfo {volume} {70}},\
  \bibinfo {pages} {101} (\bibinfo {year} {1998})}\BibitemShut {NoStop}%
\bibitem [{\citenamefont {Daley}(2014)}]{Daley2014Mar}%
  \BibitemOpen
  \bibfield  {author} {\bibinfo {author} {\bibfnamefont {A.~J.}\ \bibnamefont
  {Daley}},\ }\href {\doibase 10.1080/00018732.2014.933502} {\bibfield
  {journal} {\bibinfo  {journal} {Adv. Phys.}\ }\textbf {\bibinfo {volume}
  {63}},\ \bibinfo {pages} {77} (\bibinfo {year} {2014})}\BibitemShut {NoStop}%
\bibitem [{Sch()}]{Schulenborg2022Suppmat}%
  \BibitemOpen
  \href@noop {} {}\bibinfo {note} {See supplementary material.}\BibitemShut
  {Stop}%
\bibitem [{\citenamefont {Derakhshan~Maman}\ \emph {et~al.}(2020)\citenamefont
  {Derakhshan~Maman}, \citenamefont {Gonzalez-Zalba},\ and\ \citenamefont
  {P{\ifmmode\acute{a}\else\'{a}\fi}lyi}}]{DerakhshanMaman2020Dec}%
  \BibitemOpen
  \bibfield  {author} {\bibinfo {author} {\bibfnamefont {V.}~\bibnamefont
  {Derakhshan~Maman}}, \bibinfo {author} {\bibfnamefont {M.~F.}\ \bibnamefont
  {Gonzalez-Zalba}}, \ and\ \bibinfo {author} {\bibfnamefont {A.}~\bibnamefont
  {P{\ifmmode\acute{a}\else\'{a}\fi}lyi}},\ }\href {\doibase
  10.1103/PhysRevApplied.14.064024} {\bibfield  {journal} {\bibinfo  {journal}
  {Phys. Rev. Appl.}\ }\textbf {\bibinfo {volume} {14}},\ \bibinfo {pages}
  {064024} (\bibinfo {year} {2020})}\BibitemShut {NoStop}%
\bibitem [{\citenamefont {Lutchyn}\ \emph {et~al.}(2010)\citenamefont
  {Lutchyn}, \citenamefont {Sau},\ and\ \citenamefont
  {Das~Sarma}}]{Lutchyn2010Aug}%
  \BibitemOpen
  \bibfield  {author} {\bibinfo {author} {\bibfnamefont {R.~M.}\ \bibnamefont
  {Lutchyn}}, \bibinfo {author} {\bibfnamefont {J.~D.}\ \bibnamefont {Sau}}, \
  and\ \bibinfo {author} {\bibfnamefont {S.}~\bibnamefont {Das~Sarma}},\ }\href
  {\doibase 10.1103/PhysRevLett.105.077001} {\bibfield  {journal} {\bibinfo
  {journal} {Phys. Rev. Lett.}\ }\textbf {\bibinfo {volume} {105}},\ \bibinfo
  {pages} {077001} (\bibinfo {year} {2010})}\BibitemShut {NoStop}%
\bibitem [{\citenamefont {Oreg}\ \emph {et~al.}(2010)\citenamefont {Oreg},
  \citenamefont {Refael},\ and\ \citenamefont {von Oppen}}]{Oreg2010Oct}%
  \BibitemOpen
  \bibfield  {author} {\bibinfo {author} {\bibfnamefont {Y.}~\bibnamefont
  {Oreg}}, \bibinfo {author} {\bibfnamefont {G.}~\bibnamefont {Refael}}, \ and\
  \bibinfo {author} {\bibfnamefont {F.}~\bibnamefont {von Oppen}},\ }\href
  {\doibase 10.1103/PhysRevLett.105.177002} {\bibfield  {journal} {\bibinfo
  {journal} {Phys. Rev. Lett.}\ }\textbf {\bibinfo {volume} {105}},\ \bibinfo
  {pages} {177002} (\bibinfo {year} {2010})}\BibitemShut {NoStop}%
\bibitem [{\citenamefont {Sticlet}\ \emph {et~al.}(2012)\citenamefont
  {Sticlet}, \citenamefont {Bena},\ and\ \citenamefont
  {Simon}}]{Sticlet2012Mar}%
  \BibitemOpen
  \bibfield  {author} {\bibinfo {author} {\bibfnamefont {D.}~\bibnamefont
  {Sticlet}}, \bibinfo {author} {\bibfnamefont {C.}~\bibnamefont {Bena}}, \
  and\ \bibinfo {author} {\bibfnamefont {P.}~\bibnamefont {Simon}},\ }\href
  {\doibase 10.1103/PhysRevLett.108.096802} {\bibfield  {journal} {\bibinfo
  {journal} {Phys. Rev. Lett.}\ }\textbf {\bibinfo {volume} {108}},\ \bibinfo
  {pages} {096802} (\bibinfo {year} {2012})}\BibitemShut {NoStop}%
\bibitem [{\citenamefont {Kjaergaard}\ \emph {et~al.}(2012)\citenamefont
  {Kjaergaard}, \citenamefont {W{\ifmmode\ddot{o}\else\"{o}\fi}lms},\ and\
  \citenamefont {Flensberg}}]{Kjaergaard2012Jan}%
  \BibitemOpen
  \bibfield  {author} {\bibinfo {author} {\bibfnamefont {M.}~\bibnamefont
  {Kjaergaard}}, \bibinfo {author} {\bibfnamefont {K.}~\bibnamefont
  {W{\ifmmode\ddot{o}\else\"{o}\fi}lms}}, \ and\ \bibinfo {author}
  {\bibfnamefont {K.}~\bibnamefont {Flensberg}},\ }\href {\doibase
  10.1103/PhysRevB.85.020503} {\bibfield  {journal} {\bibinfo  {journal} {Phys.
  Rev. B}\ }\textbf {\bibinfo {volume} {85}},\ \bibinfo {pages} {020503}
  (\bibinfo {year} {2012})}\BibitemShut {NoStop}%
\bibitem [{\citenamefont {Jauho}\ \emph {et~al.}(1994)\citenamefont {Jauho},
  \citenamefont {Wingreen},\ and\ \citenamefont {Meir}}]{Jauho1994Aug}%
  \BibitemOpen
  \bibfield  {author} {\bibinfo {author} {\bibfnamefont {A.-P.}\ \bibnamefont
  {Jauho}}, \bibinfo {author} {\bibfnamefont {N.~S.}\ \bibnamefont {Wingreen}},
  \ and\ \bibinfo {author} {\bibfnamefont {Y.}~\bibnamefont {Meir}},\ }\href
  {\doibase 10.1103/PhysRevB.50.5528} {\bibfield  {journal} {\bibinfo
  {journal} {Phys. Rev. B}\ }\textbf {\bibinfo {volume} {50}},\ \bibinfo
  {pages} {5528} (\bibinfo {year} {1994})}\BibitemShut {NoStop}%
\bibitem [{\citenamefont {Clarke}(2017)}]{Clarke2017Nov}%
  \BibitemOpen
  \bibfield  {author} {\bibinfo {author} {\bibfnamefont {D.~J.}\ \bibnamefont
  {Clarke}},\ }\href {\doibase 10.1103/PhysRevB.96.201109} {\bibfield
  {journal} {\bibinfo  {journal} {Phys. Rev. B}\ }\textbf {\bibinfo {volume}
  {96}},\ \bibinfo {pages} {201109} (\bibinfo {year} {2017})}\BibitemShut
  {NoStop}%
\bibitem [{\citenamefont {Prada}\ \emph {et~al.}(2017)\citenamefont {Prada},
  \citenamefont {Aguado},\ and\ \citenamefont {San-Jose}}]{Prada2017Aug}%
  \BibitemOpen
  \bibfield  {author} {\bibinfo {author} {\bibfnamefont {E.}~\bibnamefont
  {Prada}}, \bibinfo {author} {\bibfnamefont {R.}~\bibnamefont {Aguado}}, \
  and\ \bibinfo {author} {\bibfnamefont {P.}~\bibnamefont {San-Jose}},\ }\href
  {\doibase 10.1103/PhysRevB.96.085418} {\bibfield  {journal} {\bibinfo
  {journal} {Phys. Rev. B}\ }\textbf {\bibinfo {volume} {96}},\ \bibinfo
  {pages} {085418} (\bibinfo {year} {2017})}\BibitemShut {NoStop}%
\bibitem [{\citenamefont {Knapp}\ \emph {et~al.}(2018)\citenamefont {Knapp},
  \citenamefont {Karzig}, \citenamefont {Lutchyn},\ and\ \citenamefont
  {Nayak}}]{Knapp2018Mar}%
  \BibitemOpen
  \bibfield  {author} {\bibinfo {author} {\bibfnamefont {C.}~\bibnamefont
  {Knapp}}, \bibinfo {author} {\bibfnamefont {T.}~\bibnamefont {Karzig}},
  \bibinfo {author} {\bibfnamefont {R.~M.}\ \bibnamefont {Lutchyn}}, \ and\
  \bibinfo {author} {\bibfnamefont {C.}~\bibnamefont {Nayak}},\ }\href
  {\doibase 10.1103/PhysRevB.97.125404} {\bibfield  {journal} {\bibinfo
  {journal} {Phys. Rev. B}\ }\textbf {\bibinfo {volume} {97}},\ \bibinfo
  {pages} {125404} (\bibinfo {year} {2018})}\BibitemShut {NoStop}%
\bibitem [{\citenamefont {Mishmash}\ \emph {et~al.}(2020)\citenamefont
  {Mishmash}, \citenamefont {Bauer}, \citenamefont {von Oppen},\ and\
  \citenamefont {Alicea}}]{Mishmash2020Feb}%
  \BibitemOpen
  \bibfield  {author} {\bibinfo {author} {\bibfnamefont {R.~V.}\ \bibnamefont
  {Mishmash}}, \bibinfo {author} {\bibfnamefont {B.}~\bibnamefont {Bauer}},
  \bibinfo {author} {\bibfnamefont {F.}~\bibnamefont {von Oppen}}, \ and\
  \bibinfo {author} {\bibfnamefont {J.}~\bibnamefont {Alicea}},\ }\href
  {\doibase 10.1103/PhysRevB.101.075404} {\bibfield  {journal} {\bibinfo
  {journal} {Phys. Rev. B}\ }\textbf {\bibinfo {volume} {101}},\ \bibinfo
  {pages} {075404} (\bibinfo {year} {2020})}\BibitemShut {NoStop}%
\bibitem [{\citenamefont {Khindanov}\ \emph {et~al.}(2021)\citenamefont
  {Khindanov}, \citenamefont {Pikulin},\ and\ \citenamefont
  {Karzig}}]{Khindanov2021Jun}%
  \BibitemOpen
  \bibfield  {author} {\bibinfo {author} {\bibfnamefont {A.}~\bibnamefont
  {Khindanov}}, \bibinfo {author} {\bibfnamefont {D.}~\bibnamefont {Pikulin}},
  \ and\ \bibinfo {author} {\bibfnamefont {T.}~\bibnamefont {Karzig}},\ }\href
  {\doibase 10.21468/SciPostPhys.10.6.127} {\bibfield  {journal} {\bibinfo
  {journal} {SciPost Phys.}\ }\textbf {\bibinfo {volume} {10}},\ \bibinfo
  {pages} {127} (\bibinfo {year} {2021})}\BibitemShut {NoStop}%
\bibitem [{\citenamefont {Leijnse}\ and\ \citenamefont
  {Flensberg}(2012)}]{Leijnse2012Oct}%
  \BibitemOpen
  \bibfield  {author} {\bibinfo {author} {\bibfnamefont {M.}~\bibnamefont
  {Leijnse}}\ and\ \bibinfo {author} {\bibfnamefont {K.}~\bibnamefont
  {Flensberg}},\ }\href {\doibase 10.1103/PhysRevB.86.134528} {\bibfield
  {journal} {\bibinfo  {journal} {Phys. Rev. B}\ }\textbf {\bibinfo {volume}
  {86}},\ \bibinfo {pages} {134528} (\bibinfo {year} {2012})}\BibitemShut
  {NoStop}%
\bibitem [{\citenamefont {Dvir}\ \emph {et~al.}(2023)\citenamefont {Dvir},
  \citenamefont {Wang}, \citenamefont {van Loo}, \citenamefont {Liu},
  \citenamefont {Mazur}, \citenamefont {Bordin}, \citenamefont {Ten~Haaf},
  \citenamefont {Wang}, \citenamefont {van Driel}, \citenamefont {Zatelli},
  \citenamefont {Li}, \citenamefont {Malinowski}, \citenamefont {Gazibegovic},
  \citenamefont {Badawy}, \citenamefont {Bakkers}, \citenamefont {Wimmer},\
  and\ \citenamefont {Kouwenhoven}}]{Dvir2023Feb}%
  \BibitemOpen
  \bibfield  {author} {\bibinfo {author} {\bibfnamefont {T.}~\bibnamefont
  {Dvir}}, \bibinfo {author} {\bibfnamefont {G.}~\bibnamefont {Wang}}, \bibinfo
  {author} {\bibfnamefont {N.}~\bibnamefont {van Loo}}, \bibinfo {author}
  {\bibfnamefont {C.-X.}\ \bibnamefont {Liu}}, \bibinfo {author} {\bibfnamefont
  {G.~P.}\ \bibnamefont {Mazur}}, \bibinfo {author} {\bibfnamefont
  {A.}~\bibnamefont {Bordin}}, \bibinfo {author} {\bibfnamefont {S.~L.~D.}\
  \bibnamefont {Ten~Haaf}}, \bibinfo {author} {\bibfnamefont {J.-Y.}\
  \bibnamefont {Wang}}, \bibinfo {author} {\bibfnamefont {D.}~\bibnamefont {van
  Driel}}, \bibinfo {author} {\bibfnamefont {F.}~\bibnamefont {Zatelli}},
  \bibinfo {author} {\bibfnamefont {X.}~\bibnamefont {Li}}, \bibinfo {author}
  {\bibfnamefont {F.~K.}\ \bibnamefont {Malinowski}}, \bibinfo {author}
  {\bibfnamefont {S.}~\bibnamefont {Gazibegovic}}, \bibinfo {author}
  {\bibfnamefont {G.}~\bibnamefont {Badawy}}, \bibinfo {author} {\bibfnamefont
  {E.~P. A.~M.}\ \bibnamefont {Bakkers}}, \bibinfo {author} {\bibfnamefont
  {M.}~\bibnamefont {Wimmer}}, \ and\ \bibinfo {author} {\bibfnamefont {L.~P.}\
  \bibnamefont {Kouwenhoven}},\ }\href {\doibase 10.1038/s41586-022-05585-1}
  {\bibfield  {journal} {\bibinfo  {journal} {Nature}\ }\textbf {\bibinfo
  {volume} {614}},\ \bibinfo {pages} {445} (\bibinfo {year}
  {2023})}\BibitemShut {NoStop}%
\bibitem [{\citenamefont {Nitsch}\ \emph {et~al.}(2022)\citenamefont {Nitsch},
  \citenamefont {Seoane~Souto},\ and\ \citenamefont {Leijnse}}]{Nitsch2022Nov}%
  \BibitemOpen
  \bibfield  {author} {\bibinfo {author} {\bibfnamefont {M.}~\bibnamefont
  {Nitsch}}, \bibinfo {author} {\bibfnamefont {R.}~\bibnamefont
  {Seoane~Souto}}, \ and\ \bibinfo {author} {\bibfnamefont {M.}~\bibnamefont
  {Leijnse}},\ }\href {\doibase 10.1103/PhysRevB.106.L201305} {\bibfield
  {journal} {\bibinfo  {journal} {Phys. Rev. B}\ }\textbf {\bibinfo {volume}
  {106}},\ \bibinfo {pages} {L201305} (\bibinfo {year} {2022})}\BibitemShut
  {NoStop}%
\end{thebibliography}
\end{document}